\documentclass[prc]{revtex4}
\usepackage{graphicx}
\begin{document}

\newcommand\ie {{\it i.e.}}
\newcommand\eg {{\it e.g.}}
\newcommand\etc{{\it etc.}}
\newcommand\cf {{\it cf.}}
\newcommand\etal {{\it et al.}}
\newcommand{\be}{\begin{eqnarray}}
\newcommand{\ee}{\end{eqnarray}}
\newcommand{\jp}{$ J/ \psi $}
\newcommand{\pp}{$ \psi^{ \prime} $}
\newcommand{\ppp}{$ \psi^{ \prime \prime } $}
\newcommand{\dd}[2]{$ #1 \overline #2 $}
\newcommand\noi {\noindent}

\title{Energy Dependence of Intrinsic Charm Production: \\
  What is the Best Energy for Observation?}

\author{R. Vogt}
\affiliation
{Nuclear and Chemical Sciences Division,
Lawrence Livermore National Laboratory, Livermore, CA 94551,
USA}
\affiliation
    {Department of Physics and Astronomy,
University of California, Davis, CA 95616,
USA}

\begin{abstract}
  {\bf Background:} A nonperturbative charm production contribution, known as
  intrinsic charm, was predicted in the early 1980s.
  Recent results have provided new
  evidence for its existence but further confirmation is needed.
  {\bf Purpose:} $J/\psi$ and $\overline D$ meson production are calculated
  with a combination of perturbative QCD and intrinsic charm to determine the
  best energy range to study intrinsic charm production.
  {\bf Methods:} $J/\psi$ and $\overline D$
  meson production are calculated in perturbative QCD to
  next-to-leading order in the cross section.   Cold nuclear matter effects,
  including nuclear modification of the parton densities and $p_T$ broadening
  by multiple scattering are taken into account in the production of both;
  absorption by nucleons is also included for the $J/\psi$.  Contributions from
  intrinsic charm are calculated assuming production from a
  $|uud c \overline c \rangle$ Fock state.
  {\bf Results:}  The $J/\psi$ and $\overline D$ meson rapidity and $p_T$
  distributions are calculated as a function of rapidity and transverse momentum
  $p_T$ over a wide range of center-of-mass energies with and without
  intrinsic charm in $p+p$ collisions.  The nuclear modification factor,
  $R_{pA}$, is also calculated for $p+{\rm Pb}$ interactions at appropriate
  energies.  Previous fixed-target data as a function of Feynman $x$, $x_F$,
  are also compared to calculations within the approach.  Good agreement with
  the data is found when intrinsic charm is included.
  {\bf Conclusions:}  The intrinsic charm signal may be largest at midrapidity
  for future low energy fixed target experiments such as the proposed NA60+.
\end{abstract}
\maketitle

\section{Introduction}

Heavy quark production, both of charmonium, $J/\psi$, and charm mesons such
as $D$ and $\overline D$, has been studied in a variety of environments, from
more elementary collisions such as $p+p$ and $p+\overline p$, to protons on
nuclei, $p+A$, and nucleus-nucleus collisions, $A+A$.  Many
earlier studies of $J/\psi$ production in $p+A$ collisions were made in
fixed-target configurations
\cite{NA3,E537,E772,E789,NA50,NA60,e866,herab1,herab2,SeaQuest,Ayuso}.
More recently production has predominantly been studied at hadron and nuclear
colliders, in particular at
the Relativistic Heavy-Ion Collider at Brookhaven \cite{PHENIX,STAR}
and the Large Hadron Collider at CERN \cite{ALICE,LHCb,CMS,ATLAS}.
Charm hadron production has also been studied, both at fixed-target energies
\cite{E769,E791,WA82}, where leading charm and charm hadron asymmetries in
particular were studied, and collider energies
\cite{PHENIX:2013txu,Kramarik:2020tvk,ALICEDmesonsNew,LHCbDmesonsmod}.
Several new fixed-target experiments that would include studies of heavy
flavor production
have been proposed \cite{NA60+,AFTER} or have taken data \cite{SMOG}. Some of
these experiments
would utilize the LHC for fixed-target studies \cite{AFTER,SMOG}.  The
NA60+ collaboration \cite{NA60+} has proposed making use of some of the lowest
beam energies applied to $J/\psi$ and open charm production: $p_{\rm lab} = 40$,
80 and 120~GeV.  The SMOG gas-jet target in the LHCb detector allows for
fixed-target heavy flavor studies with $\sqrt{s} = 69$, 87.7 and 110.4~GeV,
above the center of mass energies in previous fixed-target experiments
\cite{NA3,E537,E772,E789,NA50,NA60,e866,herab1,herab2,SeaQuest,Ayuso}
but below the top energy of RHIC, $\sqrt{s} = 200$~GeV.  

Several of the previous experiments as well as the new proposed
fixed-target studies have made or plan to make analyses of their data to look
for evidence of intrinsic charm production \cite{intc1,intc2,BandH}.  This
intrinsic charm should manifest itself at forward Feynman $x$, $x_F$, because
the heavy quarks carry a larger fraction of the projectile momentum than do the
light partons in these states.  In the past there have been some
tantalizing hints \cite{NA3,EMC,ISR} of intrinsic charm without any concrete
evidence.  However, recently LHCb observed an excess of $Z+c$-jets over $Z$+jets
alone at forward rapidity \cite{LHCb_intc}.  These data agree with a 1\%
intrinsic charm contribution 
in the proton.  On the other hand, an LHCb analysis with the SMOG apparatus
claimed to see no evidence of $J/\psi$ or $D^0$ production by intrinsic charm
\cite{SMOG}.

Given these seemingly contradictory results, as well as the large energy range
of past, present and future heavy flavor studies, it is worth
making an assessment of which energies and kinematic ranges would maximize the
intrinsic charm signal.  Therefore, proton-proton interactions over
the fixed target range from NA60+ and above, $p_{\rm lab} = 40$, 80, 120, 158,
450, and 800~GeV, are studied here, along with the SMOG fixed-target energies,
$\sqrt{s} = 69$, 87.7 and 110.4~GeV.  Collider energies at RHIC and the LHC are
also included with $\sqrt{s} = 200$, 500, 5000, 7000, and 13000~GeV.  The
fixed-target energies are studied in the central rapidity
region, $0 < y < 1$.  At collider energies, the focus is shifted to forward
coverage for RHIC
and the LHC, $1.1 < y < 2.2$ and $2.5 < y < 5$, respectively.
The influence of intrinsic charm on proton-nucleus collisions is assessed
with a lead target assumed at all energies for consistency.  

These calculations follow those for the $J/\psi$ recently made for the SeaQuest
Collaboration in Ref.~\cite{RV_SeaQuest}.  In that paper, the $J/\psi$ cross
section was given as
the sum of the perturbative QCD contribution, calculated in the
color evaporation model (CEM) with cold nuclear matter effects (modifications
of the parton densities in nuclei, intrinsic $k_T$ broadening, and absorption
by nucleons) included and intrinsic charm.  The intrinsic charm dependence on
the nuclear mass
was assumed to be surface-like, similar to an $A^{2/3}$ dependence \cite{BandH}.
The actual nuclear dependence employed
for intrinsic charm was $A^{0.71}$ as extracted by the NA3 Collaboration
\cite{NA3} for ``diffractive'' $J/\psi$ production at forward $x_F$ with a
proton projectile.  The $J/\psi$
calculations presented here follow this analysis \cite{RV_SeaQuest}.  The
$\overline D$ meson
calculations in this work are also a combination of perturbative QCD
production, this time of
open charm, with colid nuclear matter effects (nuclear modification of the
parton densities and intrinsic $k_T$
broadening), and production by intrinsic charm.
Rather than calculate the cold nuclear matter effects using
a $A^\alpha$ dependence to average over all nuclear target effects, as was done
in the past for charm production from multiple nuclear targets, the nuclear
suppression factor $R_{pA}$, the ratio of the per-nucleon cross section in
$p+A$ relative to $p+p$ collisions, is shown here for both $J/\psi$ and
$\overline D$ production.

The calculation of $J/\psi$ and $\overline D$ meson
production in perturbative QCD is presented in Sec.~\ref{pQCD}.
The cold nuclear matter effects included in the calculation are
introduced in Sec.~\ref{CNM}.
The rapidity and $p_T$ distributions arising from intrinsic charm production
are presented in Sec.~\ref{ICcomp}.
Section~\ref{model_comp} presents first the $p+p$ distributions over the
full energy range as a function of $y$ and $p_T$, followed by results for
modifications of $J/\psi$ and $\overline D$ production in nuclei.
The cold nuclear matter results are presented for a selected
subset of the energies considered: $p_{\rm lab} = 40$, 158, and 800~GeV and
$\sqrt{s} = 87.7$, 200 and 5000~TeV.  The $J/\psi$ results are also compared to
previous fixed-target data on the exponent $\alpha$ as a function of $x_F$.
The conclusions are presented in Sec.~\ref{conclusions}.

\section{Open Charm and $J/\psi$ Production in Perturbative QCD}
\label{pQCD}

This section describes open heavy flavor ($\overline D$ meson) and $J/\psi$
production in $p+p$ collisions in perturbative QCD.

\subsection{Open Charm Production}

There are currently two approaches to single inclusive
open heavy flavor production at
colliders:  collinear factorization \cite{FONLL,GM-VFN,Helenius:2018uul}
and $k_T$ factorization \cite{Szczurek}.  The latter approach is limited to
high energies and thus low parton momentum fractions $x$.  
The LHC data has been compared to calculations in both approaches.
Those assuming collinear factorization compare well with the LHC data.
The ALICE data \cite{ALICEDmesonsNew}, 
at $\sqrt{s} = 7$~TeV at central rapidity for $0 < p_T < 2$~GeV supports
collinear factorization.  Similarly,
the forward rapidity data of LHCb at 7 TeV \cite{LHCbDmesons}
and 13 TeV \cite{Aaij:2015bpa}, also agree well with the collinear factorization
assumption.   The $k_T$ factorization approach is incompatible with most
of the energies discussed in this work because the fixed-target regime is not
at sufficiently low $x$ for it to be applicable.
This paper will thus employ collinear factorization in all the
calculations.

Because single 
inclusive calculations cannot address $Q \overline Q$ pair
observables, an exclusive $Q \overline Q$ pair production code is used for
both $J/\psi$ and $\overline D$ production to ensure consistency. 
The HVQMNR code \cite{MNRcode} is employed in these calculations.
It includes an option to smear the parton
momentum through the introduction of intrinsic transverse momenta, $k_T$, as
described later in this section.  The open heavy flavor production cross
section calculation is described first, followed by that for
$J/\psi$ production.

The perturbative open heavy flavor (OHF) cross section can be schematically
represented as
\be
\sigma_{\rm OHF}(pp) = \sum_{i,j} 
\int_{4m^2}^{\infty} d\hat{s}
\int dx_1 \, dx_2~ F_i^p(x_1,\mu_F^2,k_{T_1})~ F_j^p(x_2,\mu_F^2,k_{T_2})~ 
\hat\sigma_{ij}(\hat{s},\mu_F^2, \mu_R^2) \, \, , 
\label{sigOHF}
\ee
where $ij = gg, q\overline q$ or $q(\overline q)g$ and
$\hat\sigma_{ij}(\hat {s},\mu_F^2, \mu_R^2)$
is the partonic cross section for initial state $ij$ evaluated at factorization
scale $\mu_F$ and renormalization scale $\mu_R$.  (Note that the
$q(\overline q)g$ process only appears at next-to-leading order in $\alpha_s$.)

The charm quark
mass, $m$, factorization scale, $\mu_F$, and renormalization scale, $\mu_R$,
were fixed by fitting the total $c \overline c$
cross section at NLO in Ref.~\cite{NVF}:
$(m,\mu_F/m_T, \mu_R/m_T) = (1.27 \pm 0.09 \, {\rm GeV}, 2.1^{+2.55}_{-0.85}, 1.6^{+0.11}_{-0.12})$.   The scale factors are defined relative to
the transverse mass of the $c \overline c$, both for a single charm meson
from a produced $c \overline c$ pair and for the $J/\psi$,
$\mu_{F,R} \propto m_T = \sqrt{m^2 + p_{T_{Q \overline Q}}^2}$ where 
$p_{T_{Q \overline Q}}$ is the pair transverse momentum, 
$p_{T_{Q \overline Q}}^2 = 0.5(p_{T_Q}^2 + p_{T_{\overline Q}}^2)$ \cite{MNRcode}.  

The default fragmentation function in HVQMNR, applied to open heavy flavor
production only, is the Peterson function \cite{Pete},
\begin{eqnarray}
  D(z) = \frac{z(1-z)^2}{((1-z)^2 + z \epsilon_P)^2} \, \, ,
  \label{Eq.Pfun}
\end{eqnarray}
where $z$ represents the fraction of the parent heavy
flavor quark momentum carried by the resulting heavy flavor hadron.
Because the default parameter $\epsilon_P$ in HVQMNR results in charm $p_T$
distributions that are too soft compared to data, even with intrinsic transverse
momentum of the partons considered \cite{MLM2}, the value of $\epsilon_P$
was modified in Ref.~\cite{RV_azi1} to match the FONLL $D$ meson $p_T$
distributions.  The same procedure is followed here and the same value of
$\epsilon_P$ is employed in
this work: $\epsilon_P = 0.008$ \cite{RV_azi1}.

The parton densities include intrinsic $k_T$, required
to keep the pair cross section
finite as $p_{T_{Q \overline Q}} \rightarrow 0$.
They are assumed to factorize into the normal
parton densities in collinear factorization and a $k_T$-dependent function,
\be
F^p(x,\mu_F^2,k_T) = f^p(x,\mu_F^2)G_p(k_T) \, \, . \label{PDFfact}
\ee
The CT10 proton parton densities
\cite{CT10} are employed in the calculations of $f^p(x,\mu_F^2)$.

Results on open heavy flavors at fixed-target energies 
indicated that some level of
transverse momentum broadening was needed to obtain agreement with the
fixed-target data once fragmentation was included \cite{MLM1}.
Broadening has typically been modeled by intrinsic transverse
momentum, $k_T$, added to the parton densities and playing the role of
low transverse momentum QCD resummation \cite{CYLO}.  

In the HVQMNR code, an intrinsic $k_T$ is added in the
final state, rather than the initial state, as in the case of
Drell-Yan production \cite{CYLO}. 
In the initial-state, intrinsic $k_T$ multiplies the parton
distribution functions for both hadrons, 
assuming the $x$ and $k_T$ dependencies factorize,
as in Eq.~(\ref{PDFfact}).  
If the $k_T$ kick is not too large, it does not matter whether
the $k_T$  is added in the initial or final state.
The effect is applied in the final state, {\it i.e.} applied to the pair
after rather than before production, with the
factors $G_p(k_T)$ in Eq.~(\ref{PDFfact}) descibed by a Gaussian distribution
\cite{MLM1}, 
\begin{eqnarray}
G_p(k_T) = \frac{1}{\pi \langle k_T^2 \rangle_p} \exp(-k_T^2/\langle k_T^2
\rangle_p) \, \, .
\label{intkt}
\end{eqnarray}
In Ref.~\cite{MLM1}, $\langle k_T^2 \rangle_p = 1$ GeV$^2$ was chosen
to describe fixed-target charm production.  

The broadening is applied by boosting the transverse momentum
of the $c \overline c$ pair
(plus light parton at NLO)
to its rest frame
from the longitudinal center-of-mass frame.
The transverse momenta of the incident partons, $\vec k_{T_1}$ and
$\vec k_{T_2}$, are
redistributed isotropically with unit modulus, according to
Eq.~(\ref{intkt}), preserving momentum conservation.
Once boosted back to the initial frame, transverse momentum of
the $c \overline c$ pair changes from $\vec p_{T_{Q \overline Q}}$ to
$\vec p_{T_{Q \overline Q}} + \vec k_{T 1} + \vec k_{T 2}$ \cite{MLM2}.  

At leading order, there is no difference between a $k_T$ kick
applied to the initial or final state.  However, at NLO, if there is a
light parton in the final state, the correspondence can be inexact.  
The difference between the two implementations is small if
$\langle k_T^2 \rangle \leq 2-3$ GeV$^2$ \cite{MLM1}, as is the case here.
While the
rapidity distributions are independent of the intrinsic $k_T$, in addition
to changes to the $p_T$ distribution itself, the $x_F$ distribution will be
somewhat modified because
$x_F = (2m_T/\sqrt{s_{NN}})\sinh y$ and $m_T = \sqrt{p_T^2 + m^2}$.

\begin{figure}
  \begin{center}
    \includegraphics[width=0.495\textwidth]{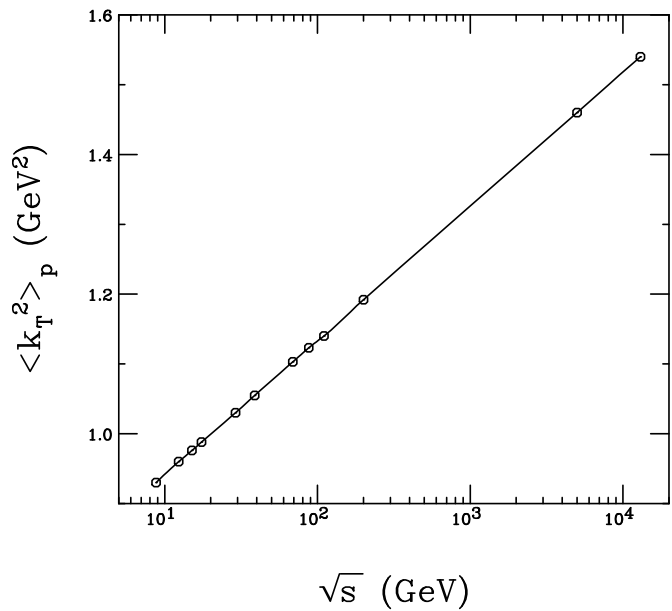}        
  \end{center}
  \caption[]{The value of $\langle k_T^2 \rangle_p$ as a function of the center
    of mass energy, $\sqrt{s}$ in $p+p$ collisions employing
    Eq.~(\ref{eq:avekt}).  The points show the energies at which
    $\langle k_T^2 \rangle_p$ is calculated.
  }
\label{fig:kt2_vs_en}
\end{figure}

The effect of the $k_T$ kick alone hardens
the single charm meson $p_T$ distribution,
particularly at low center of mass
energies.  This effect will
decrease as $\sqrt{s}$ increases because the average $p_T$ of the
$c \overline c$ pair also increases 
with energy.  The value of $\langle k_T^2 \rangle_p$ is also assumed to
increase with $\sqrt{s}$ so that effect can still be 
important for low $p_T$ heavy flavor production at higher energies.
The energy dependence of $\langle k_T^2 \rangle$ in Ref.~\cite{NVF} is
\begin{eqnarray}
  \langle k_T^2 \rangle_p = \left[ 1 + \frac{1}{n} \ln
    \left(\frac{\sqrt{s} ({\rm GeV})}{20 \,
    {\rm GeV}} \right) \right] \, \, {\rm GeV}^2 \, \, 
\label{eq:avekt}
\end{eqnarray}
with $n = 12$ for $J/\psi$ production \cite{NVF}.  The energy dependence of
$\langle k_T^2 \rangle_p$ is shown in Fig.~\ref{fig:kt2_vs_en}.  Given that
$\langle k_T^2 \rangle_p$ is defined to be 1~GeV$^2$ at $\sqrt{s} = 20$~GeV,
$\langle k_T^2 \rangle_p < 1$~GeV$^2$ for the lowest energies considered:
$p_{\rm lab} = 40$, 80, 120, and 158~GeV.  The large value of $n$ in
Eq.~(\ref{eq:avekt}) results in a slow growth of the broadening with energy
with $\langle k_T^2 \rangle_p \sim 1.55$~GeV$^2$ at $\sqrt{s} = 13$~TeV, well
below the limit of applicability proposed in Ref.~\cite{MLM1}.

\subsection{$J/\psi$ Production}

The $J/\psi$ production mechanism remains an unsettled question, with a number
of approaches having been introduced \cite{HPC,NRQCD,ICEM}.  In the calculations
presented here, the Color Evaporation Model \cite{HPC} is employed.  This model,
together with the Improved Color Evaporation Model \cite{ICEM}, can describe
the rapidity ($y$), Feynman $x$ ($x_F$), and transverse momentum ($p_T$)
distributions of $J/\psi$ production, including at low
$p_T$ where other approaches have some difficulties and may require a $p_T$
cut \cite{QWG_rev}.  

The CEM assumes that some fraction, $F_C$, of the $c \overline c$ pairs
produced in perturbative QCD with a pair mass below 
that of the $D \overline D$ pair mass threshold
will go on mass shell as a $J/\psi$,
\be
\sigma_{\rm CEM}(pp) = F_C \sum_{i,j} 
\int_{4m^2}^{4m_H^2} d\hat{s}
\int dx_1 \, dx_2~ F_i^p(x_1,\mu_F^2,k_{T_1})~ F_j^p(x_2,\mu_F^2,k_{T_2})~ 
\hat\sigma_{ij}(\hat{s},\mu_F^2, \mu_R^2) \, \, .
\label{sigCEM}
\ee
The same mass and scale parameters are employed as in Eq.~(\ref{sigOHF}).
However, now the upper limit of $4m_D^2$ is applied and the normalization factor
$F_C$ is obtained by fitting the energy dependence of the $J/\psi$ forward
cross section \cite{NVF}.
At LO in the CEM,
the $J/\psi$ $p_T$, equal to $p_{T_{Q \overline Q}}$ in the previous subsection,
is zero, requiring $k_T$ broadening to be applied to keep the
$p_T$ distribution finite as $p_T \rightarrow 0$.  The intrinsic $k_T$
broadening in $J/\psi$ production is handled the same way as for open heavy
flavor production, outlined above.  However, in this case, no fragmentation is
applied to the individual charm quarks, hadronization is implied by the factor
$F_C$.  (Note that, for simplicity, in the rest of this paper,
when open charm meson distributions are
presented, $p_T$ refers to the single charm hadron transverse momentum
distribution while, when $J/\psi$ distributions are discussed, $p_T$ refers
to the transverse momentum distribution of the $J/\psi$.)

The Improved Color
Evaporation Model was developed \cite{ICEM} and extended to
studies of quarkonium polarization \cite{CV1,CV2,CV3,CV4,CV5}
in hadroproduction.
The change in the mass integration range and in the definition of the $p_T$ of
the quarkonium state increases $F_C$ for $J/\psi$
in the ICEM by
$\sim 40$\% \cite{ICEM} due to the narrower mass integration range.
The $J/\psi$ $p_T$ distributions are compatible with
each other in the two approaches, see Ref.~\cite{CV3} for a comparison.

Other calculations have illustrated the mass and scale dependence of heavy
flavor production \cite{RV_SeaQuest,NVF}.  Because of the wide energy range
covered in this work, only the central values are shown here.  Even though the
mass uncertainty chosen is rather small, $m_c = 1.27 \pm 0.09$~GeV or 7.1\%,  
the cross section uncertainty due to the mass variation within that uncertainty
is generally significantly larger than the uncertainty due to the scale
dependence.  The dominance of the mass variation in the overall uncertainty is
because the scales were fit to the total charm cross section data, resulting in
a narrower scale range than calculations where the scales are varied by a
factor of two around a central value of $m_T$ \cite{CNV}.

At high center of mass energies, $J/\psi$ and $\overline D$ meson production
is dominated by the $gg$ initial state.  However, at the low energy end of
the fixed-target
energy range, namely that covered by NA60+, the $q \overline q$ and
$(q + \overline q)g$ channels are a non-negligible fraction of the production
cross section.

\section{Cold Nuclear Matter Effects}
\label{CNM}

This section discusses how cold nuclear matter effects:  nuclear
modifications of the parton densities, nPDF effects, Sec.~\ref{shad};
enhanced transverse momentum
broadening, Sec.~\ref{kTkick}; and $J/\psi$ absorption by nucleons,
Sec.~\ref{absorption}; are implemented in this work.  The effects of nPDF
modifications and enhanced $k_T$ broadening are common to both $J/\psi$ and
$\overline D$ meson production while absorption is assumed to impact only
$J/\psi$
production.  To simplify comparisons at different energies in most of
results presented, only calculations for a lead target will be shown.  Other
targets are considered when comparing to previous data as a function of
longitudinal momentum fraction $x_F$ later in this work.

\subsection{Nuclear Effects on the Parton Densities}
\label{shad}

The parton distribution functions in nuclei are modified from those of a free
proton.  A number of global analyses have been made by several groups
to describe the modification as a function of $x$ and factorization scale
$\mu_F$, assuming collinear factorization and starting from a minimum scale,
$\mu_{F \, 0}$.
These analyses have evolved over time, similar to global analyses of the proton
parton distribution functions, as more data become available.

Nuclear PDF effects are
generally implemented by a parameterization of the modification as
a function of $x$, $\mu_F$ and $A$ so that the $k_T$-independent
proton parton distribution
functions in Eqs.~(\ref{sigOHF}) and (\ref{sigCEM}) are replaced by the nuclear
parton distribution
functions
\be
f_j^A(x_2,\mu_F^2) = R_j(x_2,\mu_F^2,A) f_j^p(x_2,\mu_F^2) \, \, ,
\ee
where $R_j$ is the modification of the density of parton $j$ in the nucleus
relative to the density in the proton.
The
EPPS16 \cite{EPPS16} nPDF parameterization at next-to-leading order in
the strong coupling constant $\alpha_s$ is employed in the calculations
shown in this work.
Because a large energy range is covered here, only results with the
central EPPS16 set will be shown.

The EPPS16 set, like other parameterizations by Eskola and collaborators before
it, has a characteristic assumed $x$
dependence.  At high
momentum fractions, $x > 0.3$, there is a
depletion in the nucleus relative to a nucleon, known as the EMC effect.  At
smaller values of $x$, $x < 0.03$, there is also a depletion, known as
shadowing.  In the intermediate range of $x$, bridging the two regions where
$R < 1$, there is an enhancement referred to as antishadowing.  

\begin{figure}
  \begin{center}
    \includegraphics[width=0.495\textwidth]{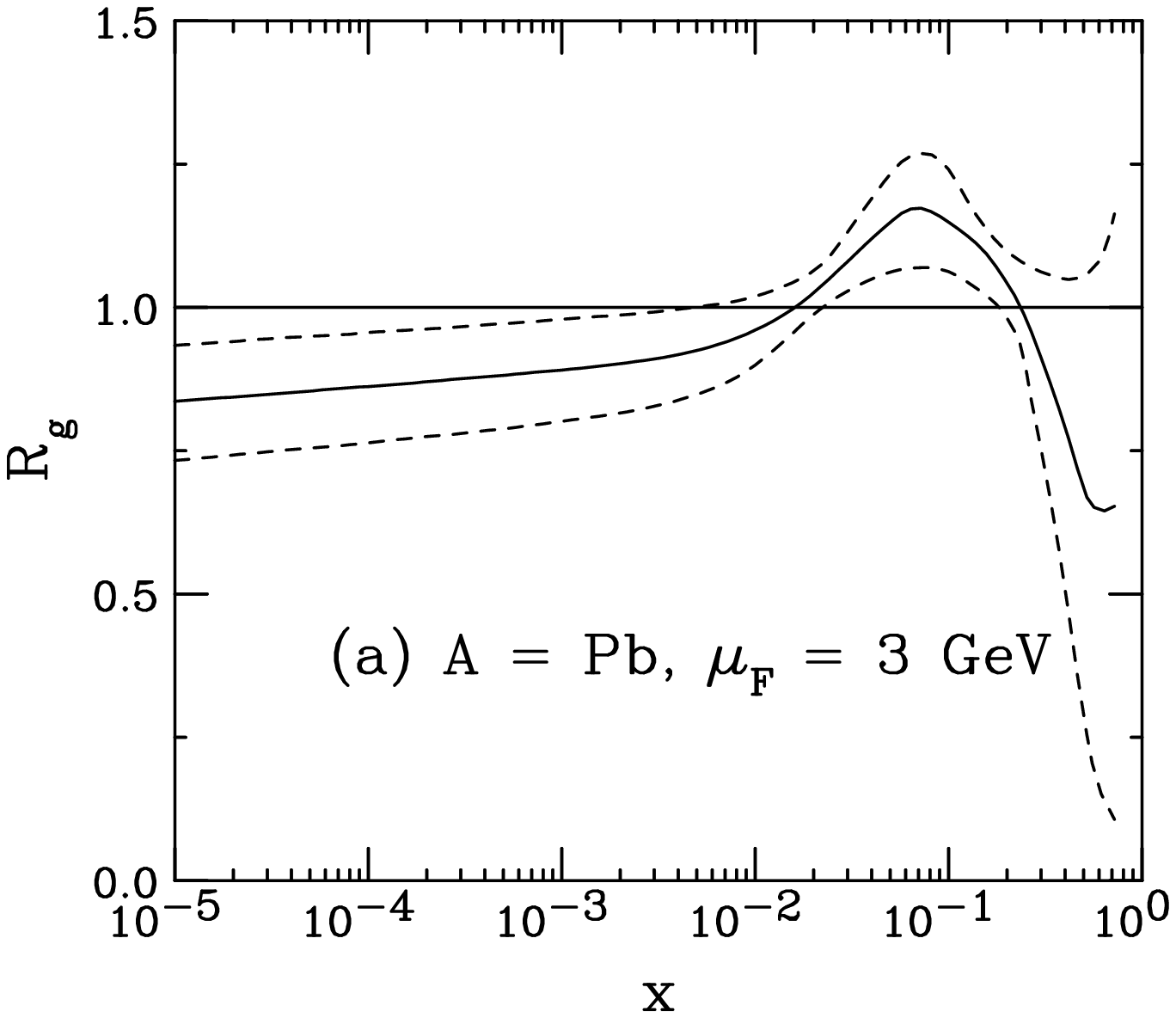}    
    \includegraphics[width=0.495\textwidth]{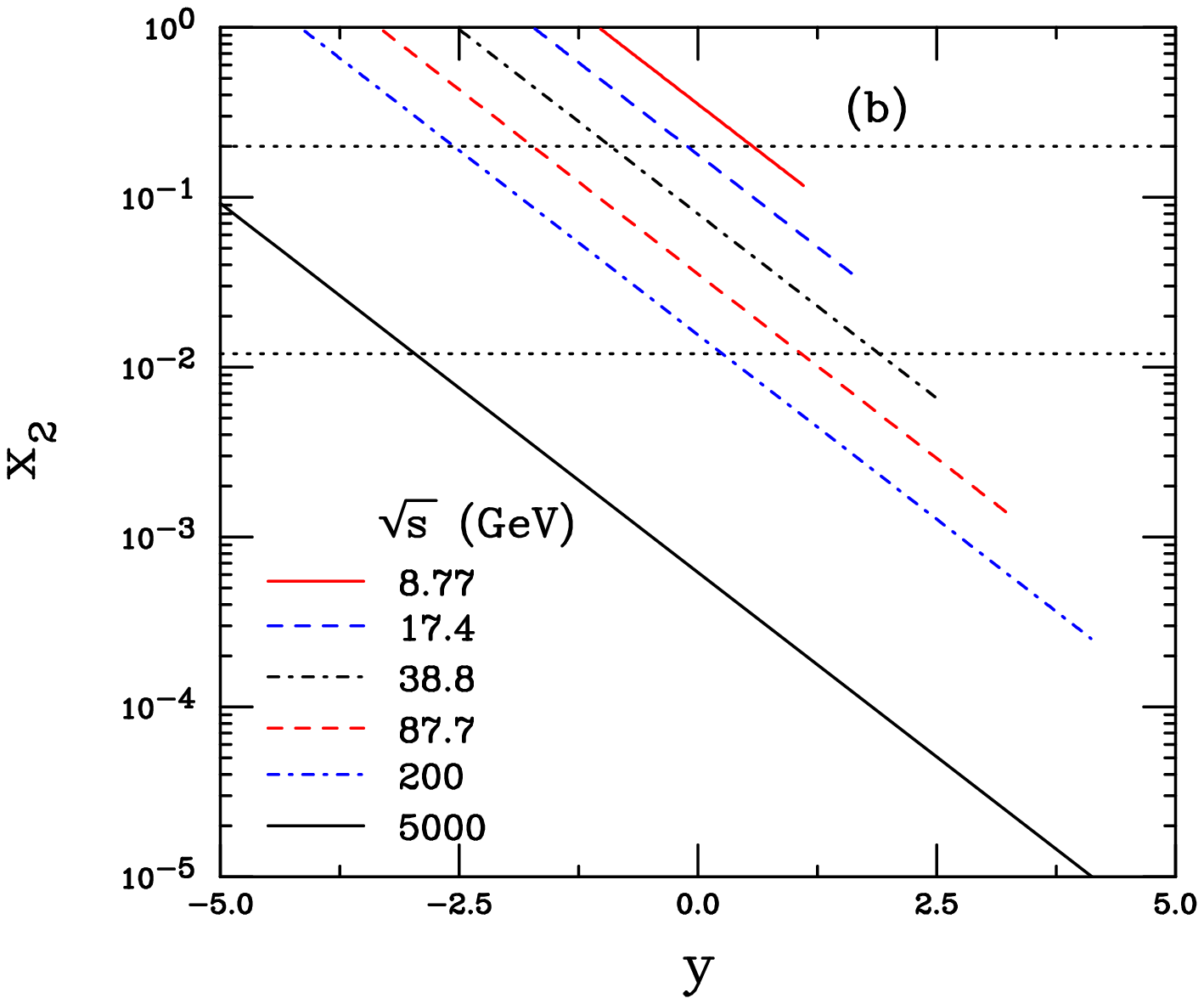}    
  \end{center}
  \caption[]{(a) The EPPS16 ratio for a lead nucleus, with uncertainties,
    is shown at the scale of the $J/\psi$ mass for gluons as a function of
    momentum fraction $x$. The central set is denoted by the solid curves
    while the dashed curves give the upper and lower limits of the uncertainty
    bands.  (b) The $x_2$ range as a function of rapidity for six values of
    $\sqrt{s_{NN}}$ covering the range of energies studied with nuclear targets:
    8.77 (solid red), 17.4 (dashed blue), 38.8 (dot-dashed black), 87.7 (dashed
    red), 200 (dot-dashed blue) and 5000 (solid black) GeV.  The upper and
    lower dotted lines at $x_2 = 0.012$ and 0.2 represent the lower and upper
    limits of the antishadowing region for $\mu_F = 3$~GeV.
  }
\label{shad_ratios}
\end{figure}

The EPPS16 ratio for gluons in a Pb nucleus is
shown at the $J/\psi$ mass scale, $\mu_F \sim 3$~GeV,
in Fig.~\ref{shad_ratios}(a).  This value of the factorization
scale is close to that of
the minimum scale employed in the EPPS16 fits, $\mu_F = 1.3$~GeV, making the
uncertainty band quite large.
The central set is shown, along with the uncertainty bands.
Thus one can expect antishadowing near $x_F \sim 0$
and shadowing at high $x_F$ in the E866 range.

To provide some scale for which values of $x_2$ are probed over the energy
range discussed in this work, $x_2$ is shown as a function of center of mass
rapidity in Fig.~\ref{shad_ratios}(b).  The range is covers the equivalent of
$-1 \leq x_F \leq 1$ at all energies except the highest value since only the
range $|y| \leq 5$ is shown.  Above $x_2 \sim 0.2$, the EMC region and Fermi
motion range ($x_2 \rightarrow 1$) are probed.  The antishadowing region lies
between the horizontal dotted lines.  At smaller $x_2$, the shadowing region
is probed.  The fixed target energies, $\sqrt{s_{NN}} = 8.77$, 17.4 and 38.8~GeV
cover the antishadowing region at midrapidity and only enter the shadowing
region for forward rapidity at 38.8~GeV. The SMOG and RHIC energies,
$\sqrt{s_{NN}} = 87.7$ and 200~GeV respectively, can cover most of $x_2$ range
although the PHENIX muon spectrometer covers only $1.1 < |y| < 2.2$, or a
minimum $x_2$ of $\sim 0.001$.  The higher LHC energy, 5~TeV, covers the
antishadowing region at negative rapidities measurable by ALICE and LHCb and
well into the low $x_2$ region for forward rapidity.

\subsection{$k_T$ Broadening}
\label{kTkick}

The effect and magnitude of intrinsic $k_T$ broadening on the heavy quark $p_T$
distribution in $p+p$ collisions was discussed in Sec.~\ref{pQCD}.
Here further broadening due to the
presence of a nuclear target is considered.
One might expect that a higher intrinsic $k_T$
is required in a nuclear medium relative to that in $p+p$ due to multiple
scattering in the nucleus, known as the Cronin effect \cite{Cronin}.  The
effect is implemented by replacing $G_p(k_T)$ in Eq.~(\ref{intkt}) by
$G_A(k_T)$ where $\langle k_T \rangle_p$ is replaced by $\langle k_T \rangle_A$.

The total broadening in a nucleus relative to a nucleon can be expressed as
\begin{eqnarray}
\langle k_T^2 \rangle_A = \langle k_T^2 \rangle_p +\delta k_T^2 \, \, .
\end{eqnarray}
The same expression for $\delta k_T^2$ used in Ref.~\cite{HPC_pA}, based on
Ref.~\cite{XNW_PRL}, is employed here,
\begin{eqnarray}
  \delta k_T^2 = (\langle \nu \rangle - 1) \Delta^2 (\mu) \, \, .
  \label{delkt2}  
\end{eqnarray}
The strength of the broadening, $\Delta^2 (\mu)$, depends on the interaction
scale \cite{XNW_PRL},
\begin{eqnarray}
  \Delta^2 (\mu) = 0.225 \frac{\ln^2 (\mu/{\rm GeV})}{1  + \ln(\mu/{\rm GeV})}
  {\rm GeV}^2 \, \, ,
\label{delta2}
\end{eqnarray}
where $\mu = 2m_c$
\cite{HPC_pA}.  The scale dependence suggests that one might expect a
larger $k_T$ kick in the nucleus for bottom quarks than charm quarks.

The size of the effect depends on the number of scatterings the incident parton
could undergo while passing through a nucleus.  This is given by
$\langle \nu \rangle -1$, the number of nucleon-nucleon
collisions, less the first collision.  The number of scatterings
strongly depends on the impact parameter of the proton with respect to the
nucleus.  However, in minimum-bias collisions, the impact parameter is
averaged over all possible paths.  The average number of scatterings is
\begin{eqnarray}
  \langle \nu \rangle
  = \sigma_{pp}^{\rm in} \frac{\int d^2b T_A^2(b)}{\int d^2b T_A(b)}
  = \frac{3}{2} \rho_0 R_A \sigma_{pp}^{\rm in}
  \label{avenu}
\end{eqnarray}
where $T_A(b)$ is the nuclear profile function,
$T_A(b) = \int_{-\infty}^{\infty} dz \rho_A(b,z)$
and $\rho_A$ is the nuclear density distribution.
An average nuclear density, $\rho_0 = 0.16$/fm$^3$, is assumed for a spherical
nucleus of radius $R_A = 1.2 A^{1/3}$~fm; and $\sigma_{pp}^{\rm in}$ is
the inelastic $p+p$ cross section, $\sim 32$~mb at fixed-target energies.  

With the given values of $\rho_0$ and $\sigma_{pp}^{\rm in}$ in Eq.~(\ref{avenu})
and $\Delta^2(\mu = 2m_c) = 0.101$~GeV$^2$ from Eq.~(\ref{delta2}),,
\begin{eqnarray}
  \delta k_T^2 \approx (0.92 A^{1/3} - 1)\times 0.101 \, {\rm GeV}^2 \, \, .
  \end{eqnarray}
The calculations shown here are for  $A = 208$ to maximize the resulting
effect with $\delta k_T^2 = 0.45$~GeV$^2$.  At low $\sqrt{s_{NN}}$, the effect
of this enhanced $k_T$ broadening in $p+A$ collisions is more significant
than at higher $\sqrt{s_{NN}}$.
The effect on the $p_T$ distributions will be seen to be quite
large, especially at low, fixed-target energies.  The enhanced
broadening has no effect on the rapidity distributions.

\subsection{Nuclear Absorption of $J/\psi$ in $pA$ Interactions}
\label{absorption}

In $p+A$ collisions, the proto-$J/\psi$ produced in the initial partonic
interactions may interact with other nucleons along its path
and be dissociated or
absorbed before it can escape the target.
The effect of nuclear absorption alone
may be expressed as \cite{rvrev}
\begin{eqnarray}
  \sigma_{pA}^{J/\psi} = \sigma_{pN}^{J/\psi}
  S_A^{\rm abs} & = & \sigma_{pN}^{J/\psi} \int d^2b \,
  \int_{-\infty}^{\infty}\, dz \, \rho_A (b,z) S^{\rm abs}(b) \\
& = & \sigma_{pN}^{J/\psi} \int d^2b \,  \int_{-\infty}^{\infty}\, dz \, \rho_A (b,z)
 \exp \left\{
-\int_z^{\infty} dz^{\prime} \rho_A (b,z^{\prime}) \sigma_{\rm abs}(z^\prime
-z)\right\} \, \, , 
\label{sigfull}
\end{eqnarray} where $b$ is the impact parameter, $z$ is the $c \overline c$
production point, $S^{\rm abs}(b)$ is the nuclear absorption survival 
probability, and $\sigma_{\rm abs}(z^\prime -z)$ 
is the nucleon absorption cross section.  Even though
the absorption cross section
is assumed to be constant at a given energy in this work, 
it is written as a function of the
path length through the nucleus in Eq.~(\ref{sigfull}) because other functional
forms have previously been assumed, see {\it e.g.}\ Ref.~\cite{RV_HeraB}.
Expanding the exponent in Eq.~(\ref{sigfull}), integrating, 
and reexponentiating the results assuming $A$ is large
leads to
\be
\sigma_{pA}^{J/\psi} = \sigma_{pN}^{J/\psi} A^\alpha \, \, \label{alfint}
\ee
with $S_A^{\rm abs} = A^\alpha$ and
$\alpha = 1 - 9\sigma_{\rm abs}/(16 \pi r_0^2)$ \cite{rvrev}.
The value of $\sigma_{\rm abs}$ can depend on the kinematics and the color state
of the $J/\psi$, see
Refs.~\cite{RV_HeraB,RV_e866,LWV,Darren} and references therein.

Note that
this $A$ dependence arises naturally through the survival probability in
Eqs.~(\ref{sigfull}) and (\ref{alfint}).  Previous data has, however, been
analyzed assuming $A^\alpha$ covers all nuclear effects.
Thus the absorption cross section extracted in previous analyses,
see {\it e.g.} \cite{NA3,e866}, is only an effective cross section.
Thus, if only absorption is assumed,
$\sigma_{\rm abs}$ could be underestimated at
low $\sqrt{s_{NN}}$ because antishadowing effects
are neglected.  If antishadowing is taken into account a larger
effective $\sigma_{\rm abs}$ is needed to match the measured
suppression in $p+A$ relative to $p+p$ collisions~\cite{LWV}.

In Ref.~\cite{LWV}, the effective $\sigma_{\rm abs}$
was extracted for each measured $x_F$ or $y$ for several nPDF
parameterizations.  Even though $\sigma_{\rm abs}$
depended weakly on the nPDF parameterization, some trends were
evident.  After accounting for nPDF effects, absorption
was generally largest at midrapidity and decreased at forward rapidity.
The Fermilab E866 experiment
\cite{e866}, at $\sqrt{s_{NN}} = 38.8$~GeV, and the NA3 experiment
\cite{NA3} at CERN, with $\sqrt{s_{NN}} = 19.4$~GeV, showed a strong increase in
effective absorption for $x_F > 0.3$ \cite{LWV}.  Although
high $x_F$ corresponds to low $x_2$ in the nucleus, in the shadowing
region, results at
different energies did not scale with $x_2$, as expected,
suggesting that shadowing
was not the dominant nuclear effect.  Other effects, such as cold matter energy
loss were suggested
to explain the observed behavior \cite{e866}.

\begin{figure}
  \begin{center}
    \includegraphics[width=0.495\textwidth]{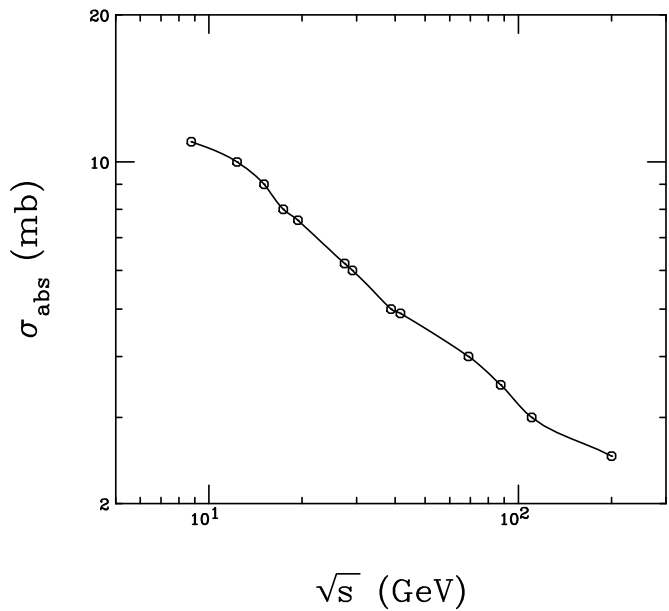}        
  \end{center}
  \caption[]{The value of $\sigma_{\rm abs}$ as a function of the center
    of mass energy, $\sqrt{s_{NN}}$.  The points show the energies
    used in this paper.  The line is meant to guide the eye.
  }
\label{fig:sigabs_vs_en}
\end{figure}

The effective absorption cross section used in this work is shown in
Fig.~\ref{fig:sigabs_vs_en}.  The dependence approximately follows the results
in Ref.~\cite{LWV} and shows a general near-linear decrease of $\sigma_{\rm abs}$
over the energy range shown.  The cross section is not shown at energies
higher than $\sqrt{s_{NN}} = 200$~GeV because nucleon
absorption is assumed to be negligible at the LHC.


Absorption by comoving particles
has been included in some calculations of cold nuclear
matter effects \cite{Elena}.  This type of suppression
is not included separately here
because it has been shown to have the same nuclear dependence in minimum
bias collisions \cite{SGRV1990}.

\section{Intrinsic Charm}
\label{ICcomp}

The proton wave function in QCD can be represented as a
superposition of Fock state fluctuations, {\it e.g.}\ $\vert uudg
\rangle$, $\vert uud q \overline q \rangle$, $\vert uud Q \overline Q \rangle$,
\ldots of the $\vert uud \rangle$ state.
When the proton projectile scatters in a target, the
coherence of the Fock components is broken and the fluctuations can
hadronize \cite{intc1,intc2,BHMT}.  These
intrinsic $Q
\overline Q$ Fock states are dominated by configurations with
equal rapidity constituents, resulting in the heavy quarks carrying a large
fraction of the proton momentum \cite{intc1,intc2}.  (While proton
projectiles are emphasized here, any hadron wave function can be so described.)
 
The frame-independent probability distribution of a $5$-particle
$c \overline c$ Fock state in the proton is 
\be
dP_{{\rm ic}\, 5} = P_{{\rm ic}\,5}^0
N_5 \int dx_1 \cdots dx_5 \int dk_{x\, 1} \cdots dk_{x \, 5}
\int dk_{y\, 1} \cdots dk_{y \, 5} 
\frac{\delta(1-\sum_{i=1}^5 x_i)\delta(\sum_{i=1}^5 k_{x \, i}) \delta(\sum_{i=1}^5 k_{y \, i})}{(m_p^2 - \sum_{i=1}^5 (\widehat{m}_i^2/x_i) )^2} \, \, ,
\label{icdenom}
\ee
where $i = 1$, 2, 3 are the light quarks ($u$, $u$, $d$)
and $i = 4$ and 5 are the $c$ and $\overline c$ quarks respectively.
Here $N_5$ normalizes the
$|uud c \overline c \rangle$ probability to unity and $P_{{\rm ic}\, 5}^0$
scales the unit-normalized
probability to the assumed probability of intrinsic charm content in the proton.
The delta functions in Eq.~(\ref{icdenom}) conserve longitudinal ($z$) and
transverse ($x$ and $y$) momentum.  The denominator of Eq.~\ref{icdenom} is
minimized when the heaviest partons carry the largest fraction of the
proton longitudinal momentum, $\langle x_Q \rangle > \langle x_q \rangle$.

Additional delta functions can be employed to form hadrons by simple
coalescence when the Fock state is disrupted.  For example, the $J/\psi$ $x_F$
distribution can be calculated
by the addition of the delta functions,
$\delta(x_F - x_c - x_{\overline c})$, for the longitudinal, $z$, direction
and $\delta(p_T - k_{x \, c} - k_{x \, \overline c}) \delta(k_{y \, c} + k_{y\, \overline c})$ in the transverse directions where the $J/\psi$ $p_T$ is chosen to be
along the $x$ direction for simplicity.  
The summed $x_c$ and $x_{\overline c}$ momentum fractions are equivalent to the
$x_F$ of the $J/\psi$ assuming that it is brought on-shell by a soft scattering
with the target.

Likewise, one can produce $\overline D$ mesons ($D^-(\overline c d)$ and
$\overline D^0 (\overline c u)$) mesons directly from the disrupted Fock state
employing $\delta(x_F - x_{\overline c} - x_i)$ for the $\overline D$ $x_F$ and
$\delta(p_T - k_{x \, \overline c} - k_{x \, i}) \delta(k_{y \, i} + k_{y\, \overline c})$ for the $p_T$
where the light parton $i$ can be either a $u$ or $d$ quark.  
The remaining
partons in the state could coalesce into a $\Lambda_c(udc)$ with an
$\overline D^0$ or a $\Sigma_c^{++}(uuc)$ with a $D^-$.

The preference for
$\overline D$ production from this state makes the $\overline D$ a ``leading''
particle relative to $D^+ (c \overline d)$ and $D^0 (c \overline u)$ which could
only be produced through standard fragmentation and would thus manifest at lower
$x_F$ than the $\overline D$ mesons.  In order to produce a $D$ meson from
coalescence, similar to the $\overline D$ in the $|uud c \overline c \rangle$
state, a higher particle-number Fock state is required, such as
$|uud c \overline c d \overline d\rangle$ for $D^+$ production \cite{tomg}.  
Previous studies of leading $D$ meson production, including asymmetries
between $D^+$ and $D^-$ production in fixed-target $\pi^- A$ interactions have
shown significant differences between leading and nonleading production
\cite{E791}.  These asymmetries have been reproduced by intrinsic charm
\cite{RVSJB_asymm} but also by string-breaking mechanisms such as those in
PYTHIA \cite{Norrbin}.  Note, however,
that while the $D$ and $\overline D$ meson distributions may differ in the
5-particle Fock state considered here, the $c$ and $\overline c$ distributions
themselves are identical: no asymmetry is assumed in their production, only
their manifestation as final-state charm hadrons.

In Ref.~\cite{RVSJB_asymm}, only the lowest $\pi^-$
Fock state, $|\overline u d c \overline c \rangle$,
was considered in the calculation of the asymmetry.  However, it is
possible for equal rapidity $D$ and $\overline D$ mesons to be produced from
higher Fock components such as
$|uud c\overline c q \overline q \rangle$ but these will reduce the average
momentum fraction of both the $D$ and $\overline D$.  They will also be produced
with lower probabilities, see {\it e.g.}\ Ref.~\cite{tomg}
for examples of charm hadron distributions from higher
Fock states.  In this work, only the 5-particle proton Fock state is
considered since it gives the most forward $x_F$ production of $J/\psi$
and $\overline D$ from intrinsic charm.  

The $x_F$ distribution, as well as the $p_T$ distribution
integrated over all phase space, is independent of the proton energy.  The
$p_T$ distribution from intrinsic charm only varies when phase space cuts are
considered, as shown in Ref.~\cite{RV_SeaQuest}.  Although fixed-target
experiments have typically reported the $x_F$ dependence of heavy flavor
hadrons, collider experiments generally report the rapidity dependence.  The
rapidity is related to $x_F$ by $x_F = (2m_T/\sqrt{s}) \sinh y$ so that, even
though the $x_F$ distribution is invariant, independent of $\sqrt{s}$, $m_i$
and $k_T$-integration range, the rapidity distribution is not.
Indeed, the $x_F$
distribution depends only weakly on the heavy quark mass with the average $x_F$
changing by only a few percent between $J/\psi$ and $\Upsilon$, as discussed
in Refs.~\cite{RVSJB_psipsi,ANDY}.

The default values of the quark masses and $k_T$ integration ranges in
Eq.~(\ref{icdenom}) are $m_c = 1.27$~GeV, $m_q = 0.3$~GeV,
$k_q^{\rm max} = 0.2$~GeV and $k_c^{\rm max} = 1.0$~GeV.
(Note that the constitutent quark masses are used for the light quarks.)
The dependence of the
rapidity and $p_T$ distributions on these values is now discussed.

\begin{figure}
  \begin{center}
    \includegraphics[width=0.4\textwidth]{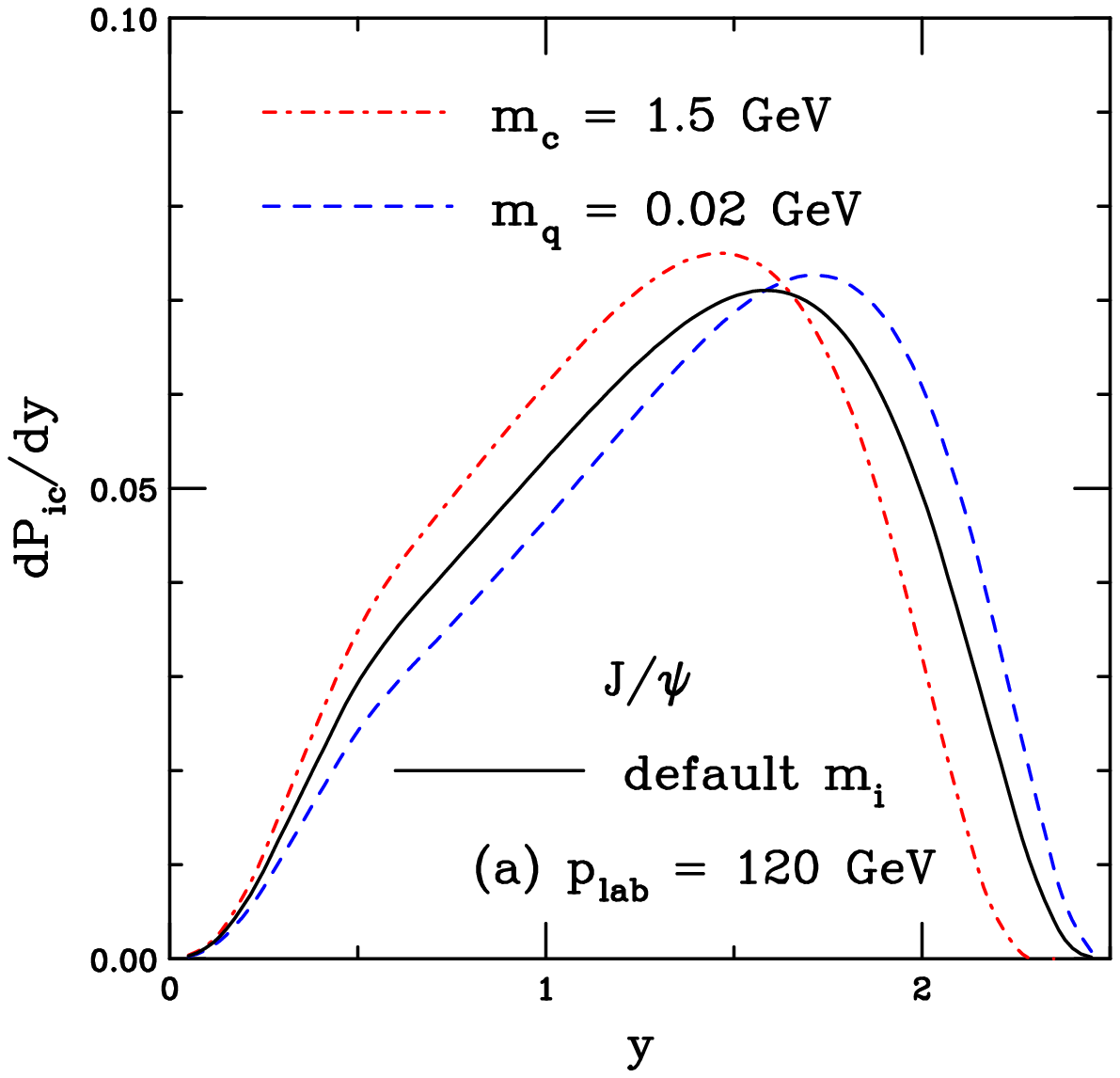}
    \includegraphics[width=0.4\textwidth]{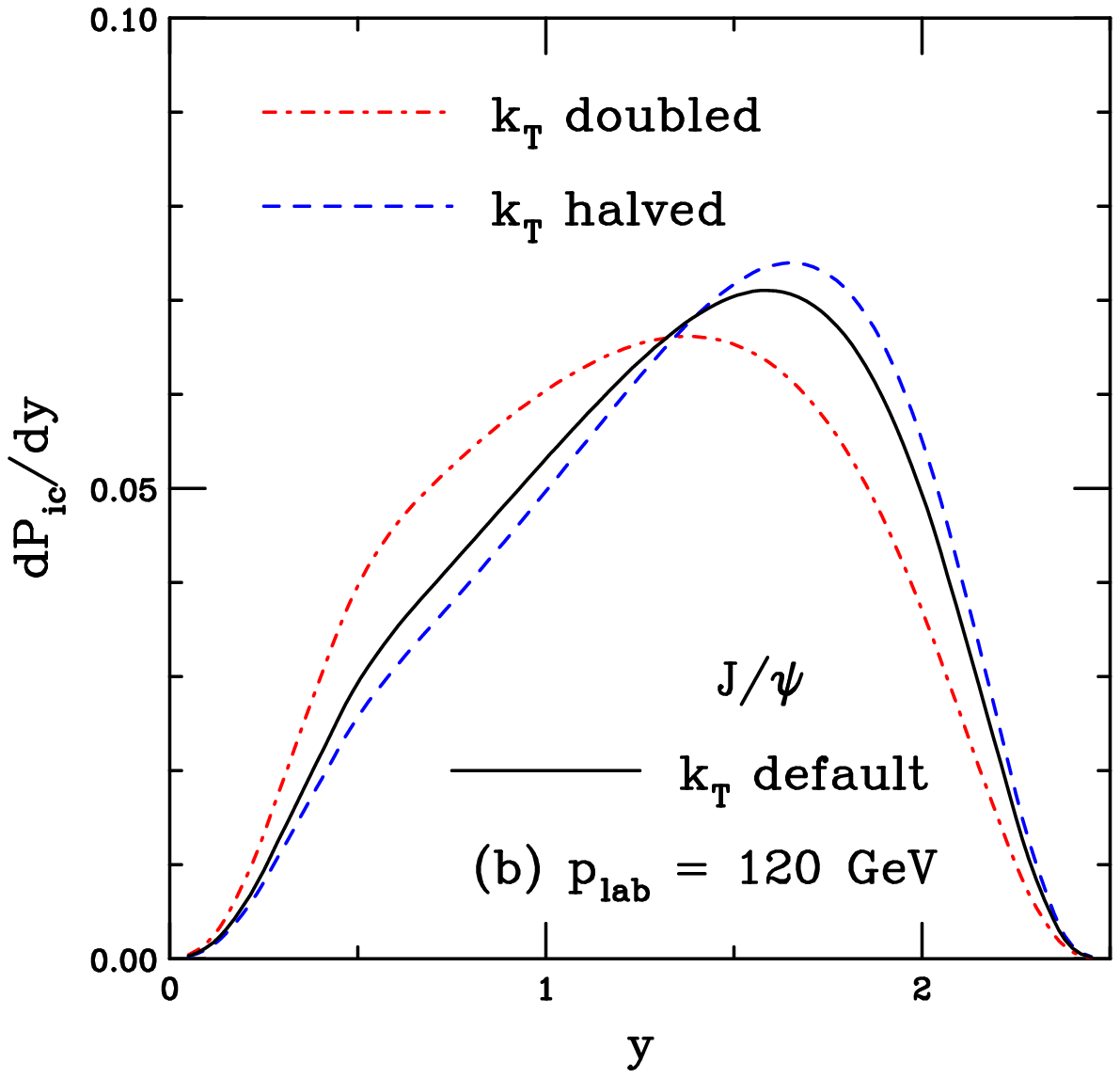} \\
    \includegraphics[width=0.4\textwidth]{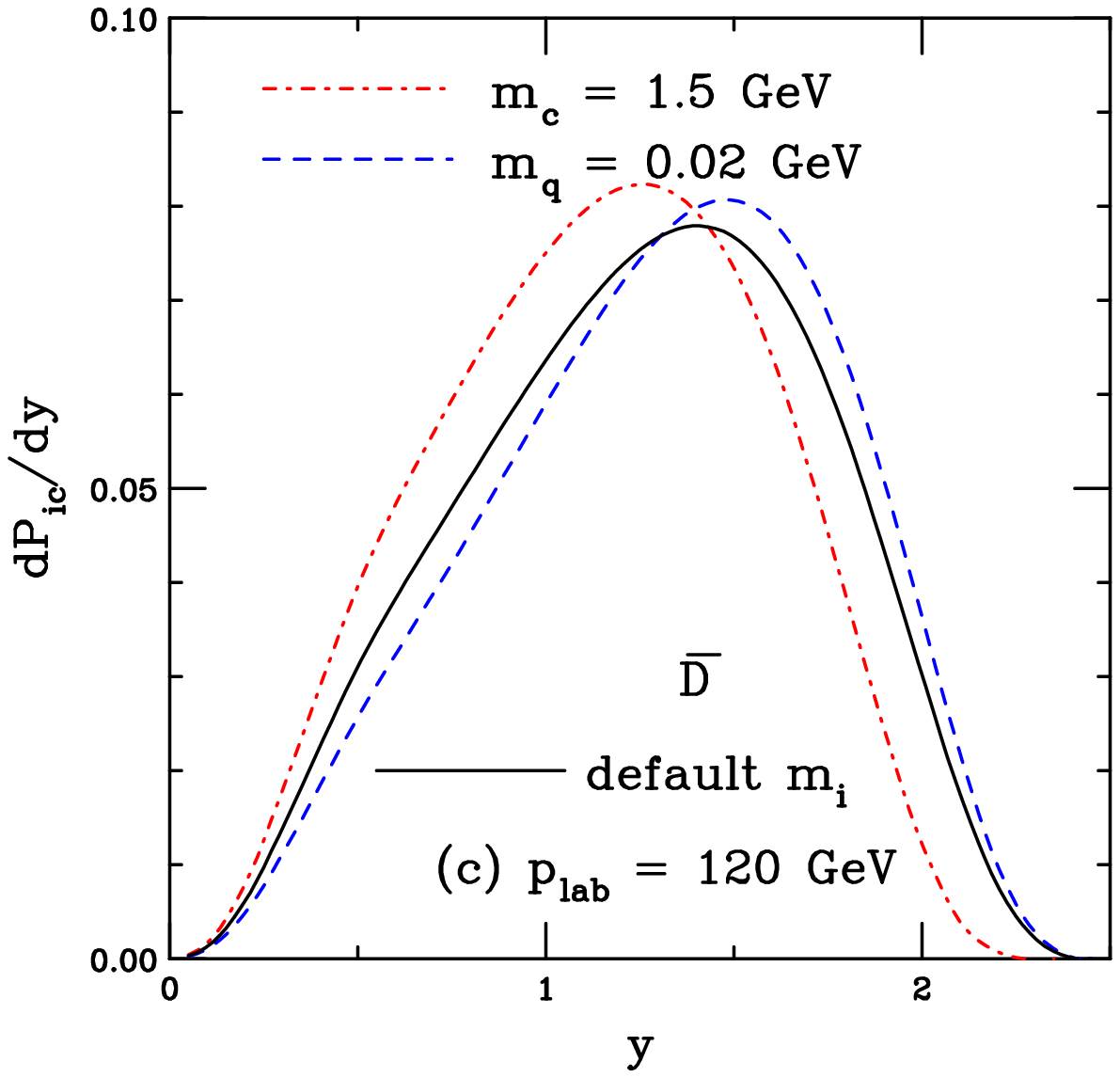}
    \includegraphics[width=0.4\textwidth]{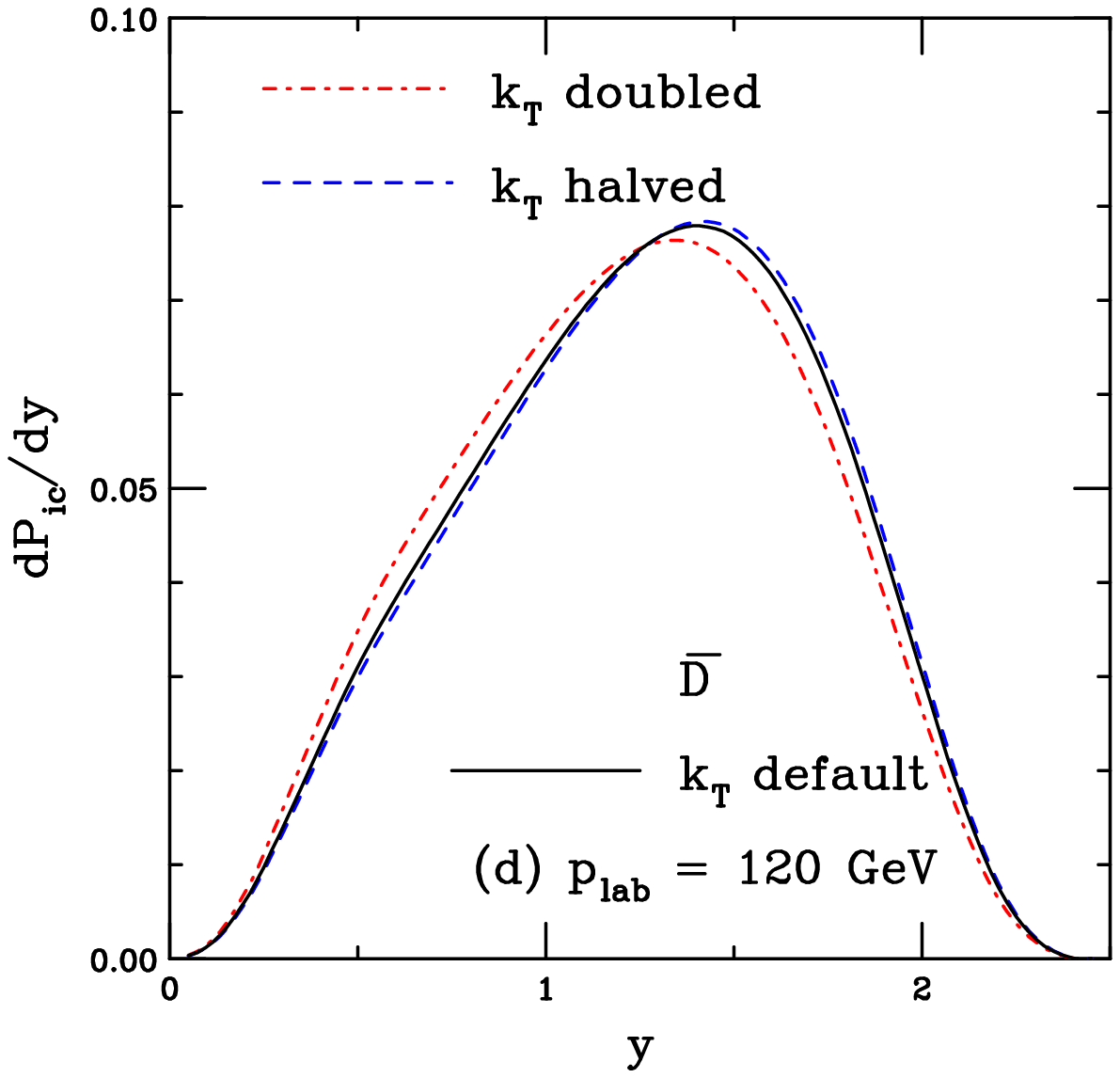}
  \end{center}
  \caption[]{ The probability distributions for $J/\psi$ (a), (b)
    and $\overline D$ (c), (d)
    production from a five-particle proton Fock state as a function of rapidity.
    In (a) and (c) the results are given for the default mass values,
    $m_c = 1.27$~GeV and $m_q = 0.3$~GeV (solid black), as well as for
    $m_c = 1.5$~GeV and $m_q = 0.3$~GeV (dot-dashed red), and 
    $m_c = 1.27$~GeV and $m_q = 0.02$~GeV (dashed blue),
    In (b) and (d) the results are shown for different values of the $k_T$
    range for the light and charm quarks.  The solid black curve employs the
    default values, $k_q^{\rm max} = 0.2$~GeV and $k_c^{\rm max} = 1.0$~GeV while
    the red dot-dashed curve increases the default values by a factor of two,
    $k_q^{\rm max} = 0.4$~GeV and $k_c^{\rm max} = 2$~GeV,
    and the blue dashed curves employs half the default values,
    $k_q^{\rm max} = 0.1$~GeV and $k_c^{\rm max} = 0.5$~GeV.
    All distributions are normalized to unity.
  }
\label{ic_ydists}
\end{figure}

The dependence on parton mass and $k_T$ range in the calculated rapidity
distributions is shown in Fig.~\ref{ic_ydists} for $p_{\rm lab} = 120$~GeV.
Although the maximum rapidity range depends on $\sqrt{s}$, the results shown
here are indicative for all energies.  The $J/\psi$ probability distributions
are shown in Fig.~\ref{ic_ydists}(a) and (b) while those for the $\overline D$
are given in Fig.~\ref{ic_ydists}(c) and (d).  The default values are given by
the solid black curves in both cases.  All distributions are normalized to unity
for the parameters employed.  Note that the $J/\psi$ rapidity
distribution is somewhat broader and covers a wider rapidity range than that
of the $\overline D$ because the $J/\psi$ contains both heavy quarks from the
Fock state while the $\overline D$ includes only one of them.  The average
rapidities in all cases shown in Fig.~\ref{ic_ydists} are given in
Table~\ref{table_ave_variations}.

Assuming a larger charm quark mass, $m_c = 1.5$~GeV, reduces the rapidity
range and shifts the overall distribution backward for both $J/\psi$ and
$\overline D$.  The backward shift is approximately the same for both cases.
Similarly, reducing the light quark mass from a constituent quark value of
$\sim 300$~MeV to nearly zero shifts the distribution forward by a similar
factor.  On the other hand, doubling or halving the $k_T$ range on both the
charm and light quarks together has a larger
effect on the $J/\psi$ distribution than on the $\overline D$ distribution.
Although doubling the $k_T$-integration range shifts the rapidity distribution
backward and halving the range shifts it forward, the average shift of the
$J/\psi$ distribution is nearly double that of the $\overline D$, likely
because the $J/\psi$ is made up of both the $c$ and the $\overline c$ while the
$\overline D$ includes only the $\overline c$ quark.

\begin{table}
  \begin{tabular}{|c|c|c||c|c|} \hline
Variation & $\langle y_{\rm ic}^{J/\psi} \rangle$ &
    $\langle y_{\rm ic}^{\overline D} \rangle$ &
    $\langle p_{T\,{\rm ic}}^{J/\psi} \rangle$ (GeV) &
    $\langle p_{T\,{\rm ic}}^{\overline D} \rangle$ (GeV) \\ \hline 
\multicolumn{5}{|c|}{default $m$, $k_T$} \\ \hline
 - & 1.344 & 1.260 & 2.067 & 1.962 \\ \hline
\multicolumn{5}{|c|}{quark mass variation} \\ \hline
$m_c = 1.5$~GeV & 1.249 & 1.149 & 2.246 & 2.185 \\
$m_q = 0.02$~GeV & 1.423 & 1.312 & 1.884 & 1.772 \\ \hline
\multicolumn{5}{|c|}{$k_t$ range variation} \\ \hline
$k_T$ doubled & 1.246 & 1.225 & 2.484 & 2.099 \\
$k_T$ halved & 1.385 & 1.276 & 1.904 & 1.920 \\ \hline
  \end{tabular}
  \caption[]{\label{table_ave_variations} The average rapidity and $p_T$ for
    $J/\psi$ and $\overline D$ meson production from a five-particle Fock state
    obtained by varying the quark mass and $k_T$ range around their default
    values of $m_c = 1.27$~GeV, $m_c = 0.3$~GeV, $k_{T, \, c}^{\rm max} = 1$~GeV,
    and $k_{T, \, q}^{\rm max} = 0.2$~GeV.  The average $p_T$ of the charm quark
    is 1.997~GeV.  The average rapidity distributions are calculated for
    $p_{\rm lab} = 120$~GeV.
    }
\end{table}

\begin{figure}
  \begin{center}
    \includegraphics[width=0.4\textwidth]{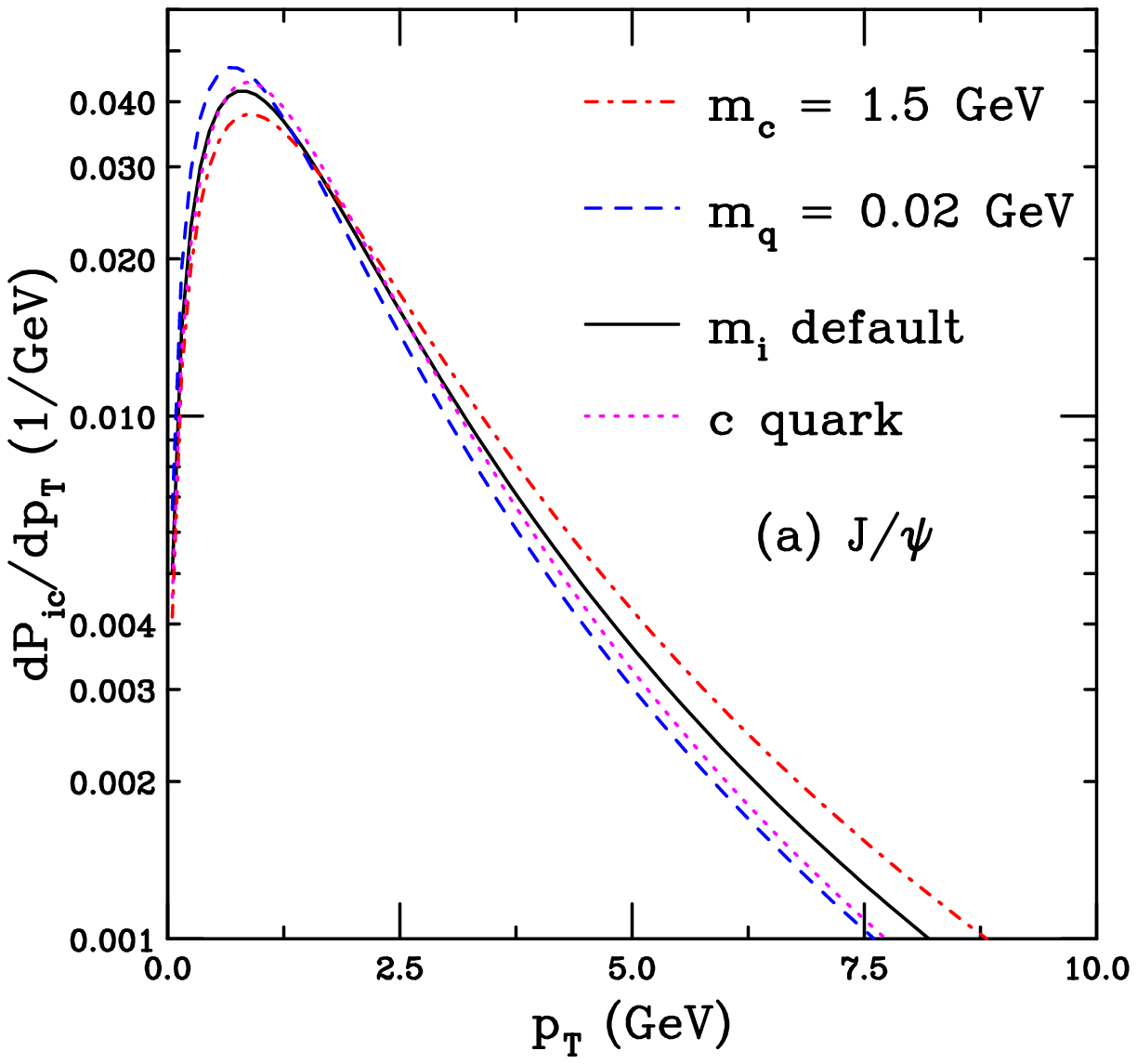}
    \includegraphics[width=0.4\textwidth]{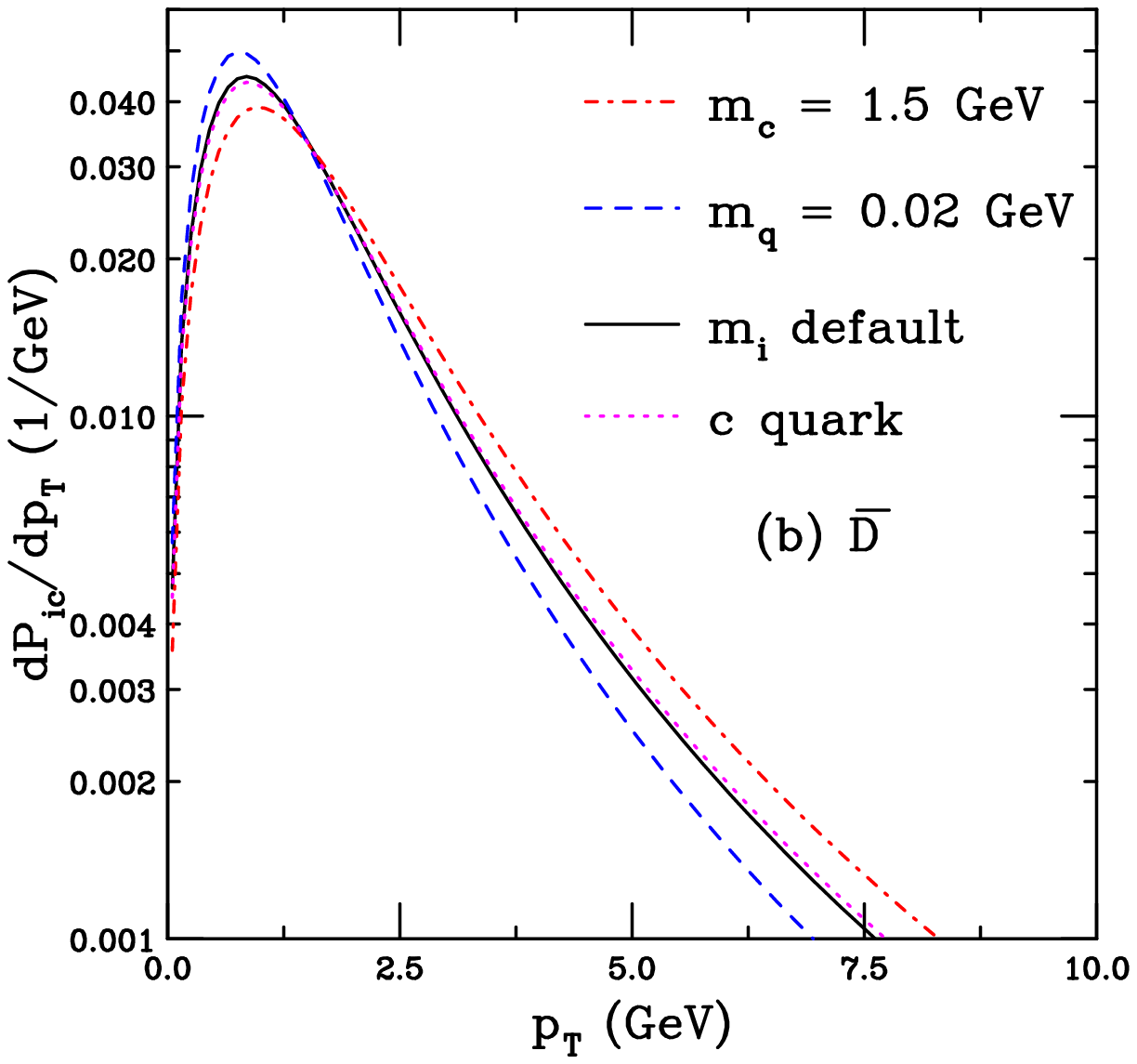} \\ 
    \includegraphics[width=0.4\textwidth]{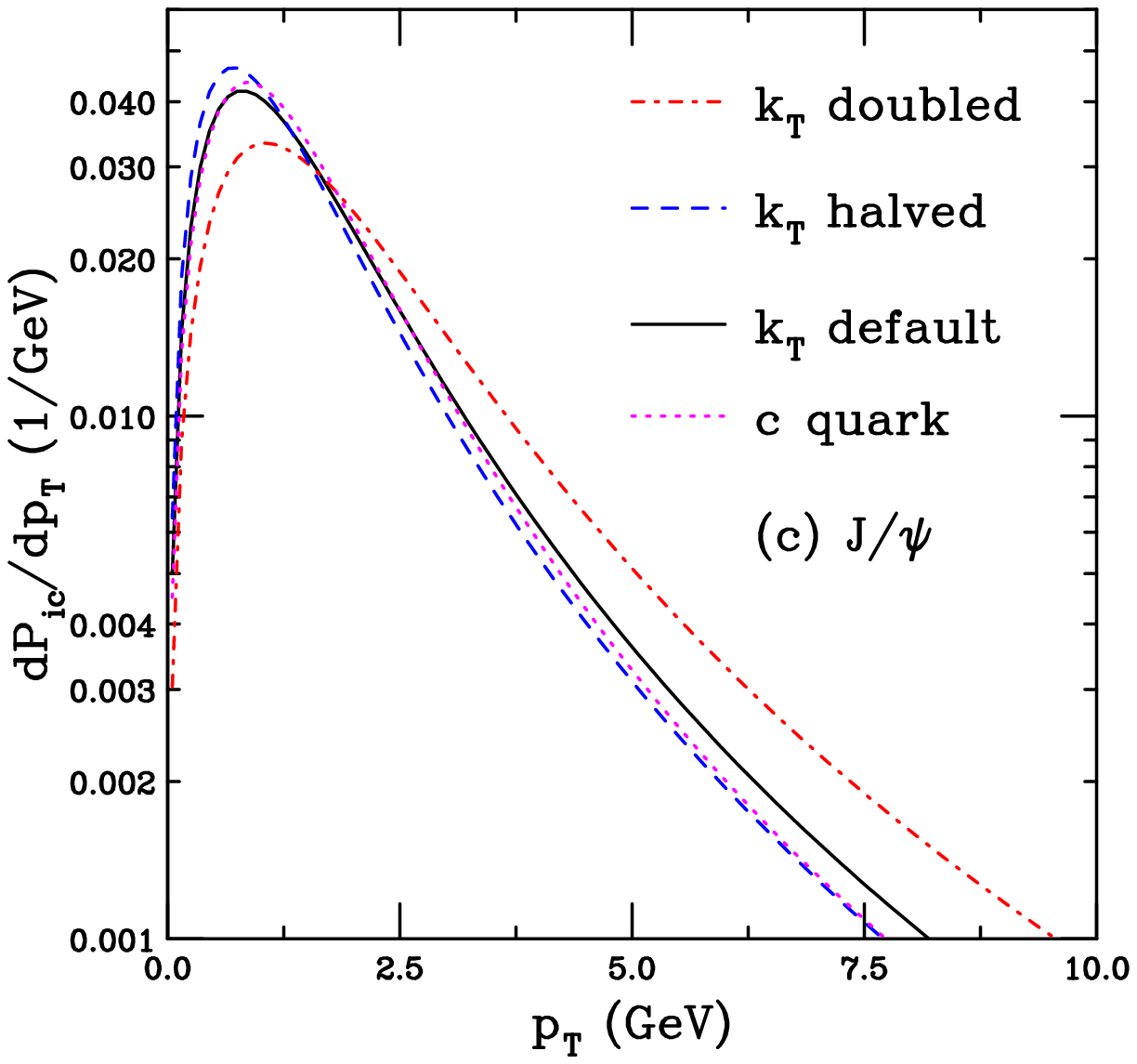} 
    \includegraphics[width=0.4\textwidth]{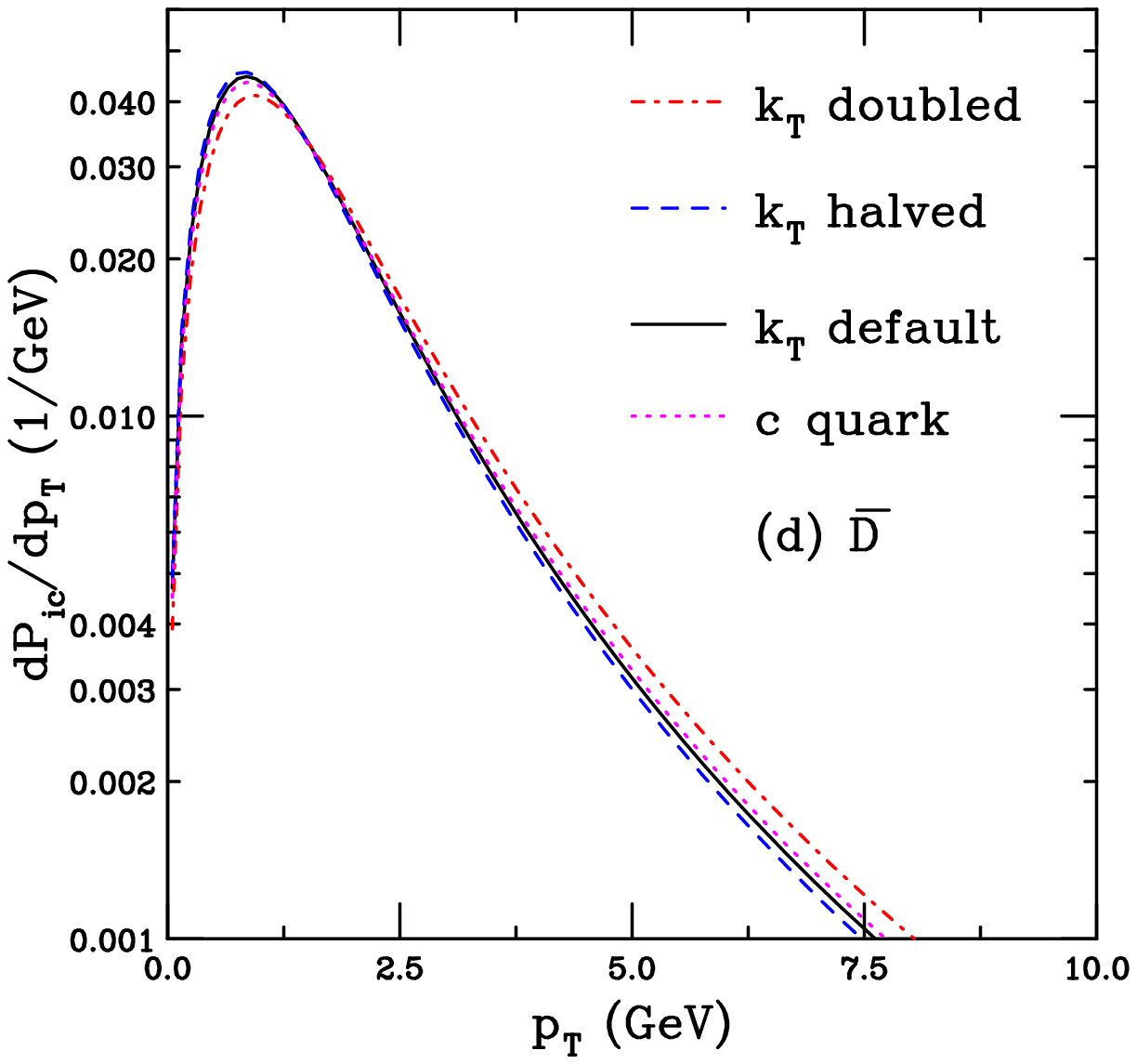}
  \end{center}
  \caption[]{ The probability distributions for $J/\psi$ (a), (c)
    and $\overline D$ (b), (d)
    production from a five-particle proton Fock state as a function of $p_T$.
    In (a) and (c) the results are given for the default mass values,
    $m_c = 1.27$~GeV and $m_q = 0.3$~GeV (solid black), as well as for
    $m_c = 1.5$~GeV and $m_q = 0.3$~GeV (dot-dashed red), and 
    $m_c = 1.27$~GeV and $m_q = 0.02$~GeV (dashed blue),
    In (b) and (d) results are shown for different values of the $k_T$
    range for the light and charm quarks.  The solid black curve employs the
    default values, $k_q^{\rm max} = 0.2$~GeV and $k_c^{\rm max} = 1.0$~GeV while
    the red dot-dashed curve increases the default values by a factor of two,
    $k_q^{\rm max} = 0.4$~GeV and $k_c^{\rm max} = 2$~GeV,
    and the blue dashed curves employs half the default values,
    $k_q^{\rm max} = 0.1$~GeV and $k_c^{\rm max} = 0.5$~GeV.
    The dotted magenta curves show the $p_T$ distributions for a single charm
    quark from the same state.  All distributions are normalized to unity.
  }
\label{ic_pTdists}
\end{figure}

The effects of the mass and scale variation on the $J/\psi$ and $\overline D$
$p_T$ distributions, integrated over all rapidity, are shown in
Fig.~\ref{ic_pTdists}.  The $J/\psi$ probability distributions
are shown in Fig.~\ref{ic_pTdists}(a) and (b) while those for the $\overline D$
are given in Fig.~\ref{ic_pTdists}(c) and (d).  The default values are given by
the solid black curves in both cases.  The $c$ quark $p_T$ distribution,
calculated with the default values, is given by the dotted magenta curve in all
plots. All distributions are normalized to unity
for the parameters employed. 

The $J/\psi$ $p_T$ distribution is slightly broader than that of the
$\overline D$ with a harder distribution at high $p_T$.  Indeed, the
$\overline D$ $p_T$ distribution is very similar to that of the charm quark
itself, a not necessarily surprising result since it is comprised of a single
charm quark and a light quark.  The average charm quark $p_T$, 1.997~GeV, is
slightly larger than that of the $\overline D$ meson, 1.967~GeV, because the
charm quark is comoving with the light quark in the meson.
The $J/\psi$ distribution, on the other hand, is harder than the charm quark
distribution.  Increasing the charm quark mass to 1.5~GeV
hardens both $p_T$ distributions while reducing the light quark mass to near
zero softens them.  Here also doubling the $k_T$ integration range hardens the
$p_T$ distributions although considerably more for the $J/\psi$ than the
$\overline D$.  Halving the $k_T$ integration range softens the $p_T$
distributions albeit not significantly.
In the rest of this work, the default values of parton mass and $k_T$ range
are assumed.

The shape of the $p_T$ distribution from intrinsic charm depends on the
chosen quark masses.  Increasing the quark masses broadens the $p_T$
distribution further.  For example, choosing $m_c = 1.5$~GeV makes the $J/\psi$
$p_T$ distribution broader than that of the charm quark distribution shown in
Fig.~\ref{ic_pTdists}(b).  On the other hand, reducing the light quark mass to
20~MeV, closer to the current quark mass, results in a more steeply falling
$J/\psi$ $p_T$ distribution than that obtained by taking $k_q^{\rm max} = 0.1$~GeV
and $k_c^{\rm max} = 0.5$~GeV.

The intrinsic charm $p_T$ distributions also depend on the $k_T$ range chosen in
the integrals in Eq.~(\ref{icdenom}).  If the
limits of the integration range are doubled to $k_q^{\rm max} = 0.4$~GeV
and $k_c^{\rm max} = 2.0$~GeV, the $p_T$ distribution is broadened, see the blue
dashed curve in Fig.~\ref{ic_pTdists}(b), while halving the integration range to
0.1~GeV and 0.5~GeV for the light and charm quarks respectively, results in
the dot-dashed magenta curve.  The average $J/\psi$ $p_T$ resulting from these
values of $k_i^{\rm max}$ are 2.06, 2.48, and 1.90~GeV for the default range,
doubling the range, and halving the range, respectively.  The average $p_T$ of
the charm quark in the default integration range is 2.00~GeV.

\begin{figure}
  \begin{center}
    \includegraphics[width=0.4\textwidth]{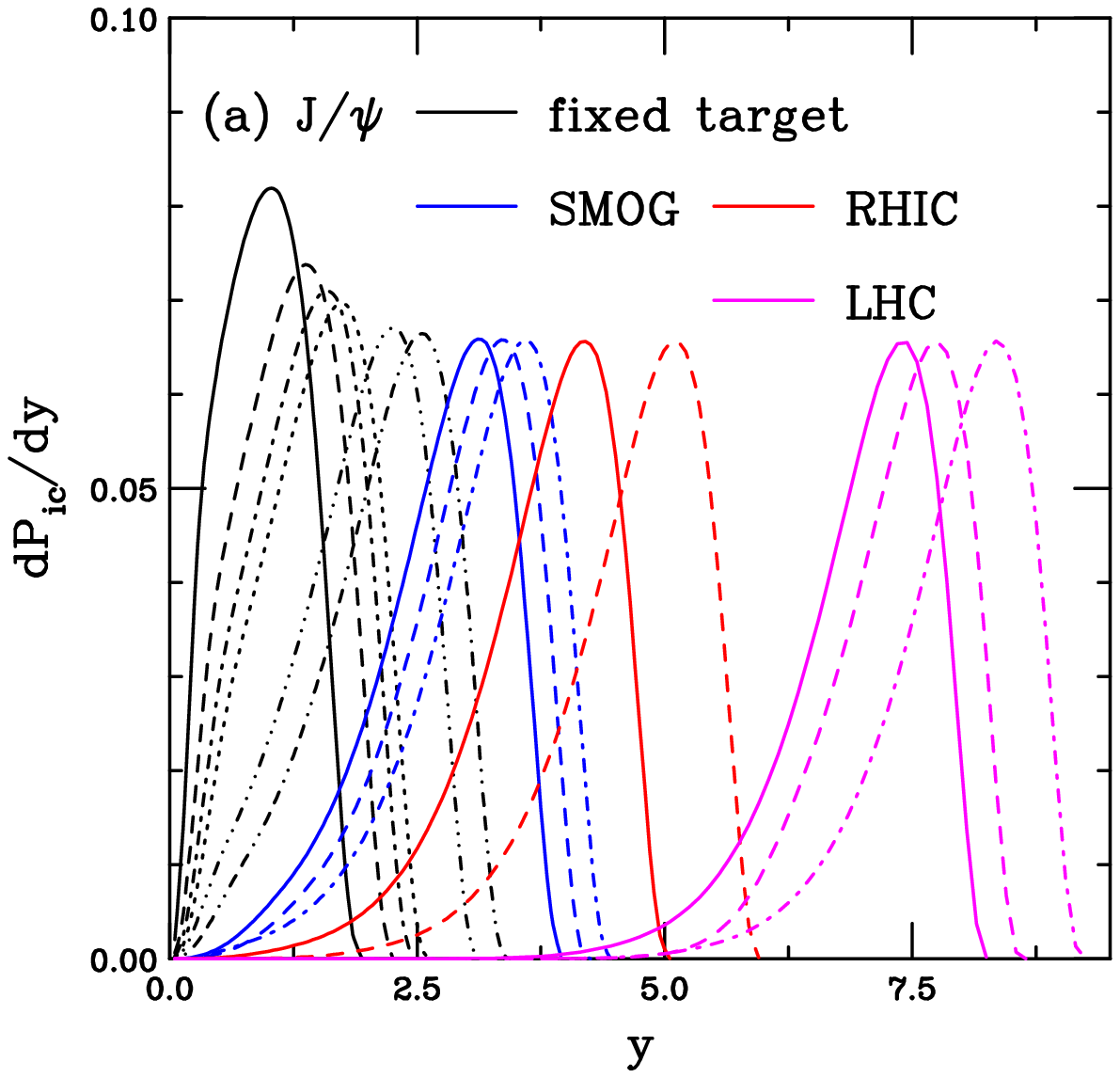}
    \includegraphics[width=0.4\textwidth]{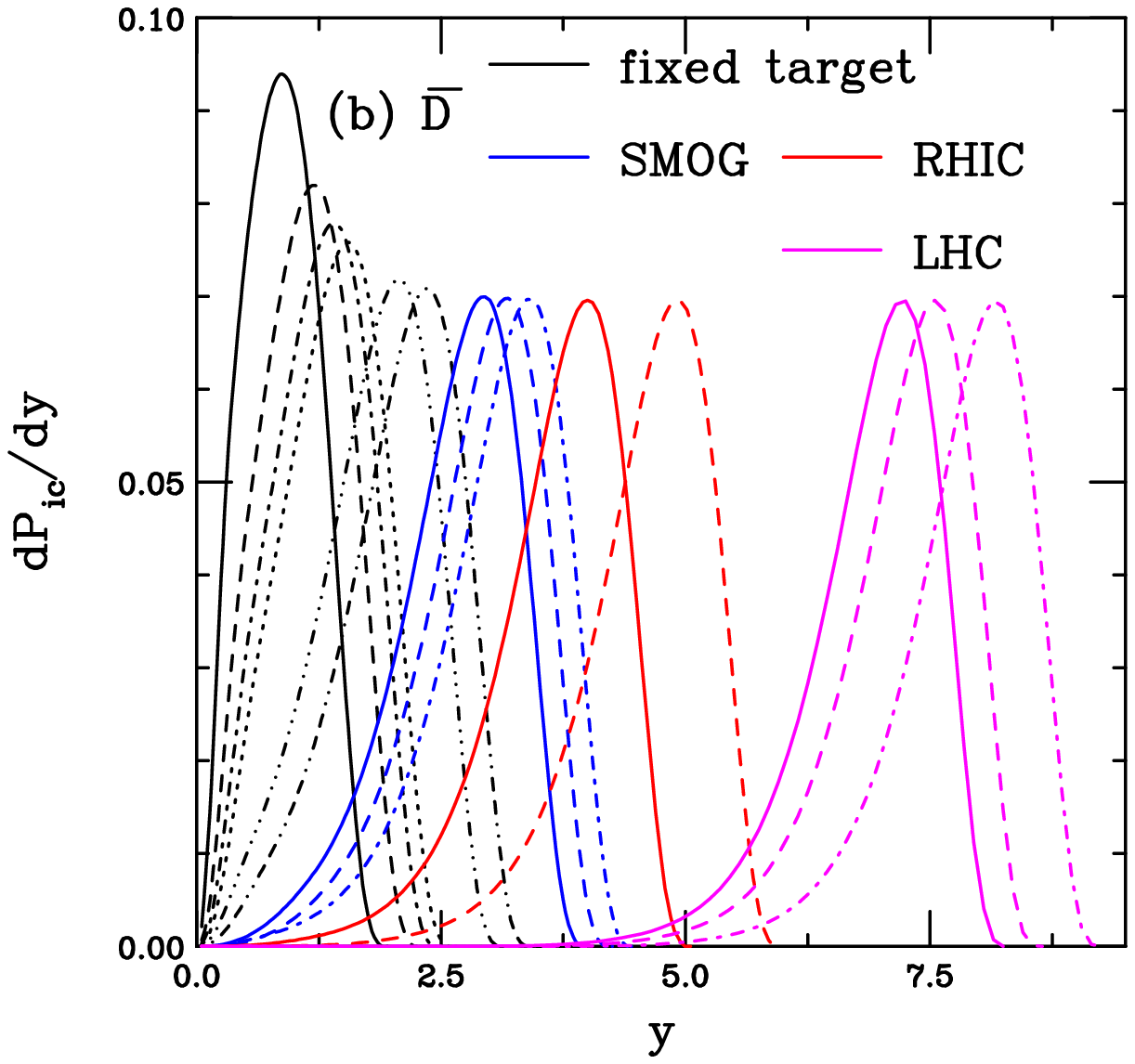}
  \end{center}
  \caption[]{ The probability distributions as a function of
    rapidity for $J/\psi$ (a) and $\overline D$ (b) mesons from a
    five-particle proton Fock state.  The black curves are for fixed-target
    energies with $p_{\rm lab} = 40$ (solid), 80 (dashed), 120 (dot-dashed),
    158 (dotted), 450 (dot-dot-dot-dashed) and 800 (dot-dot-dash-dashed) GeV.
    The blue curves correspond to SMOG energies: $\sqrt{s} = 69$ (solid),
    87.7 (dashed) and 110.4 (dot-dashed) GeV.  The red curves denote RHIC
    energies of 200 (solid) and 500 (dashed) GeV.  Finally, the magenta curves
    show the distributions for LHC energies of $\sqrt{s} = 5$ (solid), 7
    (dashed), and 13 (dot-dashed) TeV.
    All distributions are normalized to unity.
    }
\label{ic_ydists_en}
\end{figure}

The energy dependence of the rapidity distribution
from the intrinsic charm state for $J/\psi$ and $\overline D$ mesons is shown in
Fig.~\ref{ic_ydists_en} for all energies considered.  As already mentioned,
while the $x_F$ distribution of intrinsic charm in the proton is invariant with
respect to beam energy, the rapidity distribution, related to $x_F$ through
$\sqrt{s}$, is not.  Instead, the intrinsic charm rapidity distribution is
boosted forward with $\sqrt{s}$, as seen in Fig.~\ref{ic_ydists_en}.  The
$J/\psi$ distribution is boosted slightly further forward than the $\overline D$
because it has two charm quarks in the state, rather than a single charm quark
combined with a light quark like the $\overline D$ meson.
The rapidity distributions at the lowest energies
are rather close to midrapidity, $y \sim 0$.  As the energy increases, the
distributions develop a long tail at low rapidity, increasing more slowly on
the lower rapidity side of the peak and dropping off more sharply on the higher
rapidity side, as one approaches the edge of available phase space.  It is
clear from these distributions that, at high energies, $J/\psi$ and
$\overline D$ production by intrinsic charm will be difficult to detect in most
collider detectors because these detectors do not cover the far forward rapidity
region where the intrinsic charm probability is highest.

\begin{figure}
  \begin{center}
    \includegraphics[width=0.4\textwidth]{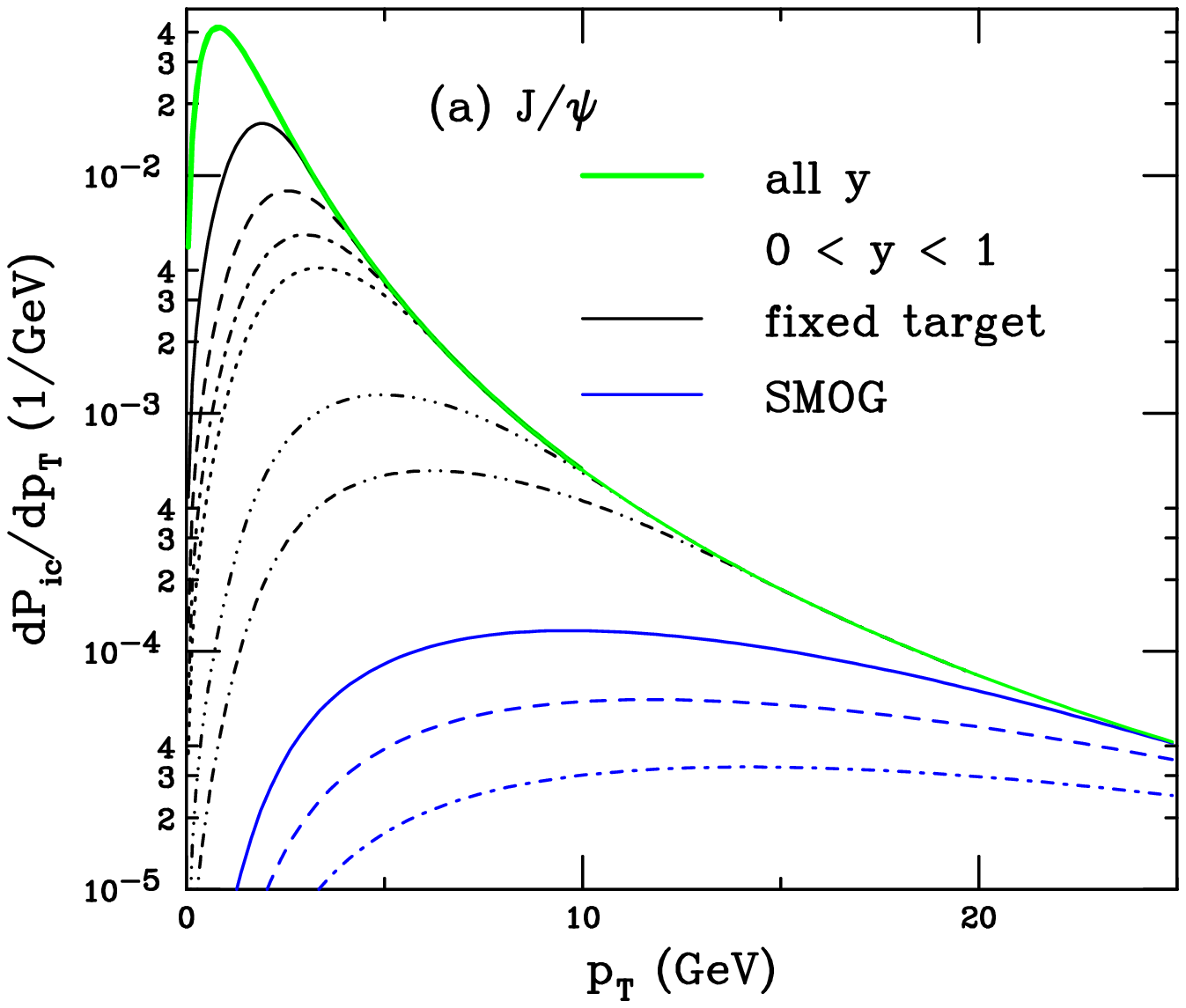}
    \includegraphics[width=0.4\textwidth]{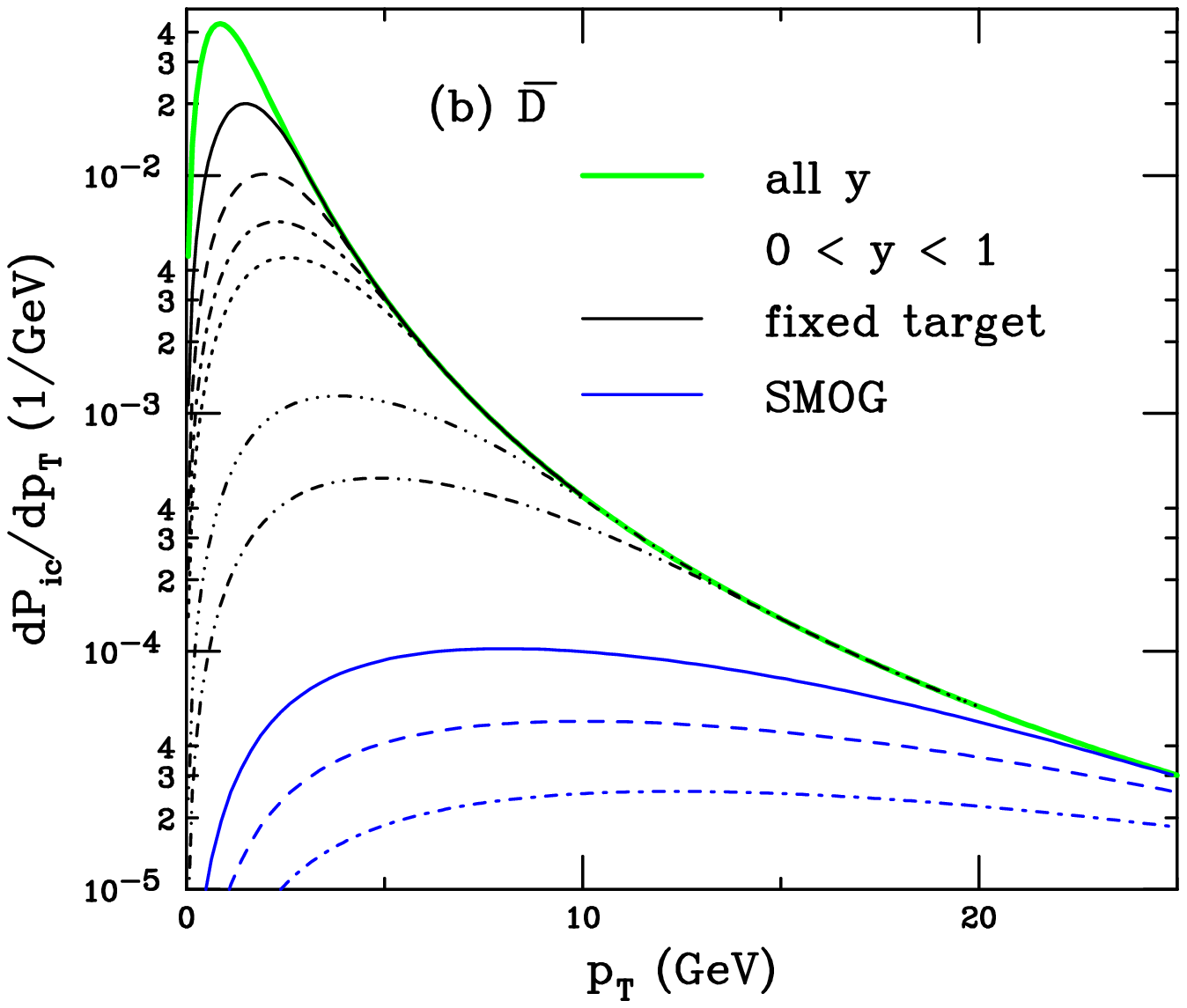} \\
    \includegraphics[width=0.4\textwidth]{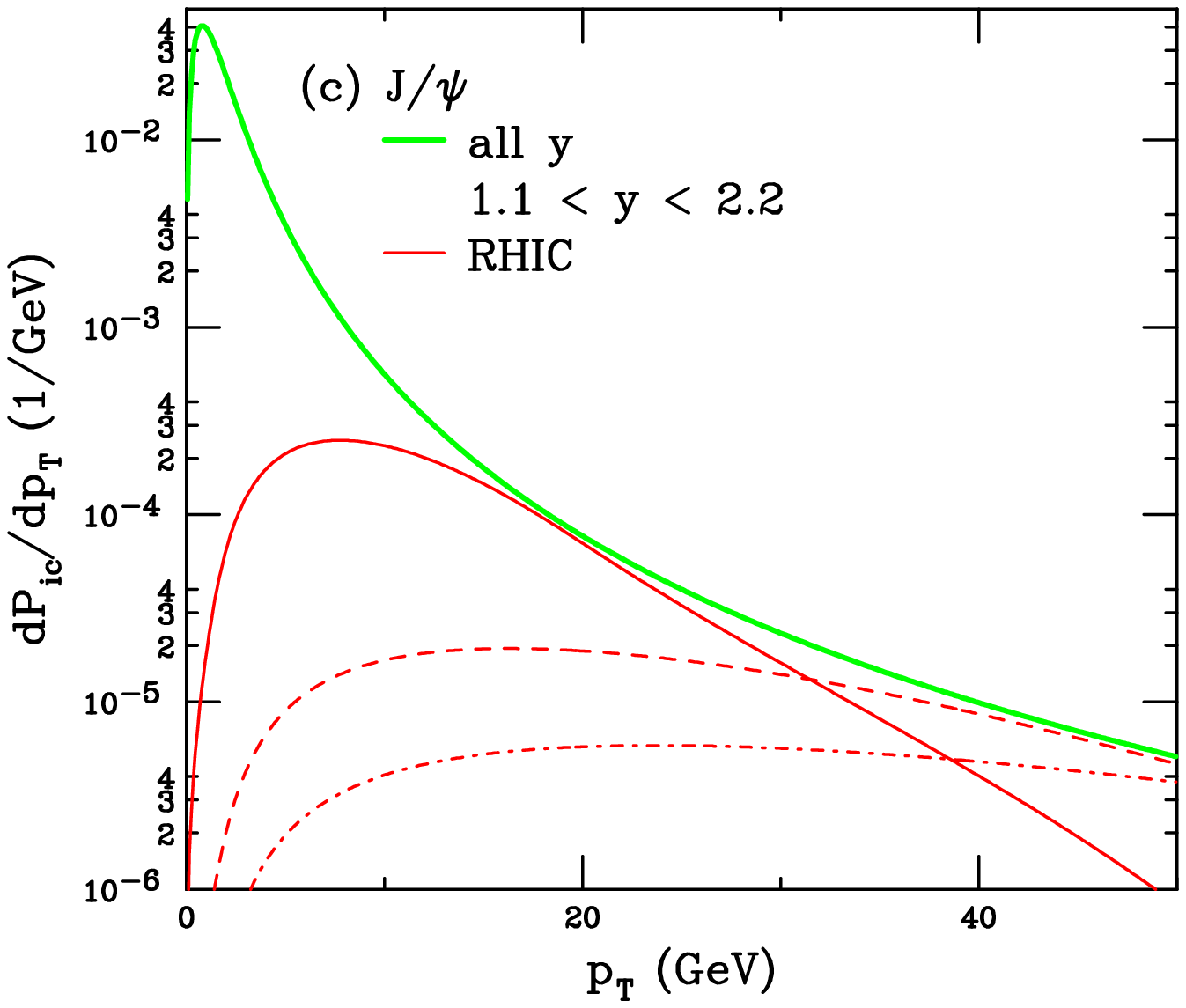}
    \includegraphics[width=0.4\textwidth]{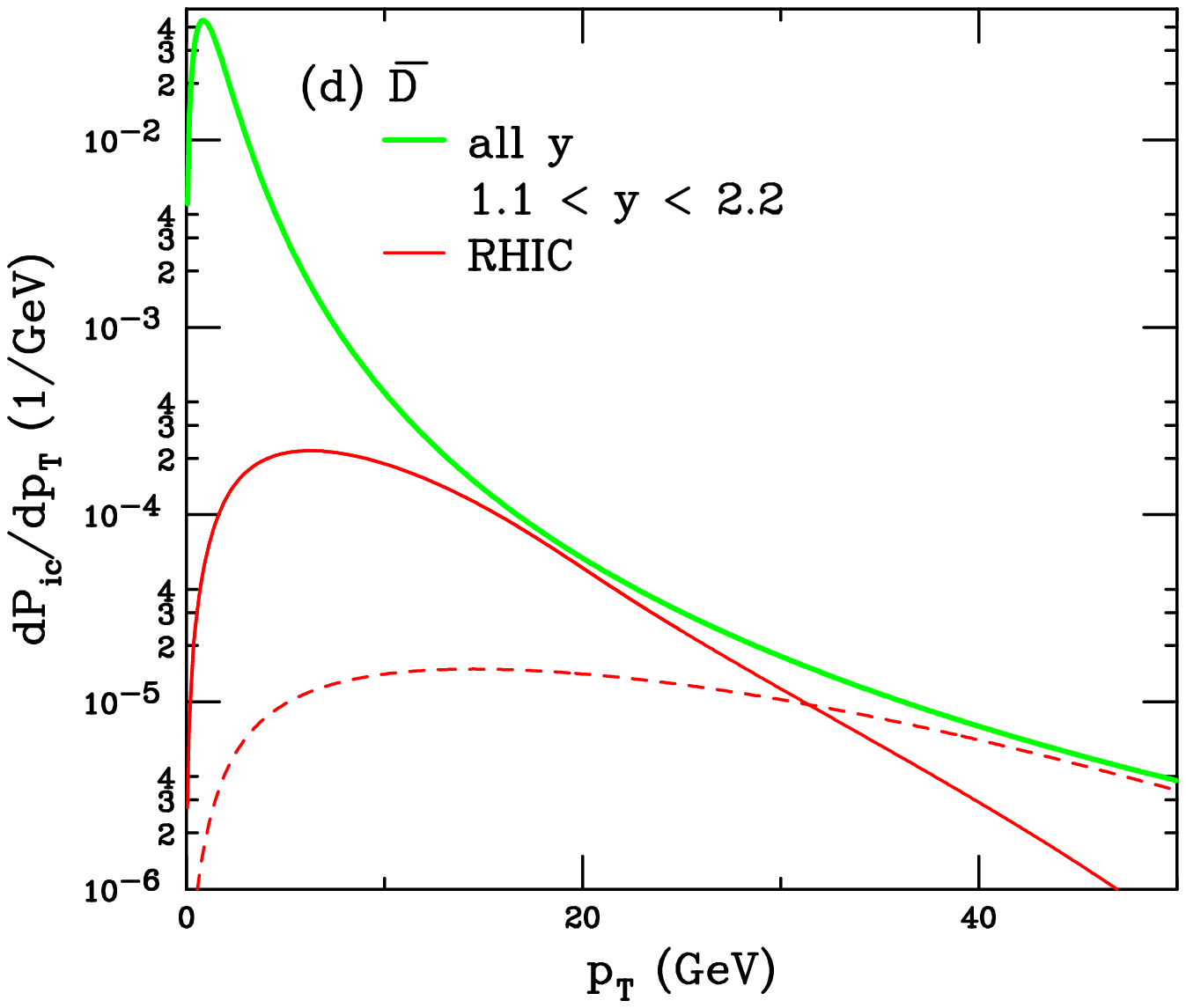} \\
    \includegraphics[width=0.4\textwidth]{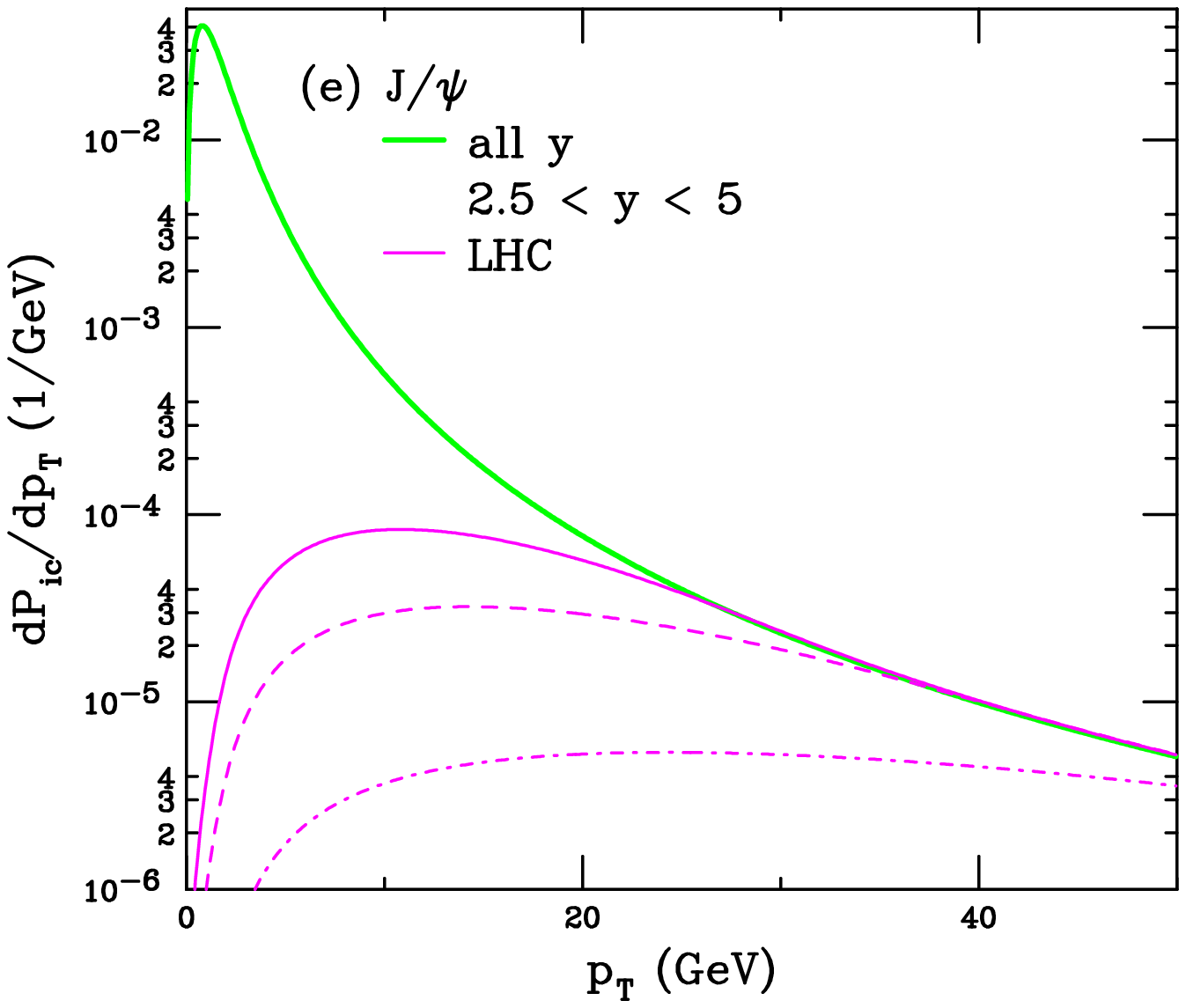}
    \includegraphics[width=0.4\textwidth]{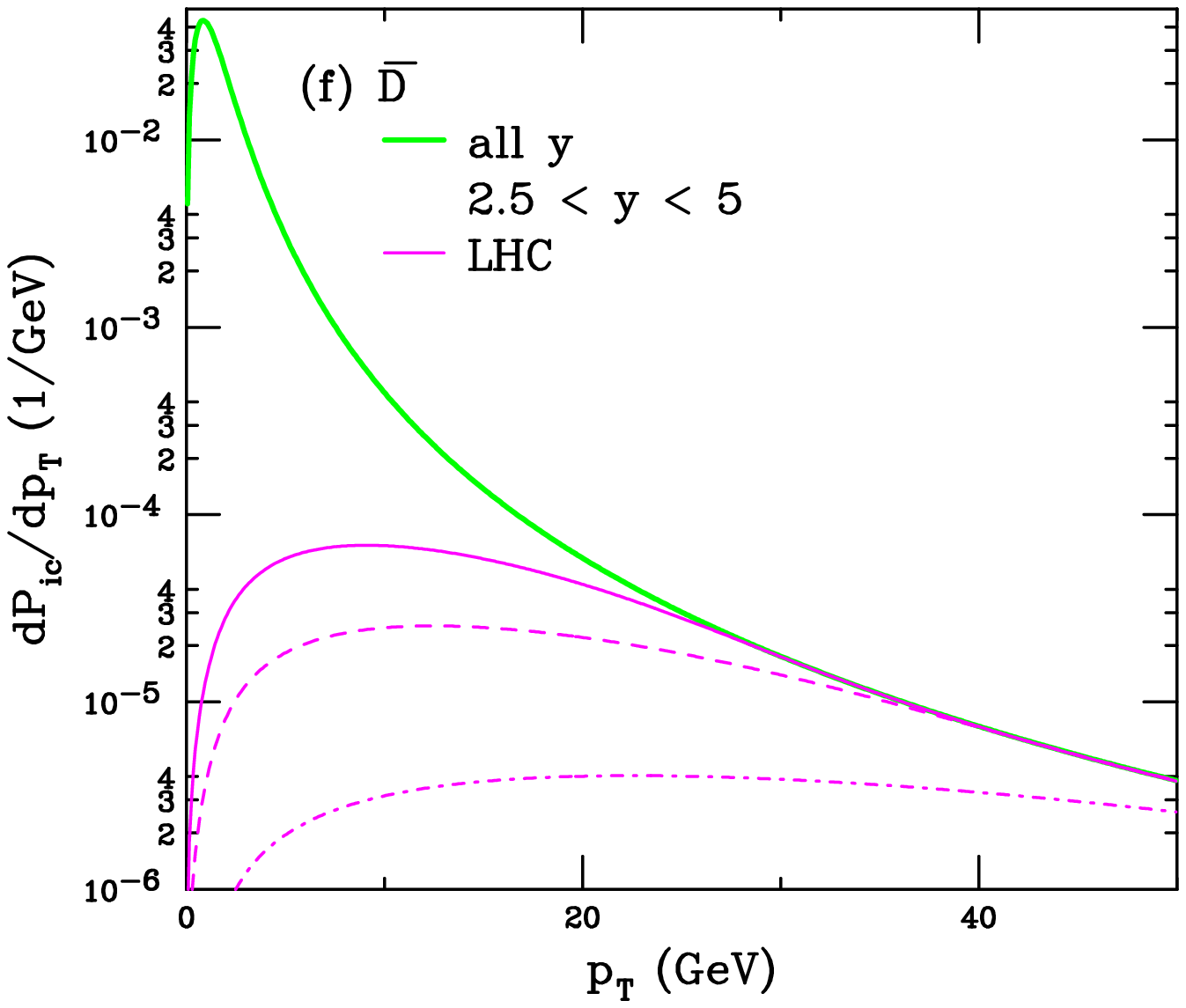}
  \end{center}
  \caption[]{ The probability distributions as a function of
    rapidity for $J/\psi$ (a), (c), (e) and $\overline D$ (b), (d), (f) mesons
    from a five-particle proton Fock state.  The green curves in all plots show
    the $p_T$ distributions integrated over all rapidity.  The calculations
    shown for other energies all include cuts in rapidity.
    The black curves in (a) and (b) are for fixed-target
    energies with $p_{\rm lab} = 40$ (solid), 80 (dashed), 120 (dot-dashed),
    158 (dotted), 450 (dot-dot-dot-dashed) and 800 (dot-dot-dash-dashed) GeV.
    The blue curves in (a) and (b) correspond to SMOG energies:
    $\sqrt{s} = 69$ (solid),
    87.7 (dashed) and 110.4 (dot-dashed) GeV.  The red curves in (c) and (d)
    denote the RHIC
    energies of 200 (solid) and 500 (dashed) GeV.  Finally, the magenta curves
    in (e) and (f)
    show the distributions for LHC energies of $\sqrt{s} = 5$ (solid), 7
    (dashed), and 13 (dot-dashed) TeV.
    All distributions are normalized to unity and integrated over all rapidity.
    }
\label{ic_pTdists_en}
\end{figure}

The $J/\psi$ and $\overline D$ $p_T$ distributions arising from intrinsic charm
in the five-particle proton Fock state are shown in Fig.~\ref{ic_pTdists_en}.
Again, the $J/\psi$ distribution is
slightly harder than that of the $\overline D$ because of the higher number of
charm quarks in the $J/\psi$.  The $J/\psi$ distribution also has a higher
average $p_T$ after rapidity cuts are applied to the distribution, as
discussed below.

Like the $x_F$ distributions, $p_T$ distributions from intrinsic charm are also
independent of energy if all of $x_F$ or rapidity space were covered by the
detectors.  The only energy limitation is the energy of the beam itself.
However, most detectors have a finite rapidity coverage which can
strongly influence the $p_T$ distributions as shown in Fig.~\ref{ic_pTdists_en}.
The $p_T$ distributions integrated over all rapidity are shown in green with
those resulting in rapidity cuts at given energies are shown in the same
colors as in Fig.~\ref{ic_ydists_en}.  Rapidity coverage in the range
$0 < y < 1$ is assumed for fixed-target energies from $p_{\rm lab} = 40$~GeV up
to the highest SMOG energy of $\sqrt{s} = 110.4$~GeV.

In this same rapidity
interval, the contribution to the intrinsic charm cross section falling within
the rapidity coverage decreases strongly with energy, by more than two orders of
magnitude as $p_T \rightarrow 0$.  The low $p_T$ part of the probability is
reduced while, at high $p_T$, the distributions join the tail of the $p_T$
distribution integrated over all rapidity space.  This is generally also the
case for collider detector coverage at forward rapidity, aside from
$\sqrt{s} = 200$~GeV at RHIC which has a different curvature at forward
rapidity (solid red curve in Fig.~\ref{ic_pTdists_en}(c)).
This curvature does not appear in the same rapidity region at
$\sqrt{s} = 500$~GeV (dashed red curve) or at midrapidity at
$\sqrt{s} = 200$~GeV (dot-dashed red curve).  This difference is because, at
large $p_T$ and the upper end of the rapidity range, $x_F > 1$, an unphysical
region.  In a range with a lower maximum rapidity at the same center of mass
energy, the $p_T$ distribution merges with the distribution integrated over all
rapidity.  On the other hand, if a region with a larger maximum rapidity is
chosen, the curvature changes more, resulting in a more steeply falling $p_T$
distribution.  This behavior is not seen in any of the other cases because
the condition $x_F < 1$ is satisfied everywhere but at $\sqrt{s} = 200$~GeV
in the range $1.1 < y < 2.2$.  The $\overline D$ distributions are subject to
the same kinematic constraint, as seen in Fig.~\ref{ic_pTdists_en}(d).

The normalization of the intrinsic charm cross section is now discussed.
The intrinsic charm production cross section from the
$|uudc \overline c \rangle$ component of the proton can be written as 
\be
\sigma_{\rm ic}(pp) = P_{{\rm ic}\, 5} \sigma_{p N}^{\rm in}
\frac{\mu^2}{4 \widehat{m}_c^2} \, \, .
\label{icsign}
\ee
The factor of $\mu^2/4 \widehat{m}_c^2$ arises from the soft
interaction which breaks the coherence of the Fock state where 
$\mu^2 = 0.1$~GeV$^2$ is assumed, see Ref.~\cite{VBH1}.  Here the inelastic
$pN$ cross section, $\sigma_{pN}^{\rm in} = 30$~mb, is employed.  Although this
quantity can change slowly with $\sqrt{s}$, no variation is assumed in the
calculations shown.

Equation~(\ref{icsign}) is used for open charm production from the Fock state,
$\sigma_{\rm ic}^{\overline D}(pp) = \sigma_{\rm ic}(pp)$. The
$J/\psi$ cross section from the same intrinsic charm stated
is calculated by scaling Eq.~(\ref{icsign}) by the 
factor $F_C$ used in the CEM calculation in
Eq.~(\ref{sigCEM}),
\be
\sigma_{\rm ic}^{J/\psi}(pp) = F_C \sigma_{\rm ic}(pp) \, \, .
\label{icsigJpsi}
\ee

The nuclear dependence of the intrinsic charm contribution is assumed to be the
same as that extracted for the nuclear surface-like component of $J/\psi$
dependence by the NA3 Collaboration \cite{NA3}.  The $A$
dependence is the same for both open charm and $J/\psi$,
\be 
\sigma_{\rm ic}^{\overline D}(pA) & = & \sigma_{\rm ic}^{\overline D}(pp) \, A^\beta \, \, , \label{icsigD_pA} \\
\sigma_{\rm ic}^{J/\psi}(pA) & = & \sigma_{\rm ic}^{J/\psi}(pp) \, A^\beta \, \, , 
\label{icsigJpsi_pA}
\ee
with $\beta = 0.71$ \cite{NA3}.

To represent the uncertainties on intrinsic charm, several values of the
intrinsic charm probability, $P_{{\rm ic}\, 5}^0$, are employed. 
The EMC charm structure function data is consistent with 
$P^0_{{\rm ic}\, 5} = 0.31$\%  for low energy virtual photons but
$P^0_{{\rm ic}\, 5}$
could be as large as 1\% for the highest virtual photon energies
\cite{EMC,hsv}.  For a lower limit, a probability of 0.1\% is used.
Generally, a subset of these three results is shown in Sec.~\ref{model_comp}
unless otherwise noted.

In this work, the formulation for intrinsic charm in the proton wave function
postulated by Brodsky and collaborators in Refs.~\cite{intc1,intc2}, the form
in Eq.~(\ref{icdenom}), has been
adapted.  Other variants of the intrinsic charm distribution in the proton
exist, including meson-cloud models where the proton fluctuates into a
$\overline D(u \overline c) \Lambda_c (udc)$ state
\cite{Paiva:1996dd,Neubert:1993mb,Steffens:1999hx,Hobbs:2013bia} and a
sea-like distribution \cite{Pumplin:2007wg,Nadolsky:2008zw}.

Intrinsic charm
has also been included in global analyses of the parton densities
\cite{Pumplin:2007wg,Nadolsky:2008zw,Dulat:2013hea,Jimenez-Delgado:2014zga,NNPDF_IC}.
The range of $P_{{\rm ic}\, 5}^0$
explored here is consistent with the results of these
global analyses.  For more details of these other works, see the review of
Ref.~\cite{IC_rev}.  (See Ref.~\cite{Blumlein} for a discussion of a possible
kinematic constraint on intrinsic charm in deep-inelastic scattering.)
New evidence for a finite charm quark asymmetry in the
nucleon wavefunction from lattice gauge theory, consistent with intrinsic
charm, was presented in Ref.~\cite{Sufian:2020coz}.  See also the recent
review in Ref.~\cite{Stan_review} for
more applications of intrinsic heavy quark states.

\section{Results}
\label{model_comp}

This section is divided into three parts.  The first shows predictions for $p+p$
collisions as a function of $\sqrt{s}$.  The second compares calculations of the average cold nuclear effects as a function of
$x_F$ to data from previous fixed-target measurements of $J/\psi$ production. 
Finally, the third presents predictions for
cold nuclear matter effects, without and with intrinsic charm.

\subsection{$p+p$ cross sections}
\label{sec:ppResults}

First, results are shown for $p+p$ collisions from
$p_{\rm lab} = 40$~GeV to $\sqrt{s} = 13$~TeV.  The rapidity and transverse
momentum distributions are shown for $J/\psi$ and $\overline D$ mesons without
and with the intrinsic charm contributions to illustrate at which center of mass
energies and for which kinematic regions intrinsic charm
may best be observable.

The cross sections for $\overline D$ and $J/\psi$ production in $p+p$ collisions
including the perturbative QCD and intrinsic charm contributions are:
\be
\sigma_{pp}^{\overline D} & = & \sigma_{\rm OHF}(pp) + \sigma_{\rm ic}^{\overline D}(pp)
\label{sig_pp_Dsum} \\
\sigma_{pp}^{J/\psi} & = & \sigma_{\rm CEM}(pp) + \sigma_{\rm ic}^{J/\psi}(pp) \, \, .
\label{sig_pp_Jsum} 
\ee
Here $\sigma_{\rm OHF}(pp)$ and $\sigma_{\rm CEM}(pp)$ are defined in Eqs.~(\ref{sigOHF}) and (\ref{sigCEM}) respectively.
The intrinsic charm contributions can be found in Eq.~(\ref{icsign}) for
$\overline D$ while
$\sigma_{\rm ic}^{J/\psi}(pp)$ is given in Eq.~(\ref{icsigJpsi}).

\subsubsection{Rapidity distributions}
\label{sec:ppRapidity}

The rapidity distributions in $p+p$ collisions are first discussed.  In
Table \ref{table:y_averages}, the average rapidity is shown for $J/\psi$ and
$\overline D$ production at all values of
$\sqrt{s}$ considered.  The averages for intrinsic charm and perturbative QCD
are shown separately.  The intrinsic charm component has a much larger average
rapidity in both cases because even though the $x_F$ distribution is independent
of beam energy, the rapidity distribution is boosted along the beam direction,
see Fig.~\ref{ic_ydists_en}.  As was noted there, the average $J/\psi$ rapidity
from intrinsic charm is higher than that for $\overline D$ from intrinsic charm
because the $J/\psi$ carries two charm quarks while the $\overline D$ only
carries one.

\begin{table}
  \begin{tabular}{|c|c|c||c|c|} \hline
    & \multicolumn{2}{c|}{$J/\psi$} & \multicolumn{2}{|c|}{$\overline D$} \\
    $\sqrt{s}$ (GeV) & $\langle y^{J/\psi}_{\rm ic} \rangle$
    & $\langle y^{J/\psi}_{\rm CEM} \rangle$
    & $\langle y^{\overline D}_{\rm ic} \rangle$
    & $\langle y^{\overline D}_{\rm OHF} \rangle$
   \\ \hline
8.77 & 0.937 & 0.387 & 0.864 & 0.455 \\
12.33 & 1.185 & 0.479 & 1.104 & 0.585 \\
15.07 & 1.344 & 0.536 & 1.260 & 0.642 \\
17.40 & 1.464 & 0.573 & 1.377 & 0.683 \\
29.1 & 1.889 & 0.692 & 1.805 & 0.806 \\
38.8 & 2.155 & 0.754 & 2.079 & 0.873 \\
69 & 2.708 & 0.966 & 2.622 & 1.079 \\
87.7 & 2.942 & 0.997 & 2.856 & 1.107 \\
110.4 & 3.154 & 1.065 & 3.072 & 1.168 \\ \hline
200 & 3.743 & 1.250 & 3.661 & 1.325 \\
500 & 4.657 & 1.661 & 4.575 & 1.678 \\ \hline
5000 & 6.960 & 2.615 & 6.877 & 2.491 \\
7000 & 7.354 & 2.755 & 7.214 & 2.616 \\
13000 & 7.914 & 3.009 & 7.833 & 2.843 \\ \hline
  \end{tabular}
  \caption[]{\label{table:y_averages}  The average rapidity for $J/\psi$ and
    $\overline D$ meson production as a function of center of mass energy,
    $\sqrt{s}$ for both intrinsic charm and perturbative QCD production
    respectively.
  }
\end{table}

\begin{figure}
  \begin{center}
    \includegraphics[width=0.4\textwidth]{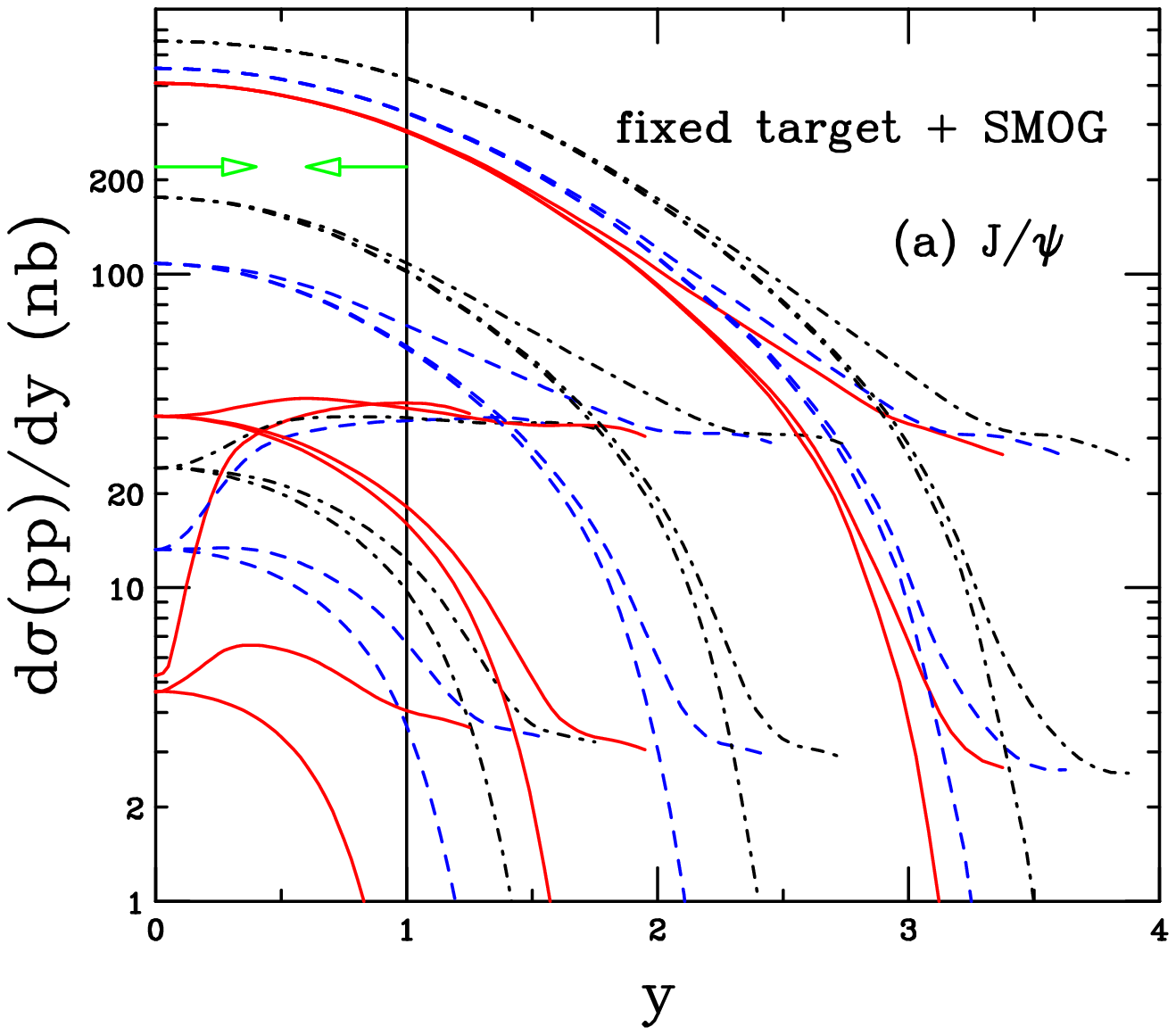}
    \includegraphics[width=0.4\textwidth]{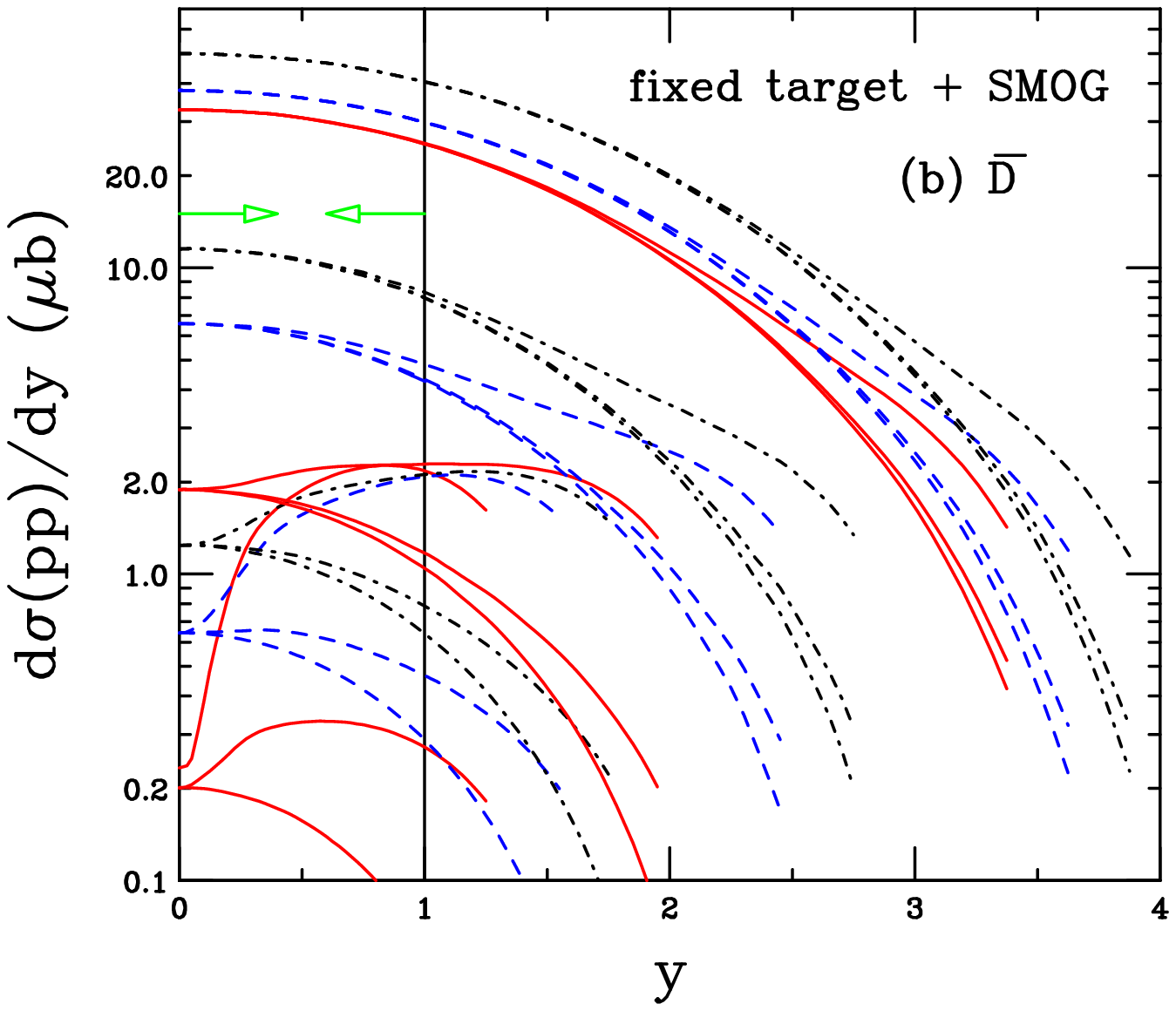} \\
    \includegraphics[width=0.4\textwidth]{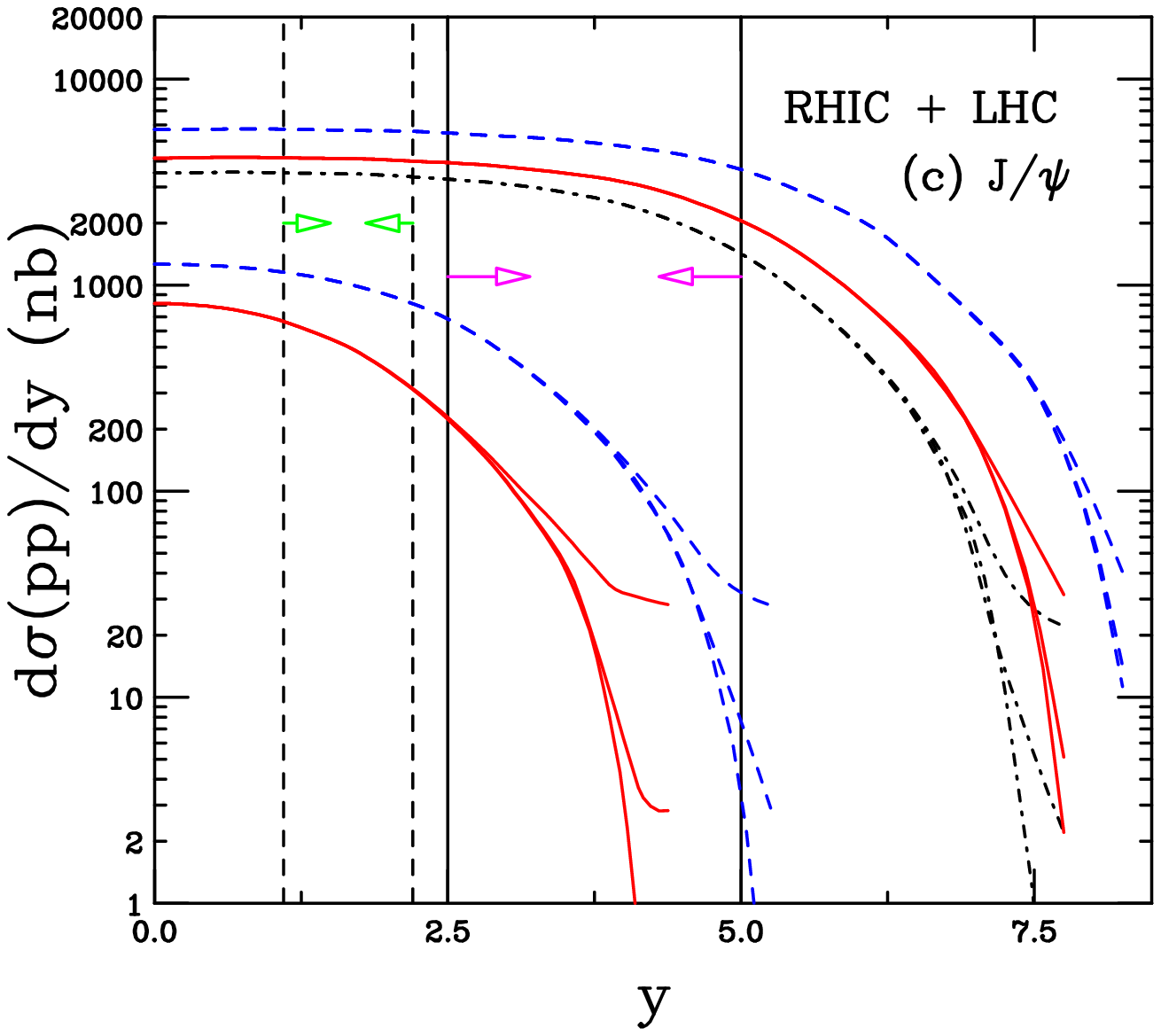}
    \includegraphics[width=0.4\textwidth]{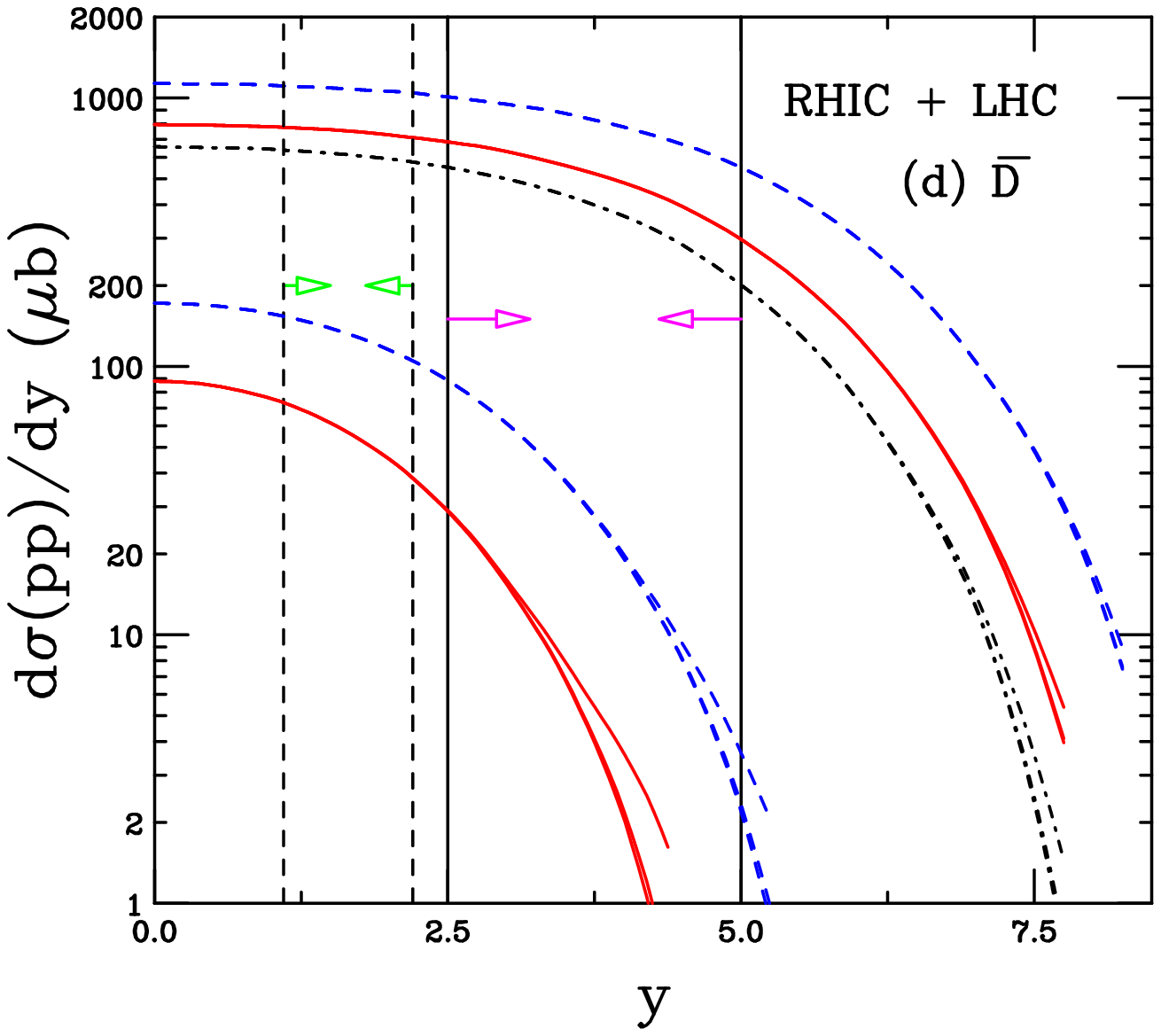} 
  \end{center}
  \caption[]{ The combined rapidity distributions for
    $J/\psi$ (a), (c) and $\overline D$ (b), (d) mesons, including
    both the typical perturbative QCD contribution and intrinsic charm from a
    five-particle proton Fock state.  Three curves are shown for each energy.
    From lowest to highest (when separable) they show: no intrinsic charm (pQCD
    only); $P_{\rm ic \, 5}^0 = 0.1$\%; and  $P_{\rm ic \, 5}^0 = 1$\%.
    In (a) and (b) the results are shown for fixed-target and SMOG energies.
    From lowest to highest the curves represent $p_{\rm lab} = 40$~GeV
    (red solid),
    80~GeV (blue dashed), 120~GeV (black dot-dashed),
    158~GeV (red solid), 450~GeV (blue dashed), 800~GeV (black dot-dashed), 
    $\sqrt{s} = 69$~GeV (solid red), 87.7~GeV (blue dashed) and 110.4~GeV
    (black dot-dashed).  The vertical line with the green arrows shows
    the assumed rapidity
    acceptance of $0 < y < 1$.  In (c) and (d) the RHIC
    energies of $\sqrt{s} = 200$~GeV (solid red) and 500~GeV (blue dashed)
    along with the LHC energies of $\sqrt{s} = 5$~TeV (red solid), 7~TeV
    (blue dashed), and 13~TeV (black dot-dashed) are given.
    The vertical lines show the assumed rapidity
    acceptance of $1.1 < y < 2.2$ for RHIC (dashed, with green arrows) and
    $2.5 < y < 5$ for LHC (solid, with magenta arrows).
    }
\label{pp_ydists_en}
\end{figure}

On the other hand, the perturbative calculations give somewhat smaller average
rapidities for $J/\psi$ than for $\overline D$ mesons, especially at lower
center of mass energies.  In this case, producing a state with a $c \overline c$
pair instead of a hadron with a single $\overline c$ quark is a disadvantage.
The $J/\psi$ rapidity distribution falls more steeply with increasing $y$, as
shown in Fig.~\ref{pp_ydists_en}.  At collider energies, the average rapidity of
the $J/\psi$ becomes equal to or somewhat larger than that of the $\overline D$,
not because the distribution is less steeply falling at forward rapidity but
because the $J/\psi$ distribution is broader over a wider range near
midrapidity.

Of course the average rapidity of charmonium and open charm calculated
perturbatively is generally less than half that found from intrinsic charm.
This is obvious because the intrinsic charm arises wholly from the projectile
proton at $y > 0$ while the perturbative calculation is maximal at $y = 0$ in
$p+p$ collisions because the $c \overline c$ pair is produced by interactions of
one parton from each proton, typically two gluons at higher energies.

Figure~\ref{pp_ydists_en} compares the $J/\psi$ and $\overline D$ meson rapidity
distributions side by side for all energies considered.  There are three curves
for each energy, all curves at a given energy are the same color and line type.
The results shown are no intrinsic charm; $P_{{\rm ic} \, 5}^0 = 0.1$\%  and 1\%.
The sum of the two distributions is shown over the rapidity range of the
perturbative contribution.  Thus the full range of the intrinsic charm
distribution is not shown in the figure.

At the lowest energy in particular, $\sqrt{s} = 8.77$~GeV
($p_{\rm lab} = 40$~GeV), not far above the
$J/\psi$ production threshold, the total
intrinsic charm cross section is compatible
with the one calculated in perturbative QCD and adding even a very small
intrinsic charm contribution produces a very large effect near $y=0$ because
the boost of the intrinsic charm rapidity distribution is small with
$\langle y \rangle < 1$ in both cases.  If the rapidity distribution could be
measured with high statistics in rapidity bins of $\Delta y = 0.1$ and an
significant increase is seen away from $y = 0$ or a relatively flat distribution
is observed rather than a decreasing distribution, this would be a strong
indication of an important contribution from intrinsic charm.

As the energy increases, the perturbative cross section grows while the
intrinsic charm cross section remains effectively constant so that, already for
$\sqrt{s} = 17.4$~GeV ($p_{\rm lab} = 158$~GeV),
instead of an increase of the cross section for $y>0$
with $P_{{\rm ic} \, 5}^0 = 1$\%, it is effectively constant in the interval
$0 < y < 1$.  At still higher $\sqrt{s}$, the boosted rapidity distribution for
intrinsic charm and the growing perturbative cross sections results in an
enhancement over the perturbative cross section that appears at ever-increasing
values of rapidity, becoming less susceptible to experimental measurement at
midrapidity.

There is a clear distinction between the $J/\psi$ results at those for
$\overline D$ near the rapidity endpoint of the perturbative calculation.  At
this point, the $J/\psi$ intrinsic charm distribution has just passed the
peak and is decreasing rather slowly on the other side.  Because the cross
section for intrinsic charm is effectively independent of energy, this value is
at approximately 2.75~nb for $P_{{\rm ic}\, 5}^0 = 0.1$\% and 27.5~nb for
$P_{{\rm ic}\, 5}^0 = 1$\%.  The inflection in the rapidity distribution at
this value of the cross section persists for almost all energies, even being
visible in the tail of the rapidity distribution at $\sqrt{s} = 5$~TeV.
As already mentioned, because the $\overline D$ distributions from intrinsic
charm, with only a single charm quark constituent, is not as forward boosted as
the $J/\psi$, at the edge of phase space for the $\overline D$ meson rapidity
distribution, the intrinsic charm contribution to the cross section is already
past the peak of the distribution and is more steeply falling.  Thus the
$\overline D$ distributions do not show such a feature and only exhibit a
slight enhancement of the rapidity distributions in the tails.

In Fig.~\ref{pp_ydists_en}, the rapidity range in which the $p_T$ distributions
are calculated is indicated by the vertical lines.  The fixed-target and SMOG
results will be compared in the range $0 < y < 1$.  This is, of course, a
convenient simplification because the Fermilab E866 collaboration covered the
entire $x_F$ range, up to $x_F \sim 0.95$.  The fixed-target LHCb experiments
employing the SMOG setup have a wider rapidity coverage in the backward
rapidity direction, i.e. at large momentum fractions $x$ in the nuclear targets.
SMOG calculations using the exact targets and rapidity ranges for $J/\psi$ and
$\overline D$ as the experiment will be discussed elsewhere
\cite{RV_SMOG_inprep}, the results here are for illustrative purposes only.

At the RHIC and LHC collider energies, the rapidity ranges indicated correspond
to those of the muon spectrometers of PHENIX, $1.1 < y < 2.2$, at RHIC and
LHCb, $2.5 < y < 5$, at the LHC.  At these more forward rapidities, the
possibility of detecting any enhacement due to intrinsic charm is unlikely.
However, the potential for a strong signal from intrinsic charm via $J/\psi$ or
$\overline D$ meson measurements can be found at midrapidity, particularly at
low center of mass energies.

\subsubsection{Transverse momentum distributions}
\label{sec:pp_pT}

In this section, the energy dependence of the $J/\psi$ and $\overline D$ $p_T$
distributions from both perturbative QCD and intrinsic charm is described.
Similar to the $x_F$ distribution from intrinsic charm, the $p_T$ distribution
is independent of energy in general, limited only by the energy of the parent
proton, as long as the distribution is integrated over all $x_F$ or rapidity.
In that case, the average $p_T$ of the $J/\psi$ and $\overline D$ from a
five-particle intrinsic charm Fock state, as in Eq.~(\ref{icdenom}), is 2.067
and 1.962~GeV respectively.

As shown in Fig.~\ref{ic_pTdists_en}, if a finite rapidity interval
(or $x_F$ range, see Ref.~\cite{RV_SeaQuest}) is considered, the $p_T$
distribution from intrinsic charm is strongly biased toward higher $p_T$. 
Consequently, the percentage of the total intrinsic charm contribution in a
fixed rapidity region decreases with $\sqrt{s}$, as shown in
Table~\ref{table:pT_averages}.  For each energy and a given rapidity interval,
the percent of the intrinsic charm cross section (labeled $f_{\rm IC}$) is given,
along with the average $p_T$ of the intrinsic charm state and from the
perturbative QCD calculation at the same energy.  Both $J/\psi$ and
$\overline D$ results are shown.  The calculations are divided
according to the assumed rapidity interval.  In the case of fixed-target
energies from the proposed NA60+ to the top LHC energy measured by SMOG,
$8.77 < \sqrt{s} < 110.4$~GeV, results are given for $0 < y < 1$.  For RHIC and
LHC energies, the more forward rapidity regions, $1.1 < y < 2.2$ and
$2.5 < y < 5$, are considered respectively.

\begin{table}
  \begin{tabular}{|c|c|c|c||c|c|c|} \hline
    & \multicolumn{3}{c|}{$J/\psi$} & \multicolumn{3}{|c|}{$\overline D$} \\
    $\sqrt{s}$ (GeV)
    & $f_{\rm IC}$ 
    & $\langle p_T \rangle_{\rm ic}^{J/\psi}$ (GeV)
    & $\langle p_T \rangle_{\rm CEM}^{J/\psi}$ (GeV)
    & $f_{\rm IC}$ 
    & $\langle p_T \rangle_{\rm ic}^{\overline D}$ (GeV) 
    & $\langle p_T \rangle_{\rm OHF}^{\overline D}$ (GeV) \\ \hline 
\multicolumn{7}{c}{$0 < y < 1$} \\ \hline
 8.77 & 54\% & 2.977 & 1.228 & 62\% & 2.555 & 0.860 \\
 12.33 & 35.6\% & 3.629 & 1.254 & 39.7\% & 3.108 & 0.913 \\
 15.07 & 26.55\% & 4.069 & 1.274 & 29.1\% & 3.505 & 0.946 \\
 17.40 & 21.17\% & 4.404 & 1.290 & 22.8\% & 3.816 & 0.969 \\
 29.10 & 10.06\% & 7.303 & 1.353 & 9.95\% & 6.411 & 1.049 \\
 38.8 & 6.12\% & 8.635 & 1.395 & 5.67\% & 7.725 & 1.095 \\
 69 & 2.02\% & 12.59 & 1.527 & 1.73\% & 11.62 & 1.228 \\
 87.7 & 1.13\% & 13.72 & 1.553 & 0.96\% & 12.80 & 1.250 \\
 110.4 & 0.61\% & 14.54 & 1.613 & 0.75\% & 20.20 & 1.300 \\ \hline
\multicolumn{7}{c}{$1.1 < y < 2.2$} \\ \hline
 200 & 3.63\% & 12.11 & 1.772 & 3.02\% & 9.85 & 1.364 \\
 500 & 0.62\% & 23.24 & 2.151 & 0.49\% & 22.07 & 1.646 \\ \hline
\multicolumn{7}{c}{$2.5 < y < 5$} \\ \hline
 5000 & 1.8\% & 17.46 & 2.904 & 1.52\% & 16.21 & 2.052 \\
 7000 & 0.93\% & 21.45 & 3.046 & 0.74\% & 20.24 & 2.158 \\
 13000 & 0.21\% & 27.52 & 3.300 & 0.16\% & 26.50 & 2.352 \\ \hline
  \end{tabular}
  \caption[]{\label{table:pT_averages} The average $p_T$ for $J/\psi$ and
    $\overline D$ meson production as a function of center of mass energy,
    $\sqrt{s}$ for both intrinsic charm and perturbative QCD production
    respectively.  Results are shown for midrapidity, $0< y <1$,
    for fixed-target energies; $1.1 < y < 2.2$ for RHIC energies;
    and $2.5 < y < 5$ for LHC
    energies.  In the case of production by intrinsic charm, the percentage of
    the $p_T$ distribution within the rapidity acceptance, $f_{\rm IC}$,
    is also given.
  }
\end{table}

\begin{figure}
  \begin{center}
    \includegraphics[width=0.4\textwidth]{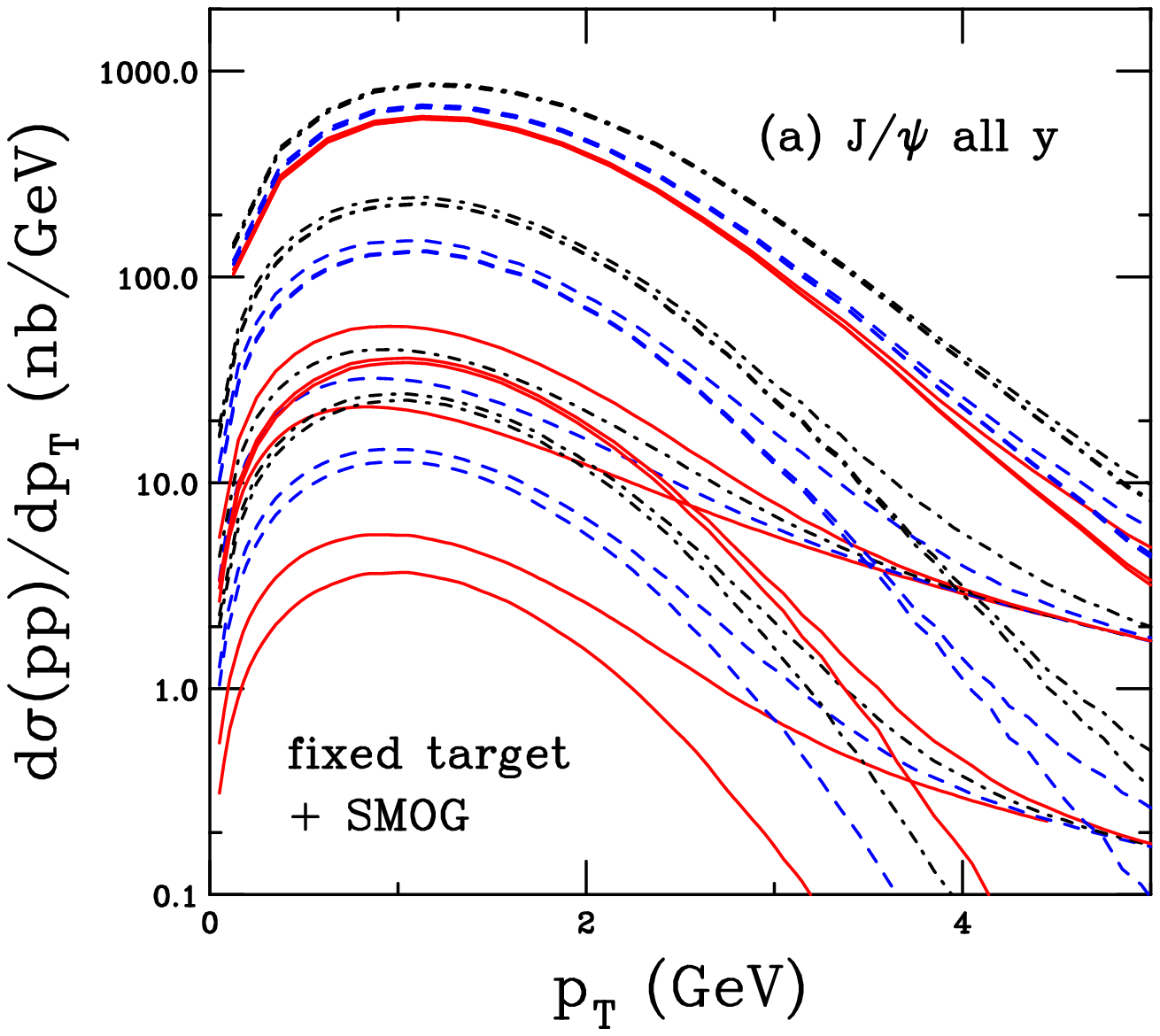}
    \includegraphics[width=0.4\textwidth]{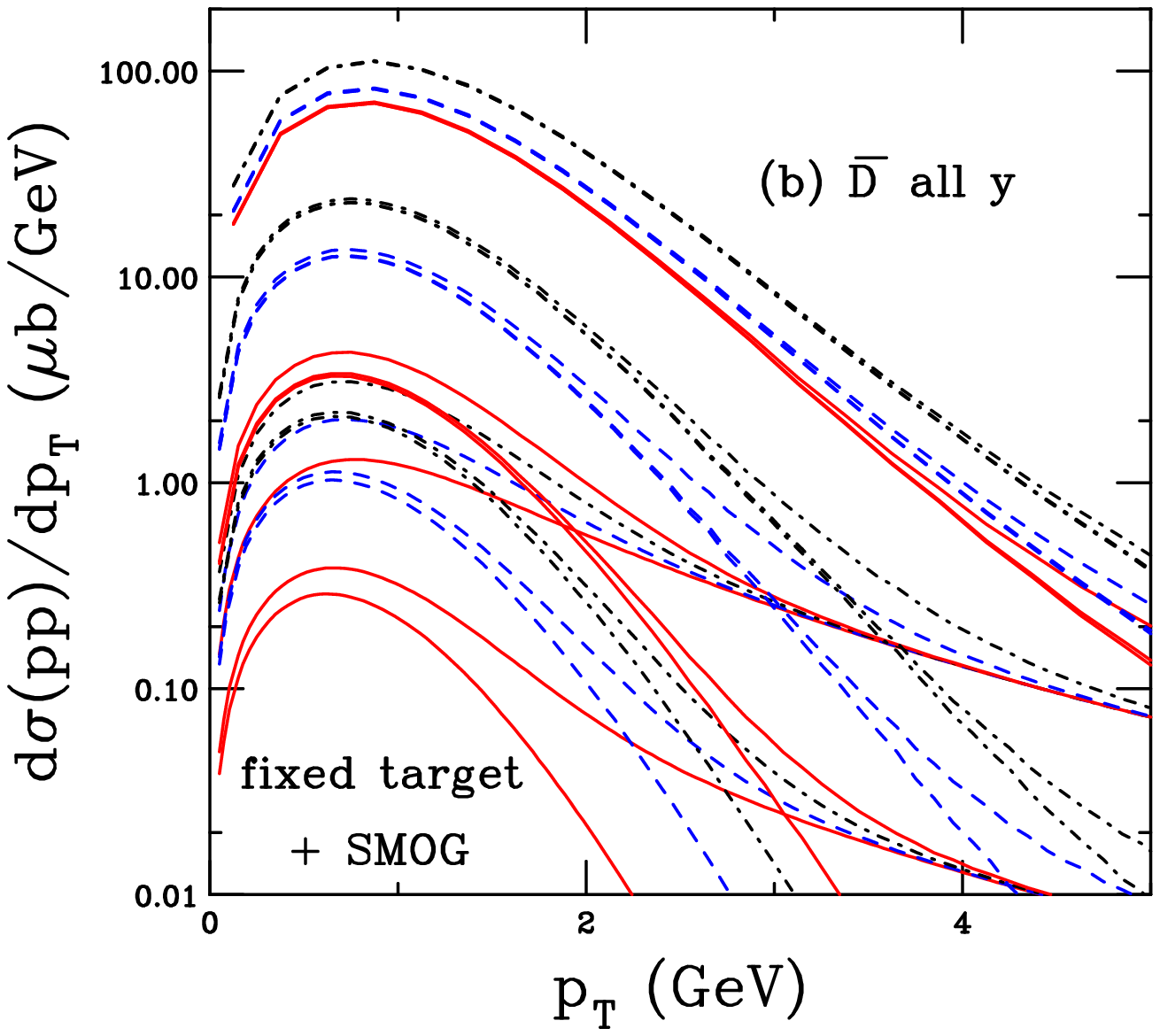} \\
    \includegraphics[width=0.4\textwidth]{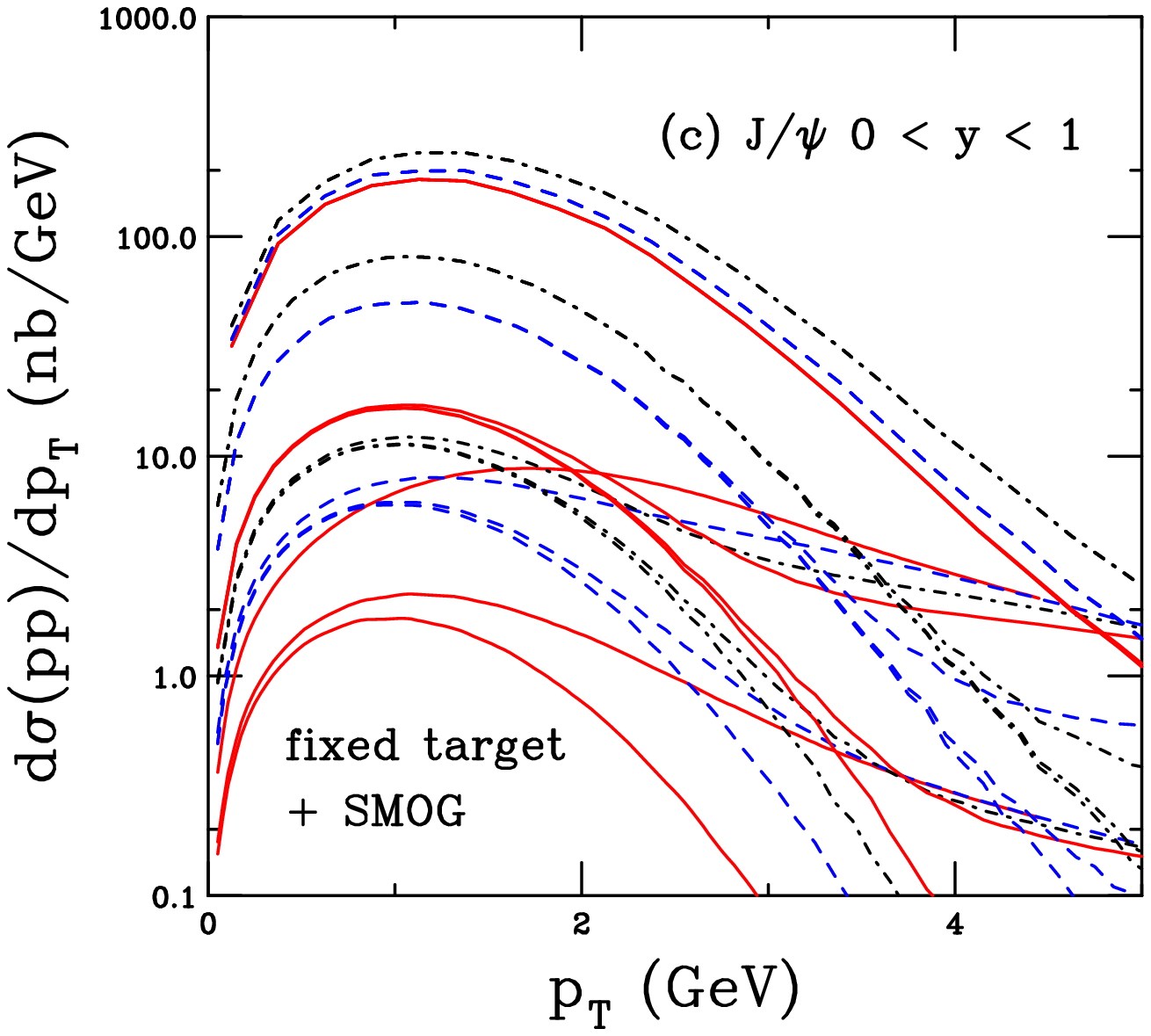}
    \includegraphics[width=0.4\textwidth]{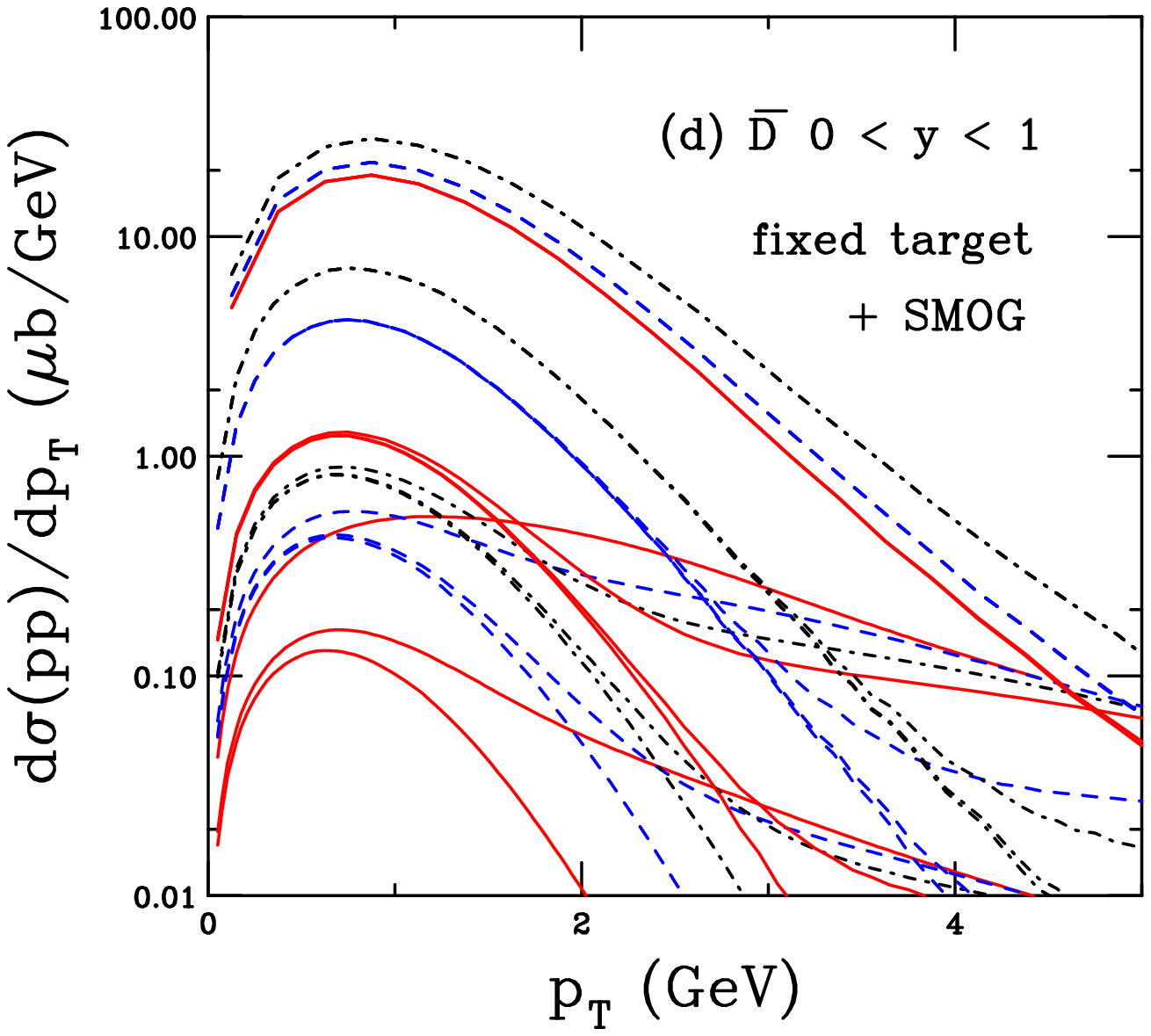} \\
  \end{center}
  \caption[]{The combined $p_T$ distributions for
    $J/\psi$ (a), (c) and $\overline D$ (b), (d) mesons, including
    both the typical perturbative QCD contribution and intrinsic charm from a
    five-particle proton Fock state.  The calculated $p_T$ distributions are
    integrated over all rapidity in (a) and (b) but limited to $0 < y < 1$ in
    (c) and (d).  Three curves are shown for each energy.
    From lowest to highest (when separable) they show: no intrinsic charm (pQCD
    only); $P_{\rm ic \, 5}^0 = 0.1$\%; and  $P_{\rm ic \, 5}^0 = 1$\%.
    The results are shown for fixed-target and SMOG energies.
    From lowest to highest the curves represent $p_{\rm lab} = 40$~GeV
    (red solid),
    80~GeV (blue dashed), 120~GeV (black dot-dashed),
    158~GeV (red solid), 450~GeV (blue dashed), 800~GeV (black dot-dashed), 
    $\sqrt{s} = 69$~GeV (solid red), 87.7~GeV (blue dashed) and 110.4~GeV
    (black dot-dashed).  
    }
\label{pp_pTdists_en}
\end{figure}

\begin{figure}
  \begin{center}
    \includegraphics[width=0.4\textwidth]{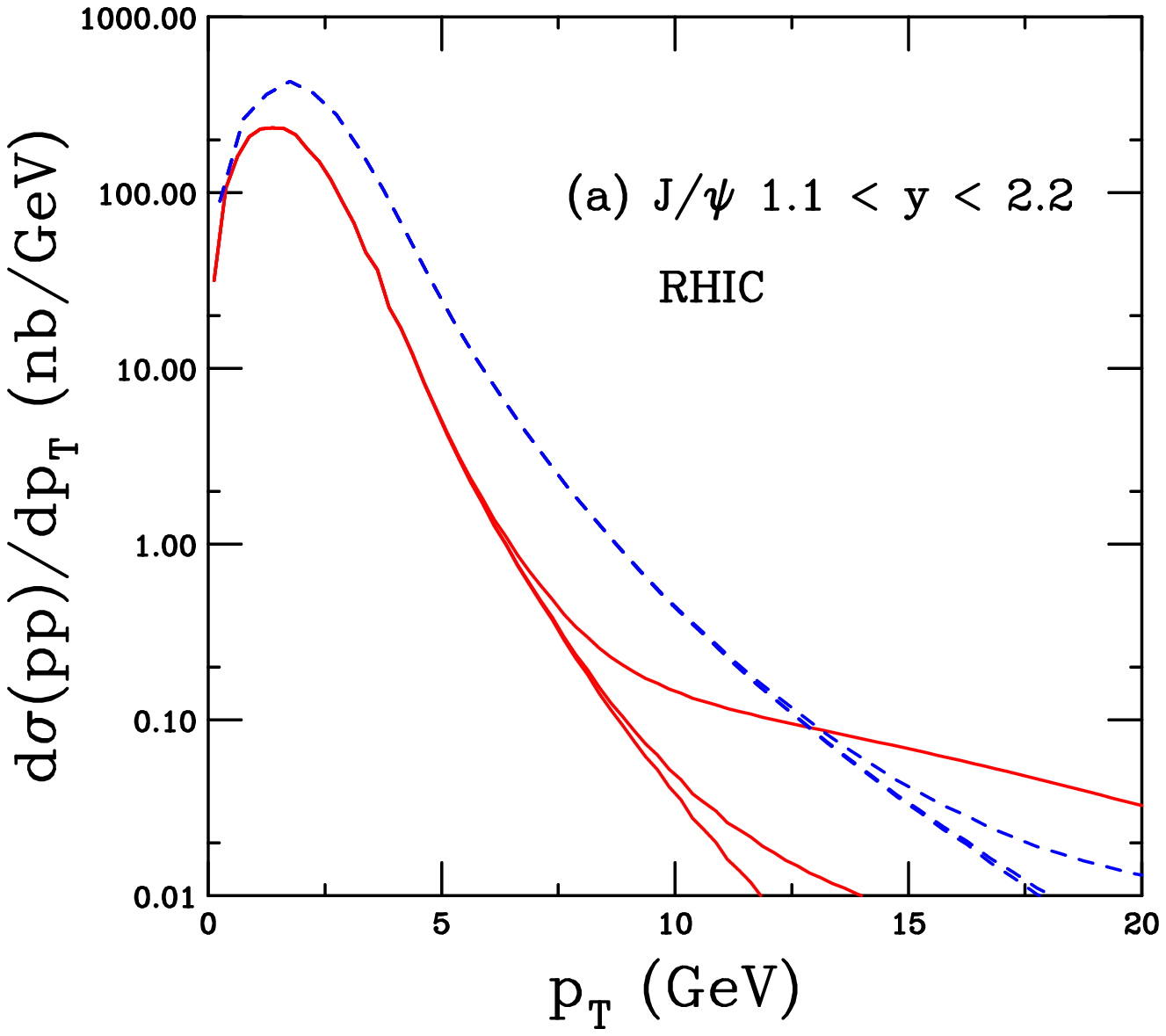}
    \includegraphics[width=0.4\textwidth]{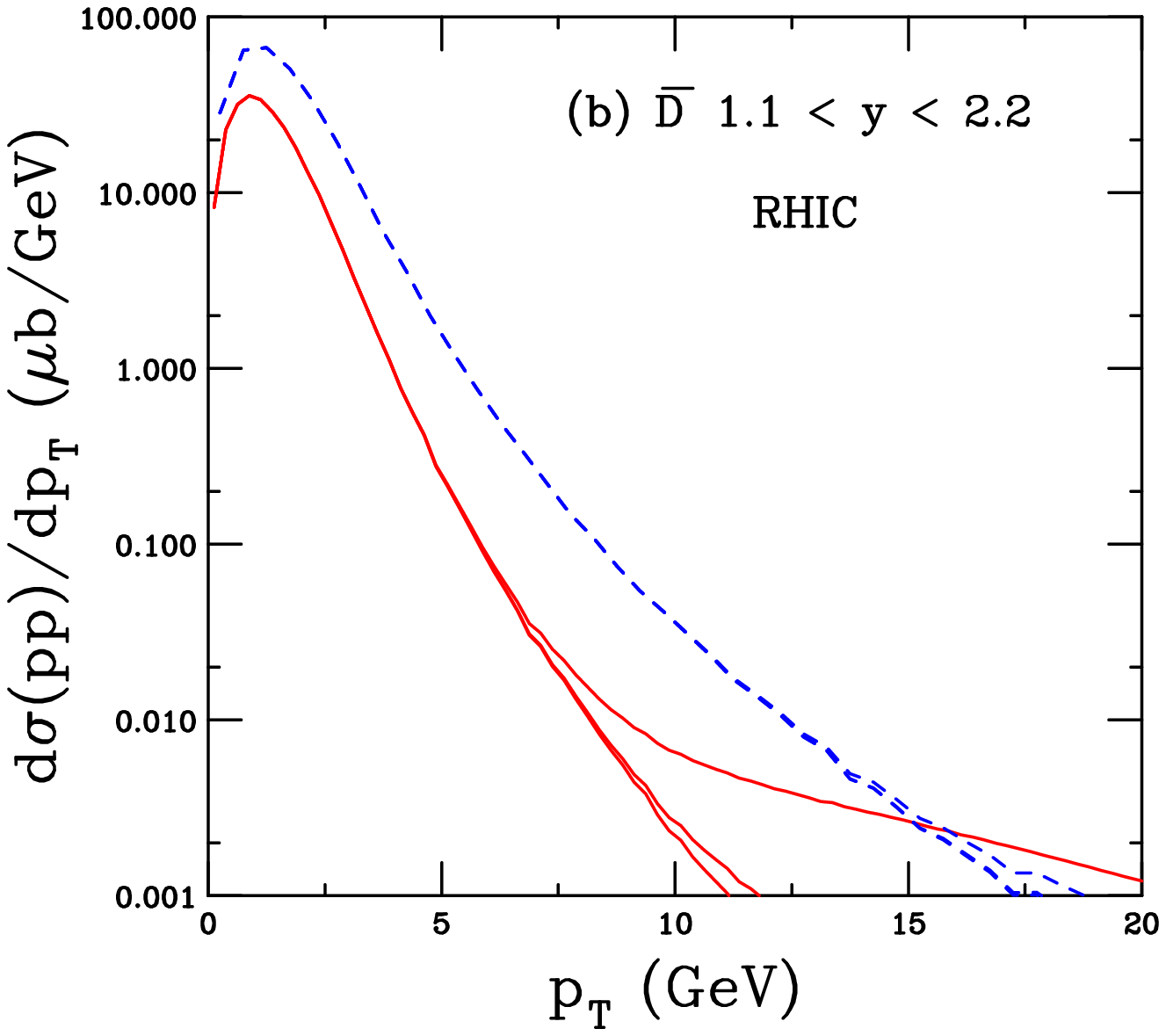} \\
    \includegraphics[width=0.4\textwidth]{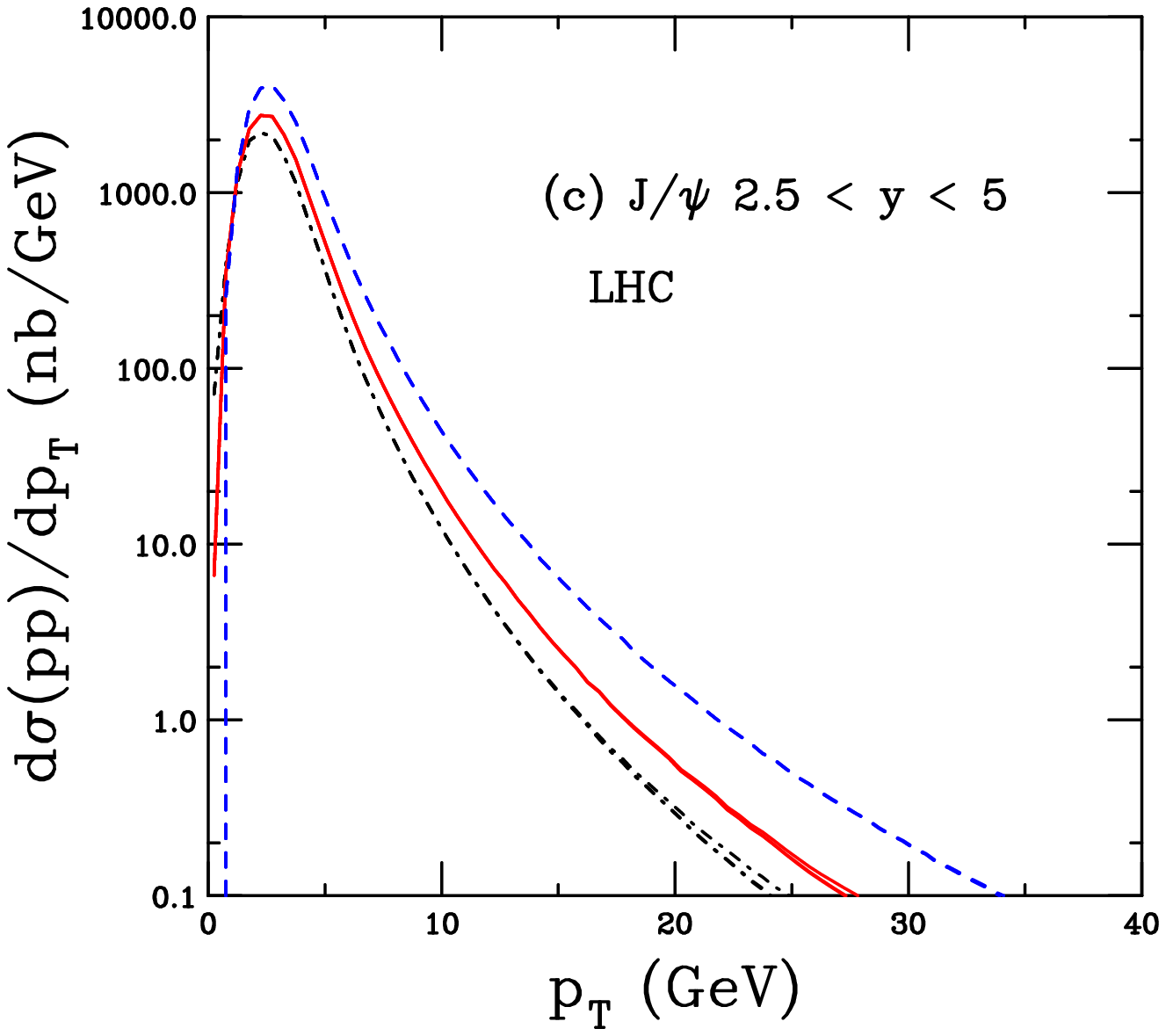}
    \includegraphics[width=0.4\textwidth]{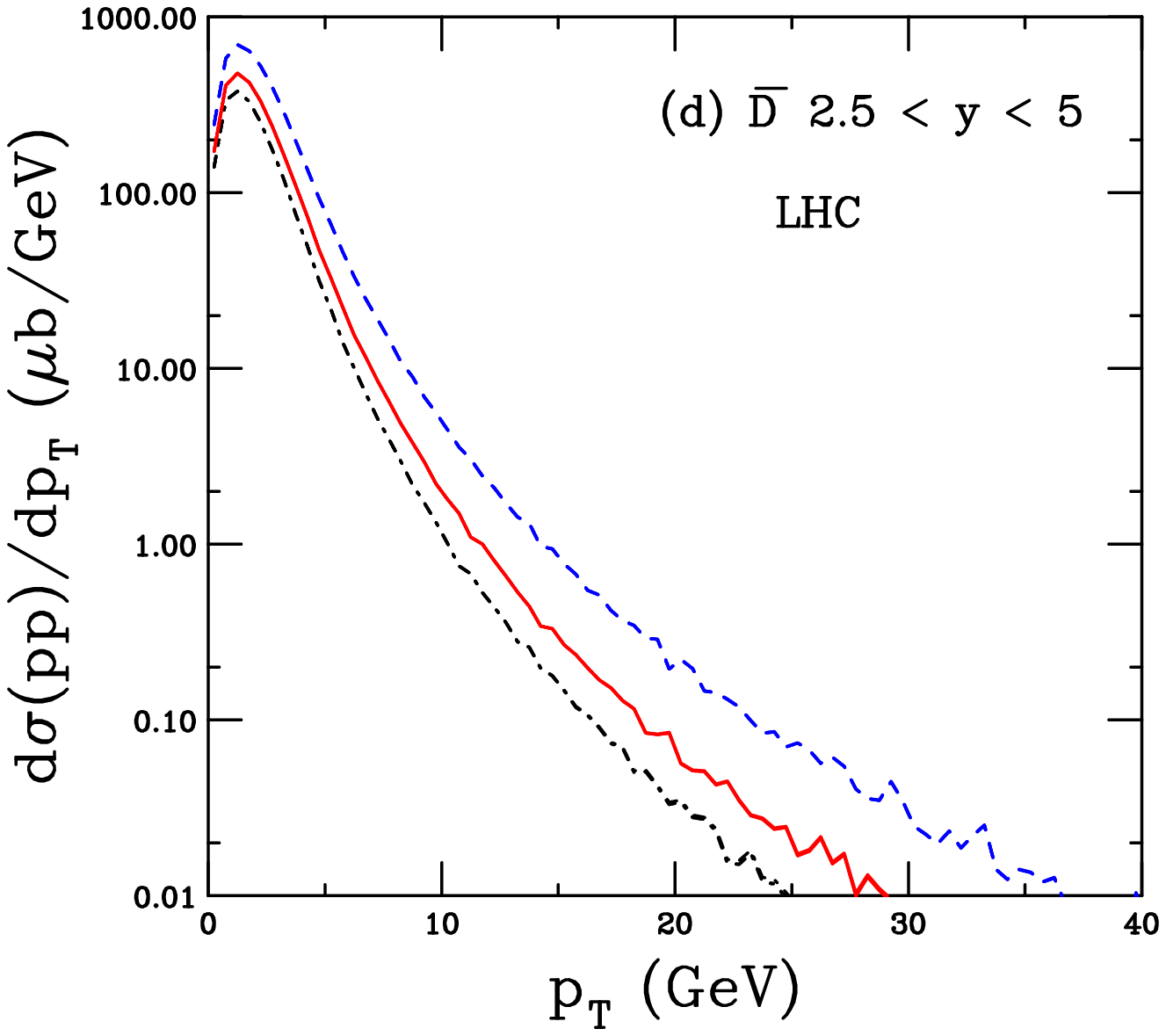}
  \end{center}
  \caption[]{The combined $p_T$ distributions for
    $J/\psi$ (a), (c) and $\overline D$ (b), (d) mesons, including
    both the typical perturbative QCD contribution and intrinsic charm from a
    five-particle proton Fock state.  Three curves are shown for each energy.
    From lowest to highest (when separable) they show: no intrinsic charm (pQCD
    only); $P_{\rm ic \, 5}^0 = 0.1$\%; and  $P_{\rm ic \, 5}^0 = 1$\%.
    In (a) and (b) the RHIC
    energies of $\sqrt{s} = 200$~GeV (solid red) and 500~GeV (blue dashed) are
    shown in the rapidity range $1.1 < y < 2.2$.  In (c) and (d)
    the distributions for LHC energies of $\sqrt{s} = 5$~TeV (red solid), 7~TeV
    (blue dashed), and 13~TeV (black dot-dashed) are given for $2.5 < y < 5$.
    }
\label{pp_pTdists_en_cuts}
\end{figure}

Several trends are immediately clear.  In the perturbative QCD calculations,
the $J/\psi$ distributions are harder than the $\overline D$ distributions.
As is generally the case for the rapidity distributions, the presence of a
second massive charm quark hardens the distribution relative to open charm with
a single charm quark.  The average $p_T$ from perturbative QCD
increases relatively slowly with 
$\sqrt{s}$.  Note that the shift to a more forward rapidity acceptance
at collider energies does not decrease the average $p_T$ of the perturbative
QCD calculation.  As can be seen in
Figs.~\ref{pp_pTdists_en} and \ref{pp_pTdists_en_cuts}, even though the
magnitude of the $p_T$ distribution grows with $\sqrt{s}$, the hardening of the
QCD distribution with $\sqrt{s}$ is also significant because the higher the
center of mass energy, the greater the potential $p_T$ reach.

No such effect is observed for the intrinsic charm distributions which are
independent of $\sqrt{s}$.  Thus the perturbative QCD cross section typically
engulfs all but the high $p_T$ tail of the distribution for $p_T$ greater than
a few GeV, as is evident from Figs.~\ref{pp_pTdists_en} and
\ref{pp_pTdists_en_cuts}.

It is notable that the $p_T$ distributions from intrinsic charm are
considerably broader than those calculated in perturbative QCD.
This can be easily understood
when one considers the two sources.

In perturbative QCD, the $p_T$ range depends on
the center of mass energy with a limit of $p_T \sim \sqrt{s_{NN}}/2$ at
$x_F = 0$ for
massless partons. The limit is lower for massive quarks
where $p_T$ is replaced by $m_T$.
Leading order $2 \rightarrow 2$ scattering is assumed for this estimate,
giving $p_T \leq 7.2$~GeV in the massless case and $p_T \leq 6.5$~GeV
for the $J/\psi$ in the CEM.  In addition, since the initial partons taking
part in the QCD interaction come from two different hadrons, one from the
projectile and the other from the target, at least one of them will carry a much
smaller fraction of the momentum than the intrinsic charm quarks in the proton
Fock state.

On the other hand, when the $J/\psi$ arises from an intrinsic charm state of
the proton, according to Eq.~(\ref{icdenom}), there is no energy limit on the
$p_T$ distribution other than that imposed by momentum conservation.  (As
discussed regarding Fig.~\ref{ic_pTdists_en}, the most important constraint
on the $p_T$ distribution is $x_F$ must be less than unity.)  The
$J/\psi$ kinematic constraints come from the incident proton alone.
A soft interaction
with the target is sufficient to disrupt the Fock state and bring the $J/\psi$
on mass shell.

The averages from intrinsic charm in Table~\ref{table:pT_averages}
are significantly higher than those from
perturbative QCD.  The difference increases substantially with $\sqrt{s}$
when a finite rapidity range is considered.  Note that if a detector could
covered all of
rapidity space, the results averaged over all rapidity would remain fixed at
$\sim 2$~GeV. Thus at
sufficiently high $\sqrt{s}$, the average $p_T$ from perturbative QCD would
become greater than that from intrinsic charm.

However, even though the average $p_T$ of the intrinsic charm distribution
is increasing, the
consequences of intrinsic charm in a measurable rapidity interval is decreasing
rapidily.  At the lowest energy considered, the interval around midrapidity
contains more than 50\% of the intrinsic charm cross section, as one might
expect from Fig.~\ref{ic_ydists_en}.  
Increasing $\sqrt{s}$ by only 4~GeV decreases the fraction of the intrinsic
charm cross
section captured at midrapidity by $\sim 20$\%, see
Fig.~\ref{ic_pTdists_en} and
Table~\ref{table:pT_averages}.  At the highest SMOG fixed-target energy,
less than 1\% of the intrinsic charm cross section is within $0 < y < 1$.
This is not surprising because of the boosted rapidity distribution, as shown
in Fig.~\ref{ic_ydists_en}.  Increasing the energy moves the average of the
intrinsic charm $p_T$ distribution to ever higher $p_T$.  Shifting the rapidity
interval forward, as at collider energies, reduces the average $p_T$ and
encompasses a larger (albeit still small) percentage of the total intrinsic
charm cross section.  Finally only about 0.2\% of the intrinsic charm
contribution remains at forward rapidity when $\sqrt{s} = 13$~TeV.

The largest differences in the total $J/\psi$ and $\overline D$
$p_T$ distributions due to intrinsic charm appear in the range
$8.77 < \sqrt{s} < 110.4$~GeV.  Therefore, Fig.~\ref{pp_pTdists_en} compares the
distributions in that energy range when integrated over all rapidity and
in the rapidity interval $0 < y < 1$.

Without the rapidity cut, the separation between the $p_T$ distributions at
$p_{\rm lab} = 40$~GeV without and with $P_{{\rm ic} \, 5}^0 = 1$\% can be a factor
of almost 10, especially near the peak of the distributions, as seen in the three lowest red curves of Fig.~\ref{pp_pTdists_en}(a) and (b).   Indeed, for
this lowest energy with  $P_{{\rm ic} \, 5}^0 = 1$\%, the peak of the $p_T$
distribution is equivalent to that of the perturbative QCD cross section
at $p_{\rm lab} = 120$~GeV (compare the highest of the lower three red curves
to the two lowest black dot-dashed curves in Fig.~\ref{pp_pTdists_en}(a) and
(b)).  At the next lowest energy, $p_{\rm lab} = 80$~GeV, the increase between
no intrinsic charm and a 1\% contribution is reduced to a factor of less than
three.  At these low fixed-target energies, the high $p_T$ tails of the
intrinsic charm distributions converge, similar to the rapidity distributions
shown in Fig.~\ref{pp_ydists_en}.  As the center of mass
energy increases, the contribution from intrinsic charm at low
$p_T$ becomes negligible by $\sqrt{s_{NN}} = 69$~GeV.  However, at $p_T$ higher
than $\sim 3$~GeV, a slight enhancement at high $p_T$ can still be observed.

When a rapidity
cut of $0 < y < 1$ is included, in Fig.~\ref{pp_pTdists_en}(c) and (d),
the separation between the calculated results
decreases.  The effect of the rapidity cut is most striking at low $p_T$ and at
the lowest energies.  There is a notable difference between the three results:
no intrinsic charm, $P_{{\rm ic}\, 5}^0 = 0.1$\%, and 1\% at $p_T \rightarrow 0$
only at $p_{\rm lab} = 40$~GeV.  The visible separation moves to higher $p_T$
with increasing energy.  Above $p_{\rm lab} = 158$~GeV, the intrinsic charm
contribution at low $p_T$ becomes negligible with the rapidity cut which only
occurs at $\sqrt{s_{NN}} = 69$~GeV without the cut.
One interesting effect of the cut is that the high $p_T$ tails of the
distributions, while still converging, do so less smoothly than without the
cut.  Because the central rapidity region, $0 < y < 1$, encompasses most of
the rapidity distribution at $p_{\rm lab} = 40$~GeV and intrinsic charm dominates
the central rapidity range, the effect of the cut is weaker at high $p_T$ than
at the higher beam energies and the order of convergence is reversed relative to
the distributions without a cut.

Because the perturbative contribution
dominates at collider energies, only the distributions captured in the forward
rapidity regions are shown in Fig.~\ref{pp_pTdists_en_cuts}.  The only
noticeable intrinsic charm contribution appears at $p_T > 15$~GeV for
$\sqrt{s_{NN}} = 200$~GeV.  Recall that this is because, in some
of the forward rapidity range at this particular energy, the resulting
$x_F$ is greater than unity, an unphysical region,
as shown also in Fig.~\ref{ic_pTdists_en}.

\subsection{$\alpha(x_F)$ in fixed-target measurements}
\label{alf_xF}

As previously mentioned,
an important validation of the need for intrinsic charm at fixed-target
energies is through comparison to previous data.  The earlier
NA60 Collaboration
compared their $J/\psi$ data collected at $p_{\rm lab} = 158$ and 400~GeV
\cite{NA60} to fixed-target data from NA3 \cite{NA3} at
$p_{\rm lab} = 200$~GeV; NA50 \cite{NA50} at $p_{\rm lab} = 450$~GeV; E866
\cite{e866} at $p_{\rm lab} = 800$~GeV; and HERA-B \cite{herab1,herab2} at
$p_{\rm lab} = 920$~GeV.

Aside from the NA3 data (in the range $x_F > 0$) taken only on a Pt
target, all the other experiments collected data from multiple nuclear targets.
The NA60 data at 158 and 400~GeV (covering $0.05 < x_F < 0.4$ and
$-0.075 < x_F < 0.125$ respectively) were taken on Be, Al, Cu, In, W, Pb, and U
targets.  The NA50 data (in the midrapidity range $-0.1 < x_F < 0.1$),
used Be, Al, Cu, Ag, W and Pb targets.  The E866 data, available
from $-0.09 < x_F < 0.95$, used Be, Fe, and W targets.
The HERA-B data, in the region $-0.34 < x_F < 0.14$, used C, Ti and W
targets. 

The value of $\alpha$ is obtained by assuming that the cross section in $p+A$
collisions can be described as growing relative to the $p+p$ cross section
by the target mass to the power $\alpha$,
\be
\sigma_{pA} = \sigma_{pp} A^\alpha \, \, \, , \label{alpha_def}
\ee
where $\alpha$ includes all cold nuclear matter effects.  
In this case, $\alpha$ can be calculated by averaging $\sigma_{pA}$ over
all nuclear targets, as described in the following section, Sec.~\ref{sec:pPb}.
It can be calculated based on a corresponding $p+p$
measurement or relative to a light nuclear target.  In the
latter case, the $p+p$ cross section is not needed and one has
\be
\left(\frac{\sigma_{pA_1}}{\sigma_{pA_2}} \right) = \left(\frac{A_1}{A_2}
\right)^\alpha \, \, \, . \label{alpha_rel_def}
\ee
Note that $A_2$ is generally a light nuclear target to reduce the cold nuclear
matter effects.  When data are taken on several targets, $\alpha$ is obtained by
averaging over all targets.  Such averaging makes it difficult to extract
subtle differences between nuclear targets, for example between Pb and U targets
where the lead nucleus (with $A = 208$) is doubly magic in proton and neutron
numbers and therefore spherical in shape while the uranium nucleus is very
deformed, almost cigar-like in shape.  It is generally more convenient to use
a light nuclear target rather than a proton target in fixed-target experiments.
Solid targets are easier to work with and generally give higher statistics
data.  This problem does not exist in experiments with colliding beams because
data are taken with colliding $p+p$ and $p + A$ beams.  However, due to run time
requirements, data are generally taken with fewer nuclear beams.

In all of the experiments except NA3, which took $p+p$ and $p+{\rm Pt}$ data,
$\alpha(x_F)$ was formed relative to the lightest target, C for HERA-B and Be
for NA50, NA60 and E866.  (In Ref.~\cite{RV_SeaQuest}, the E866
calculations were
shown with $\alpha$ as a function of $x_F$ and $p_T$ with a proton target
assumed for the base.)  Assuming that $\alpha$ is calculated with respect to
a proton or the lightest nuclear target changes $\alpha$ by an average of 1-2\%.

The calculations shown here were done at the same energy as the data were
taken with the same values of mass number $A$ for each experimental setup as
far as possible.  This statement is qualified because, unlike earlier
iterations, EPPS16 is only available for certain mass numbers, with
no extrapolation between them.  There are four targets employed by the
experiments, Ti ($A = 48$), Ag ($A = 107$), In ($A = 114$), and U ($A = 238$)
that do not have EPPS16 data files associated with their average mass numbers
\cite{EPPS16_web}.  The closest mass number is used in each case, $A = 50$ for
Ti and $A = 117$ for Ag and In.  The largest mass included in the EPPS16
data tables is $A = 208$, thus the perturbative QCD calculation for the Pb
target is used in the calculations instead.  The only difference between the Pb
and U calculations is then simply the mass number.  Note also that some of
the beam energies here, namely the NA3 energy of $p_{\rm lab} = 200$~GeV; the
upper NA60 energy of $p_{\rm lab} = 400$~GeV; and the HERA-B energy of
$p_{\rm lab} = 920$~GeV, were not included in the $p+p$ distributions shown
previously.  Separate calculations were made for these energies to match the
experimental energies.

The calculations were made over the full $x_F$ range and
integrated over all $p_T$.  All the calculations include the central EPPS16
set, enhanced $k_T$ broadening, and nuclear absorption.
Since the intrinsic charm component is invariant
with energy when calculated as a function of $x_F$, its contribution is the
same at all energies, only the perturbative QCD contribution changes.  The
$J/\psi$ production cross section increases with energy while its absorption
cross section decreases with energy \cite{LWV}.  While $\sigma_{\rm abs}$
changes with incident energy, it is held fixed to the value at
$x_F = 0$.  It is worth noting
that Ref.~\cite{LWV} also extracted the effective absorption cross section as
a function of $x_F$ and found that it did vary with $x_F$ but that this
dependence was approximately independent of the parameterization of the
nuclear parton distribution functions used in the extraction.  This shape may
not necessarily be attributable to absorption but to other cold nuclear matter
effects, including intrinsic charm or parton energy loss or some combination
thereof.

\begin{figure}
  \begin{center}
    \includegraphics[width=0.4\textwidth]{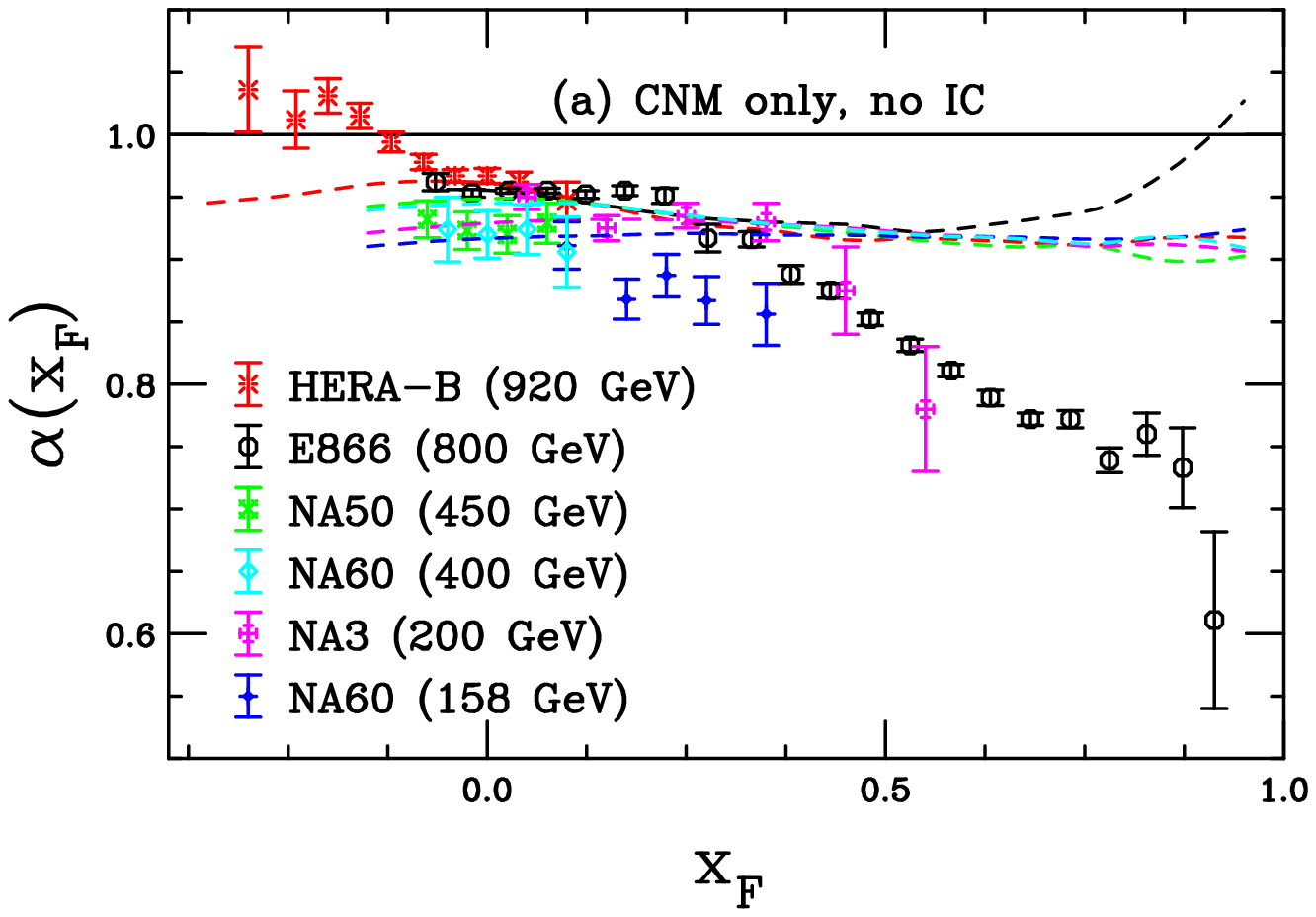}
    \includegraphics[width=0.4\textwidth]{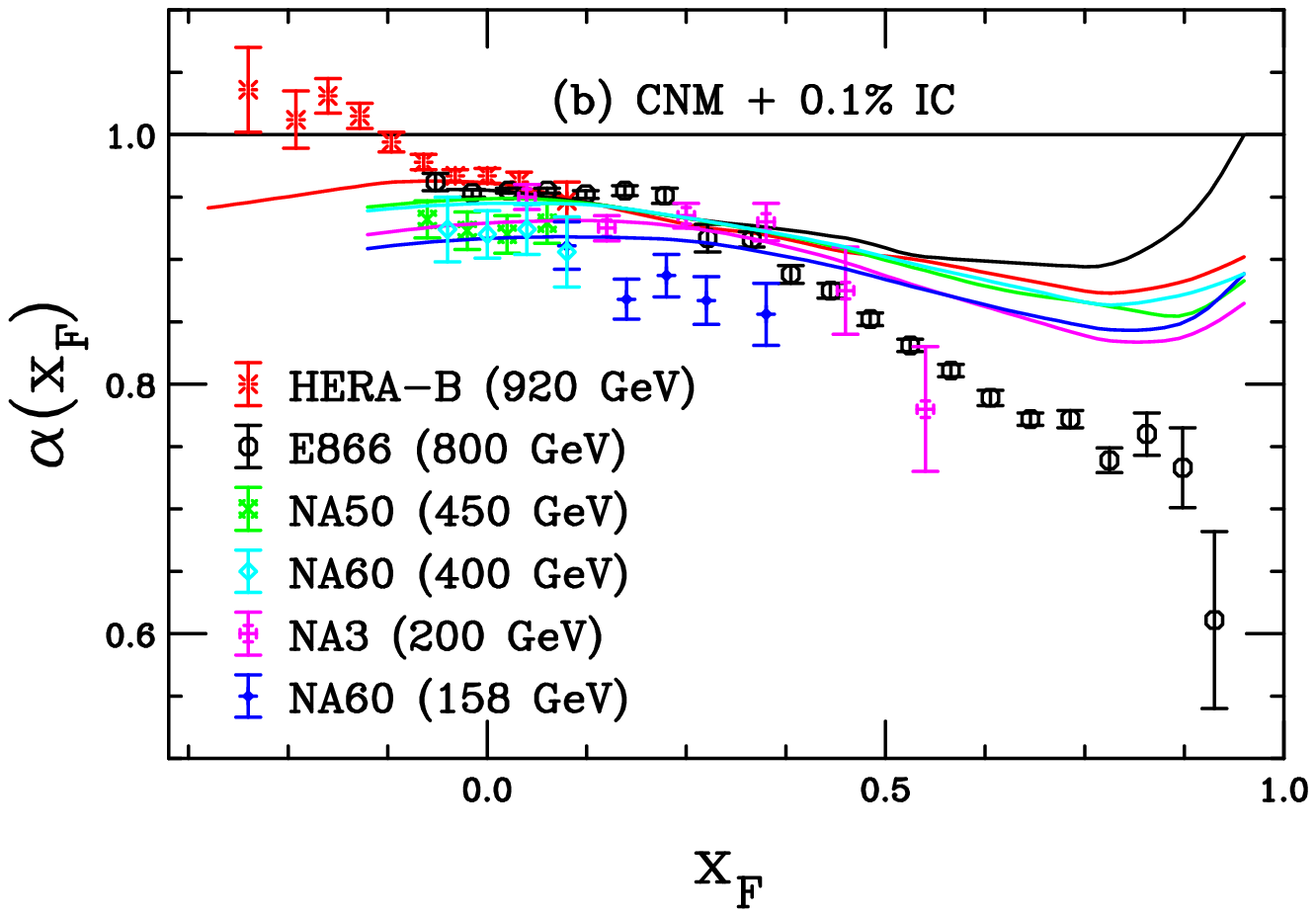} \\
    \includegraphics[width=0.4\textwidth]{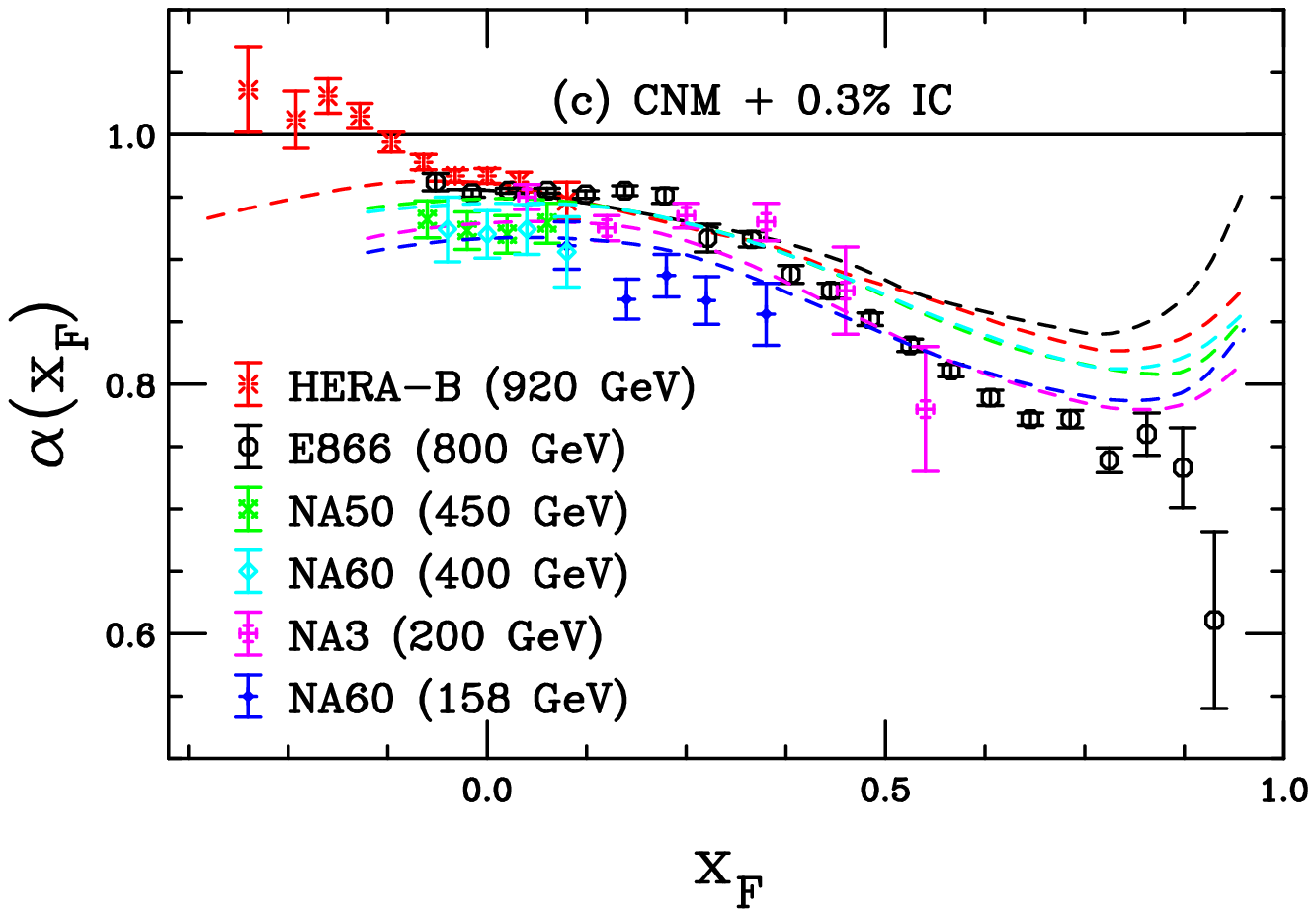}
    \includegraphics[width=0.4\textwidth]{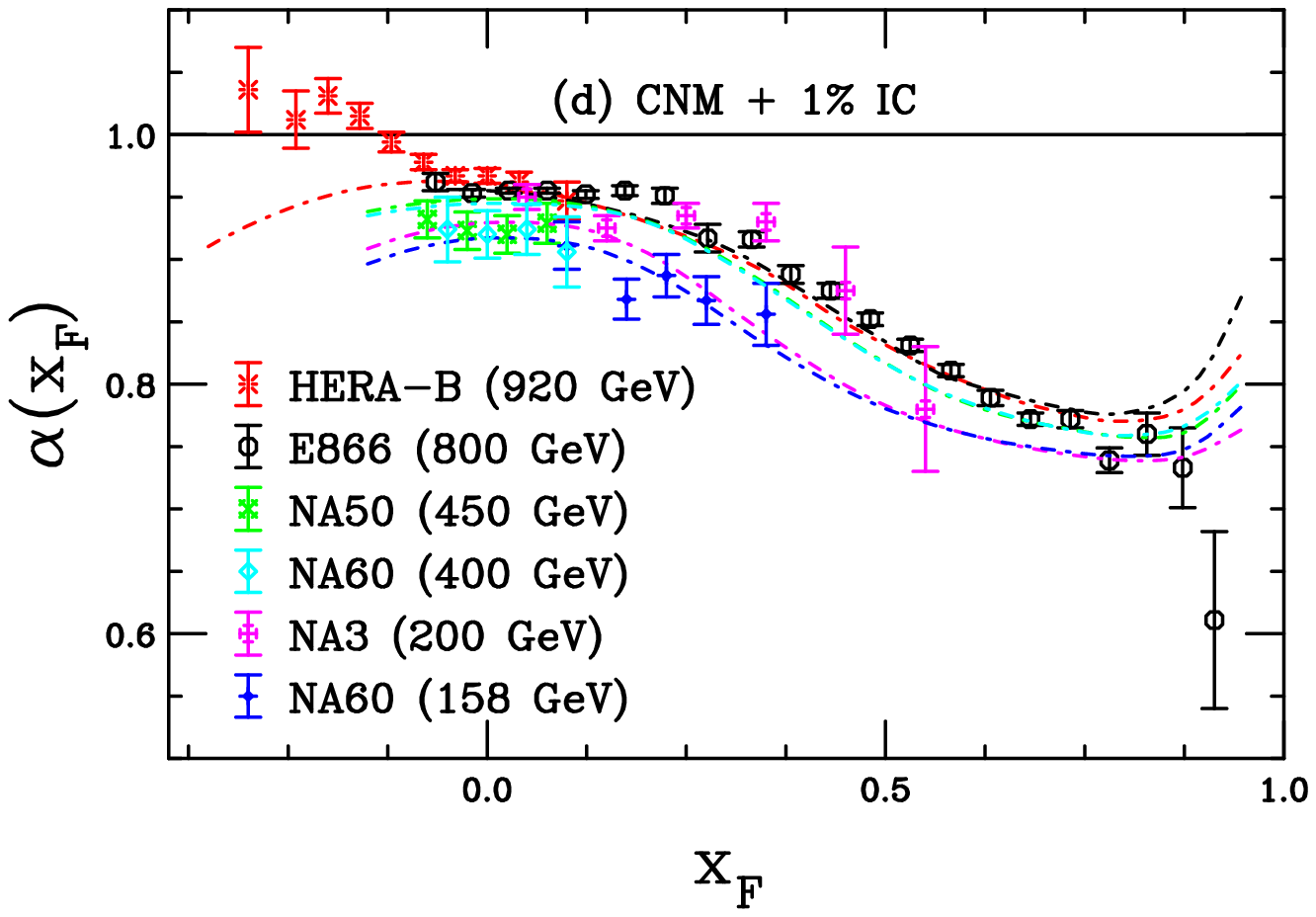} 
  \end{center}
  \caption[]{ The exponent $\alpha$ as a function of $x_F$
    for $J/\psi$ production at fixed-target energies: NA60 at
    $p_{\rm lab} = 158$~GeV \cite{NA60} (blue), NA3 at $p_{\rm lab} = 200$~GeV
    \cite{NA3} (magenta), NA60 at $p_{\rm lab} = 400$~GeV
    \cite{NA60} (cyan), NA50 at $p_{\rm lab} = 450$~GeV \cite{NA50} (green),
    E866 at $p_{\rm lab} = 800$~GeV \cite{e866} (black), and HERA-B at
    $p_{\rm lab} = 920$~GeV \cite{herab2} (red).  The points are the experimental
    data while the curves of the same color are calculations made to match the
    energy of the measurement.  Calculations without intrinsic charm are shown
    in (a) while calculations with $P_{\rm ic \, 5}^0 = 0.1$\%, 0.3\%, and 1\%
    are shown in (b)-(d) respectively.
    }
\label{fig:alpha_xF}
\end{figure}

The results are shown in Fig.~\ref{fig:alpha_xF}, along with the data.  
When plotted together as a function of $x_F$, the data appear to suggest a
relatively steady decrease of $\alpha(x_F)$ from $x_F \sim -0.25$ to $\sim 1$.
The HERA-B and E866 data are consistent with each other where they overlap, as
are the E866 and NA3 data although mostly due to the larger uncertainties of the
lower energy NA3 data.
There are some notable differences, however.  The NA50 and the 400~GeV NA60
data overlap with each other but are below the measured $\alpha(x_F)$ at higher
energy and the NA60 158~GeV data are below all the others, suggesting stronger
nuclear effects at the lower energy although this is not clearly seen for the
200~GeV NA3 data, perhaps due to the fact that only a single target was
utilized.

The calculations, shown without intrinsic charm in Fig.~\ref{fig:alpha_xF}(a)
and with increasing levels of intrinsic charm, $P_{\rm ic \, 5}^0 = 0.1$\%,
0.3\%, and 1\% in Fig.~\ref{fig:alpha_xF}(b)-(d) respectively, reflect the
general hierarchy with energy due to the energy dependence of $\sigma_{\rm abs}$.
Without intrinsic charm, the curves are all rather flat and become quite similar
at high $x_F$.  They do not show any significant decrease due to nuclear
modifications of the parton densities even though the nuclear suppression factor
$R_{pA}$ shows a distinct modulation as a function of rapidity.  The reason why
$\alpha(x_F)$ shows a much smaller effect is because $\alpha$ depends on the
logarithm of the cross section ratio rather than the ratio itself.

Nuclear
effects are further washed out when the ratio is between two per nucleon cross
sections rather than $p+A$ relative to $p+p$.  Only after intrinsic charm is
included, with its different nuclear dependence, does some separation of the
results at larger $x_F$ become apparent.  The larger the contribution from
intrinsic charm, the greater the curvature at large $x_F$.  With
$P_{\rm ic \, 5}^0 = 1$\%, good agreement with almost all the data is achieved.
A notable exception is the negative $x_F$ HERA-B data which has been difficult
to describe other than with energy loss models, see {\it e.g.}
Refs.~\cite{herab2,Arleo,Arleo:2012rs},
generally implemented as a shift in either the parton
momentum fraction in the incident proton, $x_1$, or in $x_F$.  

Taken together, these data seem to be consistent with an intrinsic charm
contribution in the proton on the order of 1\%.  This is also consistent with
the LHCb $Z + {\rm charm}$ jet data \cite{LHCb_intc}
and the recent NNPDF evaluation
\cite{NNPDF4}.

\subsection{$p+{\rm Pb}$ Interactions and the Nuclear Modification Factor}
\label{sec:pPb}

In this section, examples are given for the $p+{\rm Pb}$ rapidity and $p_T$
distributions compared to those in $p+p$.
Results are also presented for the nuclear suppression factor,
\be
R_{pA} = \frac{1}{A} \frac{\sigma_{pA}}{\sigma_{pp}} \, \, . \label{RpA_def}
\ee
While the target mass is explicitly given in Eq.~(\ref{RpA_def}), the per
nucleon cross section is displayed in the figures presenting the individual
distributions.  A lead target is
chosen for convenience for all energies to both maximize the cold nuclear
matter effects and facilitate comparison between energies.
The nuclear modification factor, $R_{pA}$, is a more direct comparison of two
systems at the same energy than the exponent $\alpha$ discussed previously.
It has been used most often to present collider results since most of these
data are taken with a single nuclear beam.  Note, however,
that when multiple targets are employed, as in the case of many earlier
fixed-target experiments, averaging data over many targets using $\alpha$ is a
convenient means of displaying all the data collectively.  As previously noted,
it can, however, obscure individual mass-dependent nuclear effects.  

The combined cold nuclear matter effects on perturbative QCD production of
open heavy flavor and $J/\psi$,
described in Sec.~\ref{pQCD}, are
\be
\sigma_{pA}^{\overline D} & = & \sigma_{\rm OHF}(pA) = \sum_{i,j} 
\int_{4m^2}^\infty d\hat{s}
\int dx_1 \, dx_2~ F_i^p(x_1,\mu_F^2,k_T)~ F_j^A(x_2,\mu_F^2,k_T)~ 
\hat\sigma_{ij}(\hat{s},\mu_F^2, \mu_R^2) \, \, ,  \label{sigOHF_pA} \\
\sigma_{pA}^{J/\psi} & = & \sigma_{\rm CEM}(pA) = S_A^{\rm abs} F_C \sum_{i,j} 
\int_{4m^2}^{4m_H^2} d\hat{s}
\int dx_1 \, dx_2~ F_i^p(x_1,\mu_F^2,k_T)~ F_j^A(x_2,\mu_F^2,k_T)~ 
\hat\sigma_{ij}(\hat{s},\mu_F^2, \mu_R^2) \, \, , 
\label{sigCEM_pA}
\ee
where
\be
F_j^A(x_2,\mu_F^2,k_T) & = & R_j(x_2,\mu_F^2,A) f_j(x_2,\mu_F^2) G_A(k_T) \, \, \\
F_i^p(x_1,\mu_F^2,k_T) & = & f_i(x_1,\mu_F^2) G_p(k_T) \, \, .
\ee
The total $k_T$ broadening in the nuclear target is applied 
as discussed in Sec.~\ref{pQCD} with the enhanced broadening in the nuclear
target introduced in Sec.~\ref{kTkick}.

When intrinsic charm is included, the cross sections are now
\be
\sigma_{pA}^{\overline D} & = & \sigma_{\rm OHF}(pA) + \sigma_{\rm ic}^{\overline D}(pA)
\label{sig_pA_Dsum} \\
\sigma_{pA}^{J/\psi} & = & \sigma_{\rm CEM}(pA) + \sigma_{\rm ic}^{J/\psi}(pA)
\label{sig_pA_Jsum}
\ee
where $\sigma_{\rm OHF}(pA)$ and $\sigma_{\rm CEM}(pA)$
were defined in Eqs.~(\ref{sigOHF_pA}) and (\ref{sigCEM_pA}) above while
$\sigma_{\rm ic}^{\overline D}(pA)$ $\sigma_{\rm ic}^{J/\psi}(pA)$
are given in Eqs.~(\ref{icsigD_pA}) and (\ref{icsigJpsi_pA}).

Although the results have been calculated for all three values of
$P_{{\rm ic}\, 5}^0$ in
Eq.~(\ref{icdenom}), only results with 0.1\% and 1\% are
shown, providing a range of uncertainty on the intrinsic charm contribution to
the nuclear modification factor.  The calculations of
$\alpha(x_F)$ shown in Sec.~\ref{alf_xF} seem to have a clear preference for
the larger value of $P_{{\rm ic}\, 5}^0$.  On the other hand, the calculations of
$\alpha(p_T)$ at $p_{\rm lab} = 800$~GeV compared to the E866 \cite{e866} data,
shown in Ref.~\cite{RV_SeaQuest}, seemed to prefer a smaller
contribution.  Thus there is some possible tension between the results.
Further data as a function of $p_T$ at more energies may help clarify the
situation.

The $p + {\rm Pb}$ calculations are shown for a few selected
energies for clarity of display while still covering the full energy range:
$p_{\rm lab} = 40$, 158, and 800~GeV and $\sqrt{s} = 87.7$,
200, and 5~TeV.  

\subsubsection{Rapidity dependence}
\label{sec:pPbRapidity}

\begin{figure}
  \begin{center}
    \includegraphics[width=0.4\textwidth]{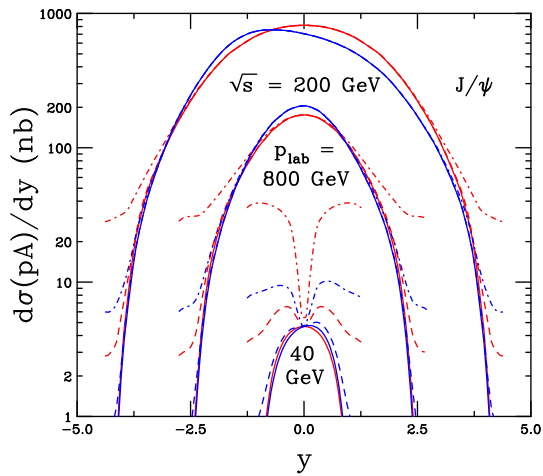}
  \end{center}
  \caption[]{ The $p+p$ and $p+{\rm Pb}$ (per nucleon)
    $J/\psi$ distributions at
    $p_{\rm lab} = 40$ and 800~GeV and $\sqrt{s} = 200$~GeV as a function of
    rapidity.  The red curves show the results for $p+p$ collisions while the
    blue curves show the $p+{\rm Pb}$
    distributions.  Three curves are shown: no intrinsic charm (pQCD
    only, solid); $P_{\rm ic \, 5}^0 = 0.1$\% (dashed); and
    $P_{\rm ic \, 5}^0 = 1$\% (dot-dashed).  No $J/\psi$ absorption by nucleons
    is considered in the $p+{\rm Pb}$ calculation.
    }
\label{pp_pPb_ydists}
\end{figure}

The rapidity dependence is shown first, beginning with the individual $J/\psi$
distributions at $p_{\rm lab} = 40$ and 800~GeV and $\sqrt{s_{NN}} = 200$~GeV in
Fig.~\ref{pp_pPb_ydists}.  The $p+p$ calculations are shown in red while the
$p + {\rm Pb}$ distributions are in blue.  The intrinsic
charm contributions are assumed to be symmetric around $y=0$.  This is true for
$p+p$ collisions, it is also assumed to hold for the $A$ dependence of intrinsic
charm in $p+A$ collisions as well.  The effects of antishadowing (for
$p_{\rm lab} = 40$~GeV) and shadowing (for $\sqrt{s_{NN}} = 200$~GeV) are clearly
illustrated by the differences in the solid curves.  In addition, the strong
suppression of intrinsic charm relative to the perturbative QCD $A$ dependence
in Eq.~(\ref{sig_pA_Jsum}) is clear in, for example, the difference between
the red and blue dot-dashed curves for $P_{{\rm ic} \, 5}^0 = 1$\%.  It is obvious
that, for low energies, intrinsic charm can play a significant role in the
nuclear suppression factor at midrapidity but that it becomes less important at
higher energies.

It is also clear that the tails of the rapidity distributions
will be dominated by intrinsic charm, whether or not this part of the
distribution is measurable.  Recall that the only energy dependence in the
total intrinsic charm cross section is from the inelastic cross section,
$\sigma_{pN}^{\rm in}$, see Eq.~(\ref{icsign}).  This quantity changes only slowly
with center-of-mass energy and has been left fixed in this study.  Thus when
intrinsic charm is added to the perturbative QCD cross section, the high
rapidity part of the cross section, where the perturbative part is steeply
falling due to phase space, the intrinsic charm cross section at high rapidity
remains the same,
independent of collision energy, see Fig.~\ref{pp_pPb_ydists}.
The intrinsic charm contribution to the $p+{\rm Pb}$ distribution has an $A$
dependence of $A^\beta$, see Eq.~(\ref{icsigJpsi_pA}), resulting in a lower
per nucleon cross section at high rapidity for the same value of
$P_{{\rm ic}\, 5}^0$.  (Compare {\it e.g.} the dot-dashed red and blue curves in
Fig.~\ref{pp_pPb_ydists} for $P_{{\rm ic}\, 5}^0 = 1$\%.)
Therefore, there
will be a limiting value of $R_{pA}$ at the edge of the rapidity range
regardless of the energy, $R_{pA} \rightarrow A^{\beta -1}$.  This effect will
be evident in the calculations of $R_{pA}(y)$
shown later in this section and also for $R_{pA}(p_T)$.  Although the result in
Fig.~\ref{pp_pPb_ydists} is shown for $J/\psi$, the result will be similar for
$\overline D$ production.

\begin{figure}
  \begin{center}
    \includegraphics[width=0.4\textwidth]{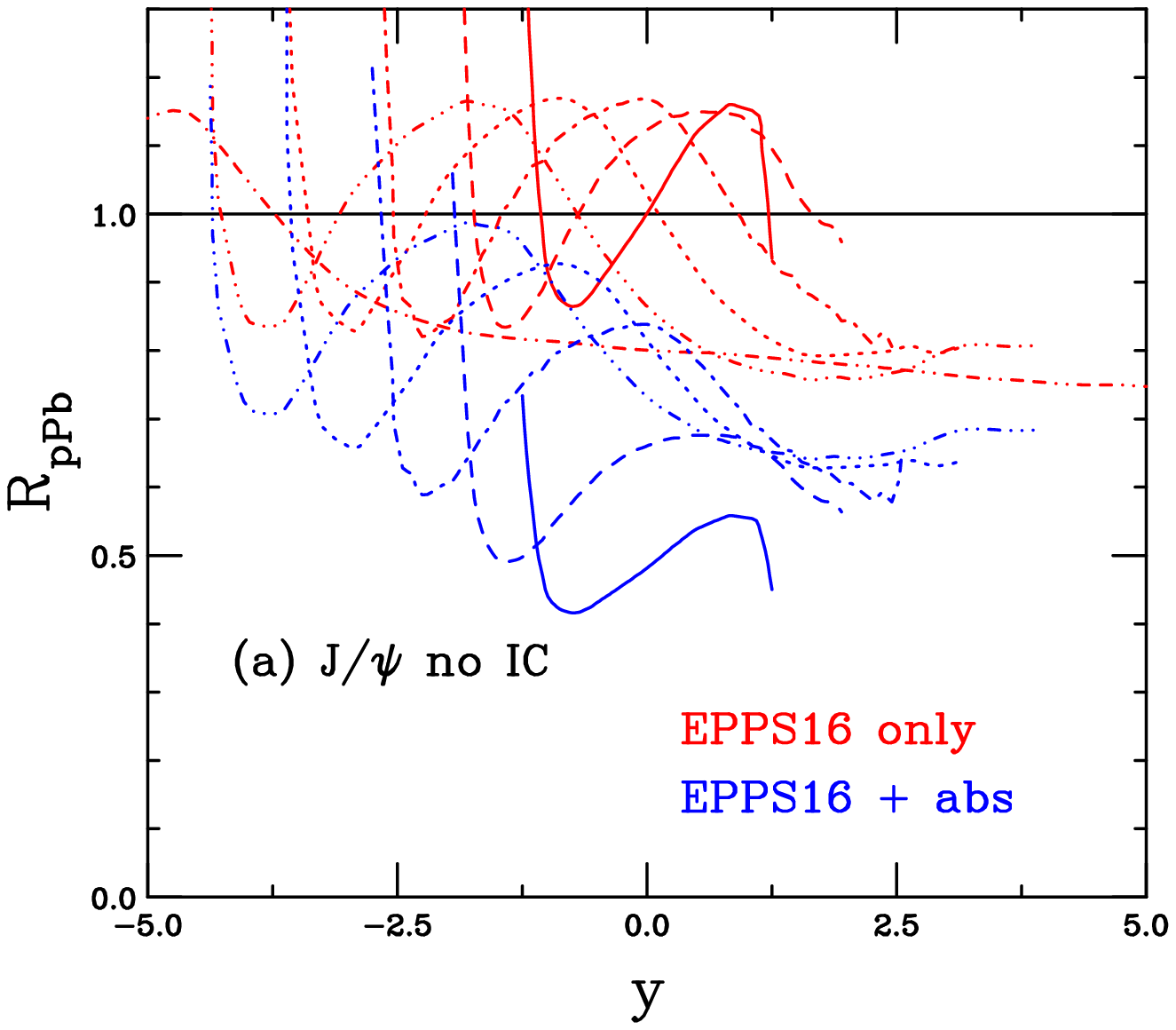}
    \includegraphics[width=0.4\textwidth]{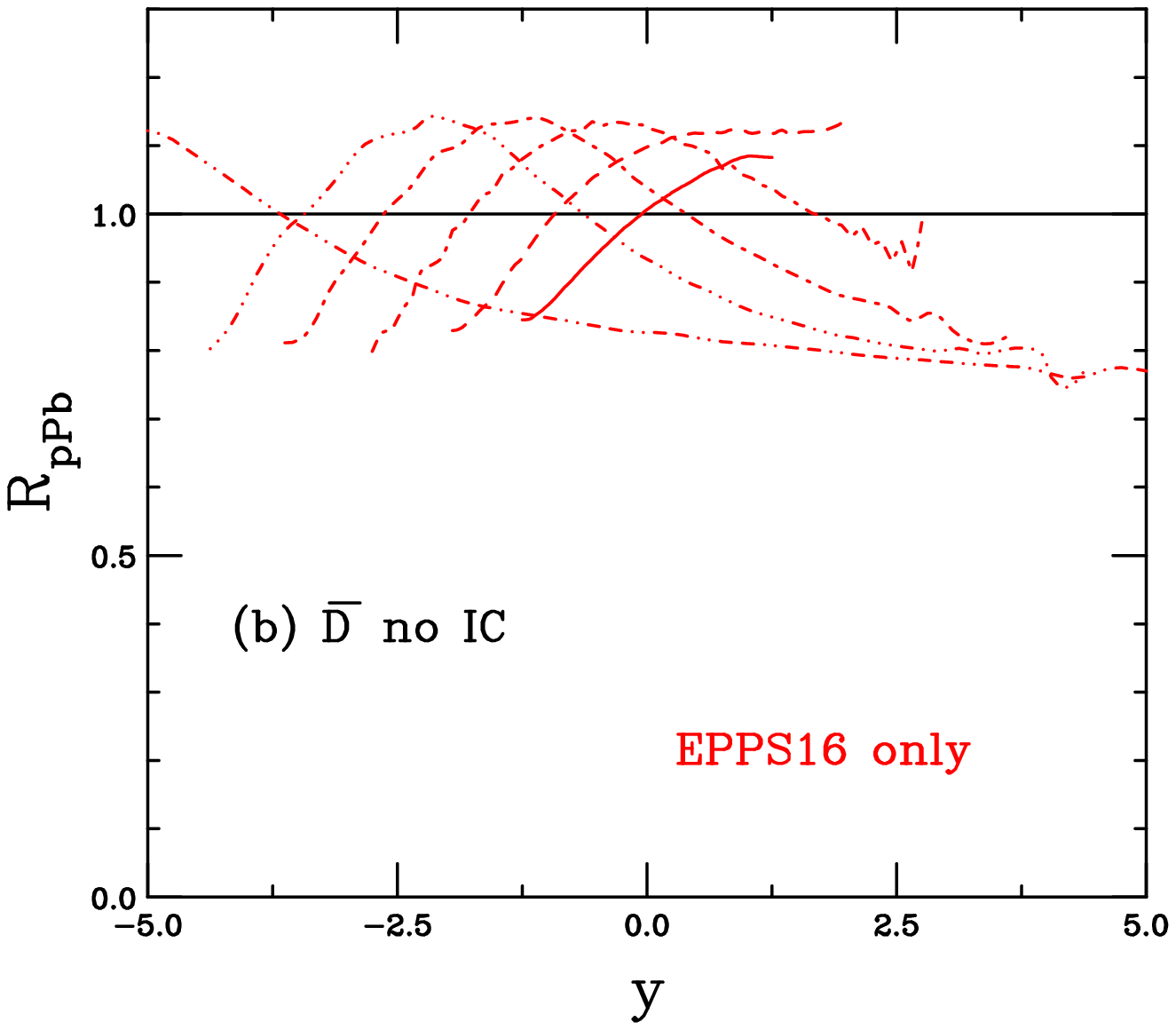} \\
    \includegraphics[width=0.4\textwidth]{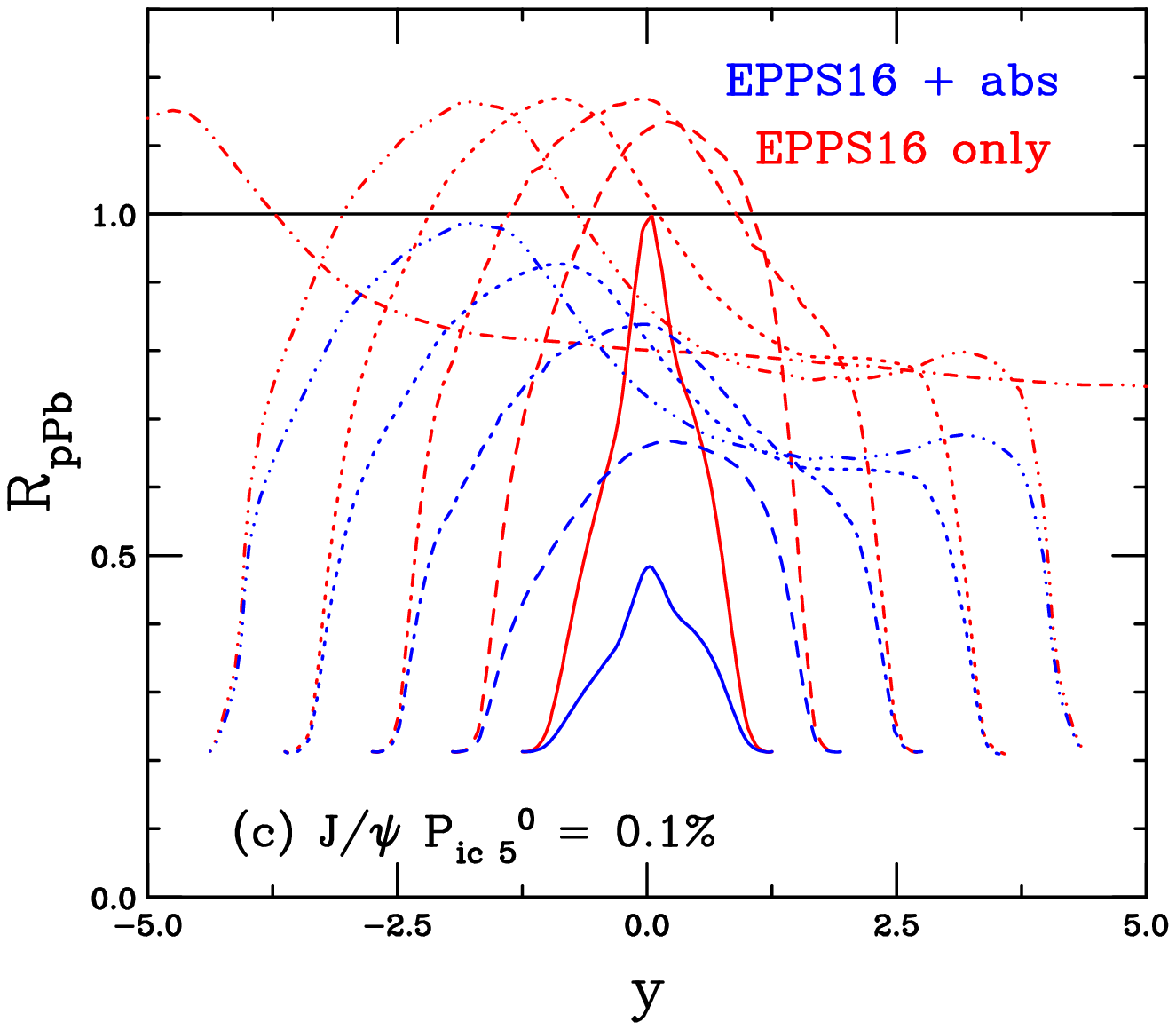}
    \includegraphics[width=0.4\textwidth]{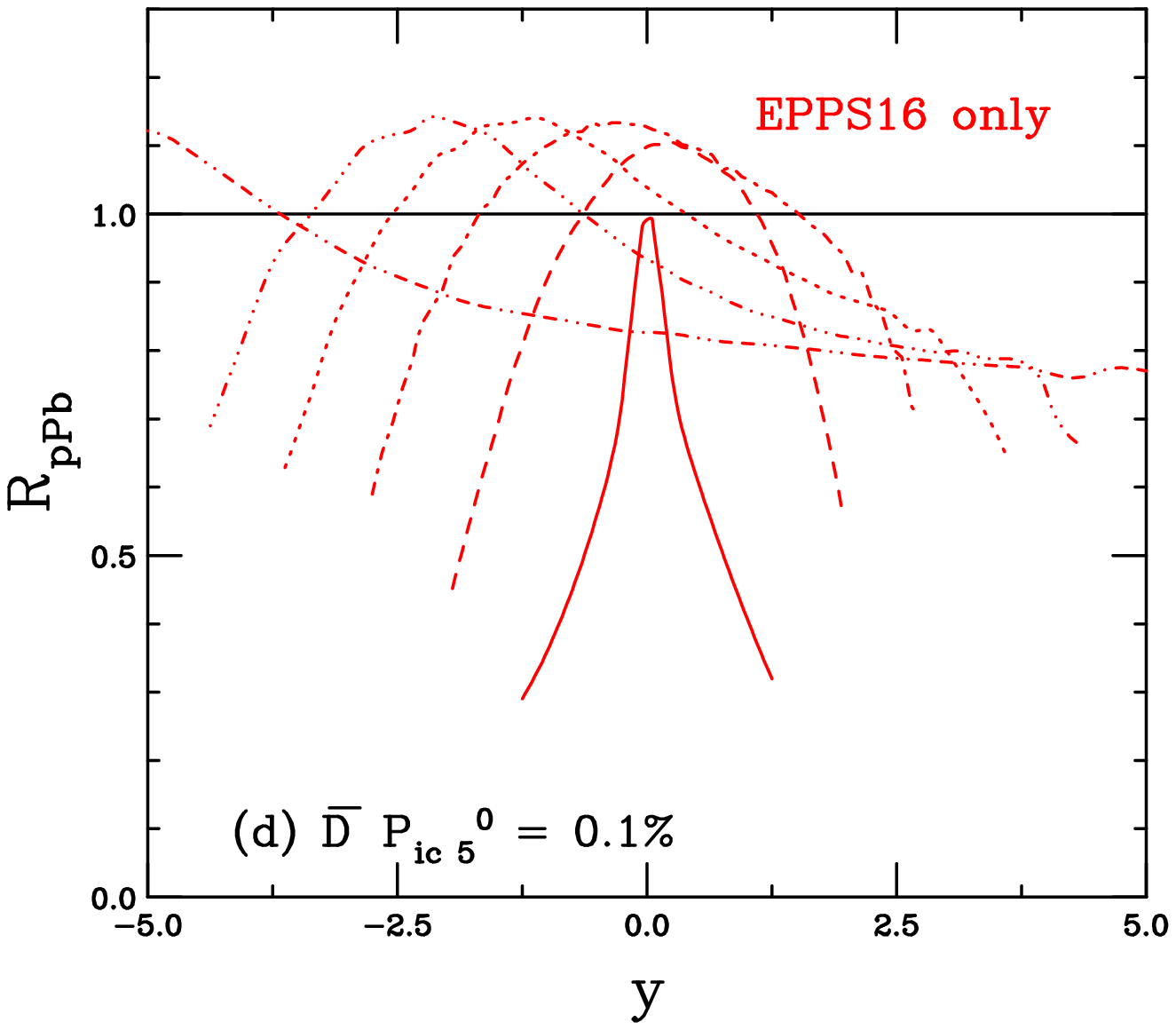} \\
    \includegraphics[width=0.4\textwidth]{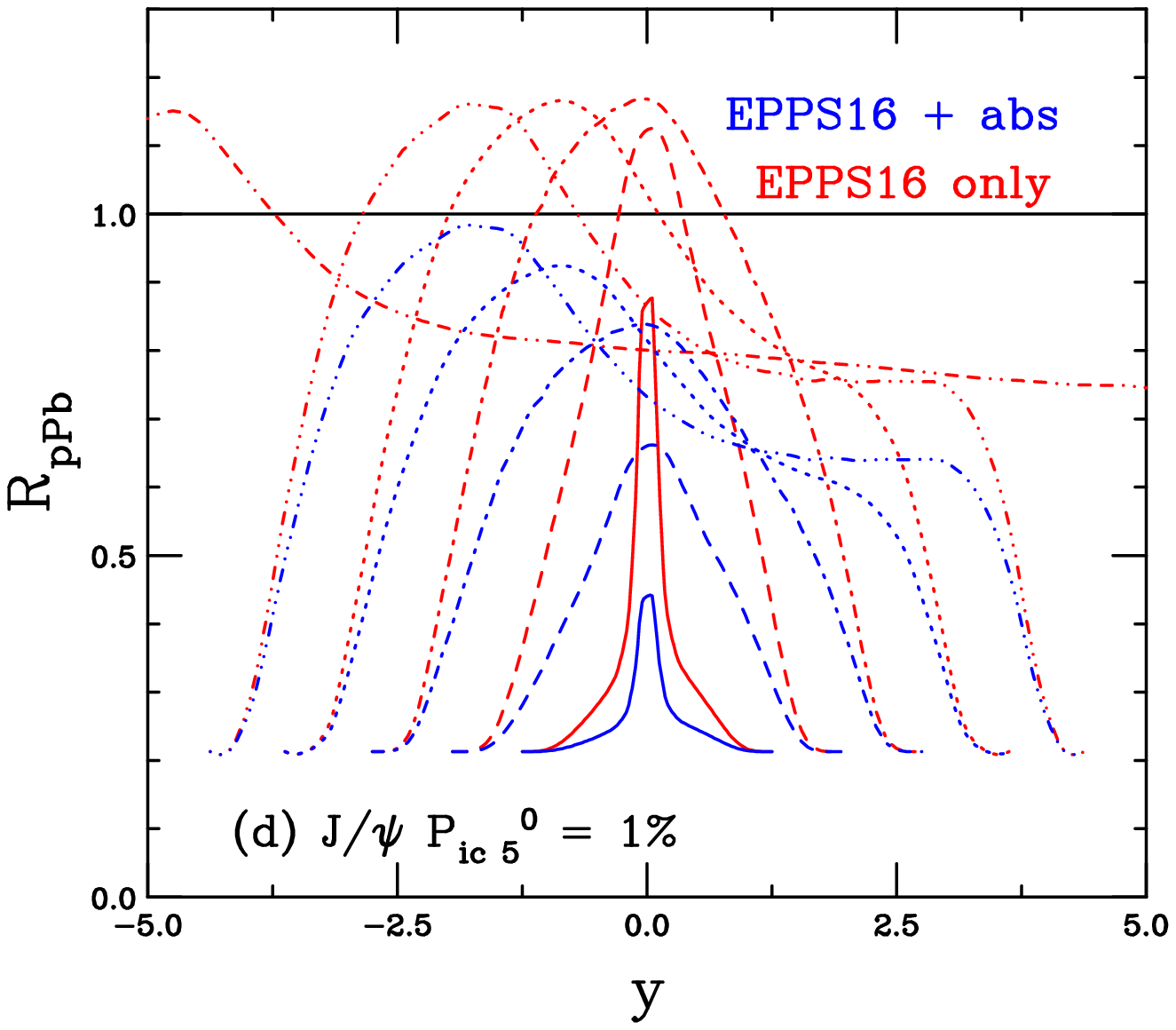}
    \includegraphics[width=0.4\textwidth]{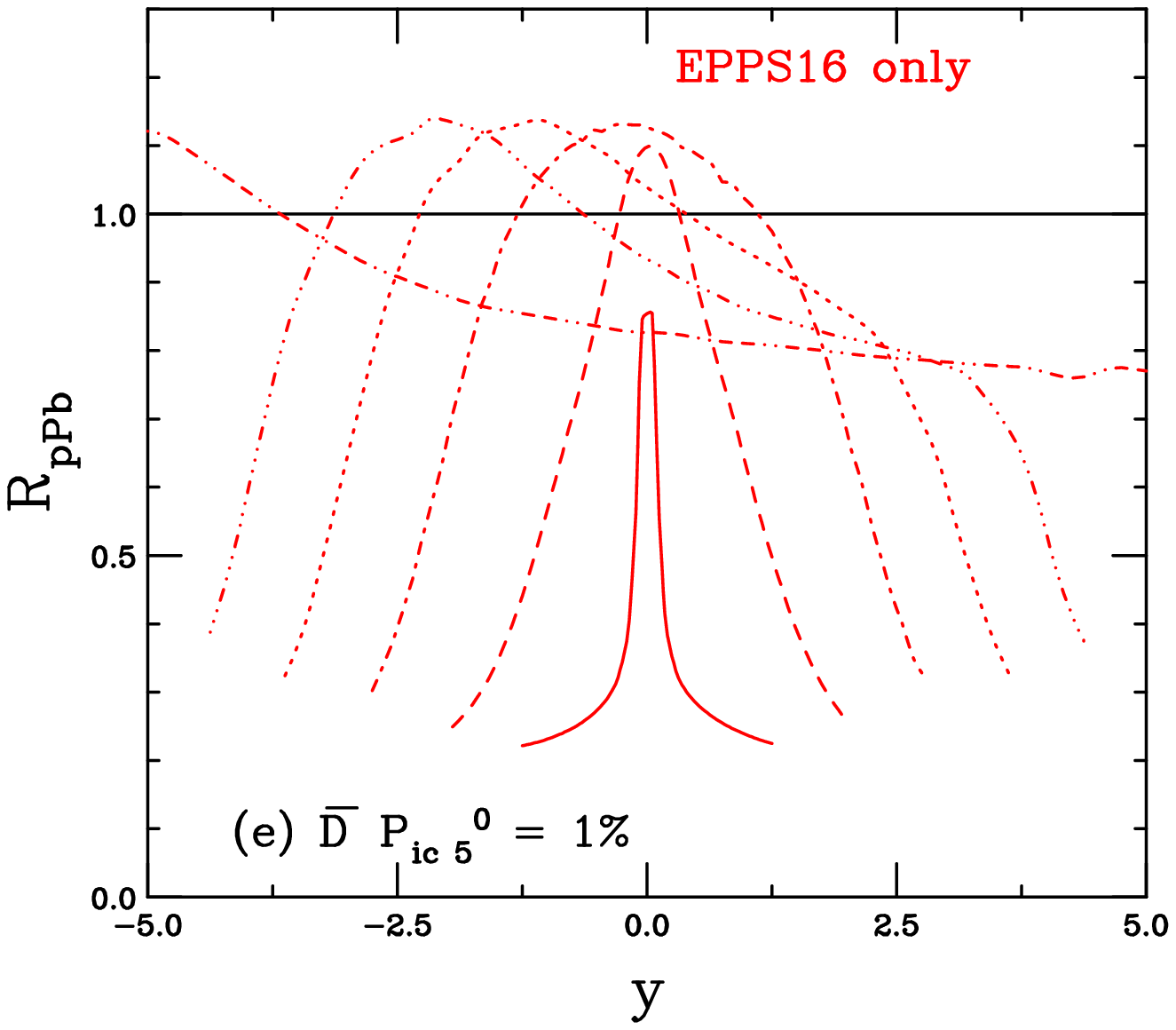}
  \end{center}
  \caption[]{ The nuclear suppression factor $R_{p {\rm Pb}}$ as a
    function of rapidity for 
    $J/\psi$ (a), (c), (e) and $\overline D$ (b), (d), (f) mesons, including
    both the typical perturbative QCD contribution and intrinsic charm from a
    five-particle proton Fock state.  The intrinsic charm contribution is
    varied in the panels from no intrinsic charm (pQCD
    only) (a) and (b); $P_{\rm ic \, 5}^0 = 0.1$\% (c) and (d); and
    $P_{\rm ic \, 5}^0 = 1$\% (e) and (f).  The red curves include EPPS16
    modifications of the parton densities only while the blue curves include
    nuclear absorption of the $J/\psi$.  (There is no absorption of the
    $\overline D$ mesons in cold nuclear matter.)  The line types denote
    different energies: $p_{\rm lab} = 40$~GeV
    (solid), 158~GeV (dashes), 800~GeV (dot-dashed), 
    $\sqrt{s} = 87.7$~GeV (dotted), 200~GeV (dot-dot-dot-dashed) and 5~TeV
    (dot-dot-dash-dashed).  
    }
\label{pPb_yrats}
\end{figure}

With this basic understanding of the behavior of the rapidity distributions in
mind, the behavior of $R_{p{\rm Pb}}(y)$ in Fig.~\ref{pPb_yrats} will be more
comprehensible.  Results are shown for $J/\psi$ on the left-hand side while
those for $\overline D$ mesons are on the right-hand side.  There are two sets
of curves for the $J/\psi$ depending on which cold nuclear matter effects are included in the perturbative QCD calculation:
in red, labeled ``EPPS16 only'' that includes 
nuclear modifications of the parton densities, and in blue, ``EPPS16 + abs'',
including absorption in nuclear matter, a finite $\sigma_{\rm abs}$.  Because no
absorption is included in the $\overline D$ calculations, only red curves are
shown.  The rapidity distribution does not depend on $k_T$ broadening.

The top plots, Fig.~\ref{pPb_yrats}(a) and (b), do not include intrinsic charm.
The ratio $R_{p{\rm Pb}}(y)$ generally follows the inverse of the ratio of the
nuclear modifications as a function of $x$ shown in Fig.~\ref{shad_ratios} with
low $x$ shadowing at forward rapidity, an antishadowing peak, an EMC region at
negative rapidity and, in the case of the $J/\psi$, the peak for Fermi motion
at the largest backward rapidity.  The $J/\psi$ rapidity distribution in
perturbative QCD is narrower than that of the $\overline D$ on average, thus
tending to reach both smaller and larger $x$ in the positive and negative
rapidity tails of the distributions respectively.  At the lowest energies shown,
the antishadowing peak appears at forward rapidity and moves backward to more
negative rapidity as the energy increases.

With no other cold nuclear matter
effects, the antishadowing peak remains at the same value of $R_{p{\rm Pb}}$ and
is only shifted backward in rapidity with increasing $\sqrt{s_{NN}}$.  However,
when $J/\psi$ absorption by nucleons is included, the entire ratio $R_{p{\rm Pb}}$
is shifted lower based on the value of $\sigma_{\rm abs}$ employed.  Because
$\sigma_{\rm abs}$ decreases with increasing $\sqrt{s_{NN}}$, the curves with and
without absorption move closer together at higher energies until, at the LHC
energy of 5~TeV, the two curves are on top of each other.

Intrinsic charm is included in the calculated $R_{pA}$ in
the lower four plots of Fig.~\ref{pPb_yrats}
with $P_{\rm ic \, 5}^0 = 0.1$\% in (c) and (d); and $P_{\rm ic \, 5}^0 = 1$\% in
(e) and
(f).  The perturbative QCD contribution remains unchanged in these plots.  The
change in $R_{pA}$ after intrinsic charm is included
is rather dramatic for the lowest energies where the intrinsic
charm contribution is large at midrapidity, particularly for
$p_{\rm lab} = 40$~GeV.  Already by $p_{\rm lab} = 158$~GeV, however, the
antishadowing peak in $R_{p {\rm Pb}}$ is almost at the same level as it was
without intrinsic charm.  Increasing the
value of $P_{\rm ic \, 5}^0$ from 0.1\% to
1\% tends to narrow the distribution without strongly affecting the peak
position except for $p_{\rm lab} = 40$~GeV.  Depending on the range of the
rapidity coverage of the experiment, even though the effect of including
intrinsic charm appears dramatic over all rapidity, the potential effect in the
measured region may, in fact, be very small or even negligible, as demonstrated
by the fact that the LHC result remains unchanged at $y = \pm 5$.  

It is worth noting that all the ratios tend to the same minimum value of
$R_{p{\rm Pb}}$ after intrinsic charm is dominant and the steeply-falling
perturbative QCD result becomes negligible at the edge of rapidity space.  As
previously discussed, that
minimum is simply the ratio of the intrinsic charm dependence in a nucleus
relative to the proton, $A^{\beta-1} = 0.213$ for a lead target.
It would be interesting to see if there would be a strong narrowing of the
$A$ dependence in the fixed-target region of $p_{\rm lab} = 40$, 80 and 120~GeV
expected to be studied by the NA60+ Collaboration \cite{NA60+}.  However, for
the result to be better quantified, there should be sufficient statistical
significance for the rapidity distribution in the central unit of rapidity,
$|y| < 1$ to make several bins with small enough uncertainties to discern
whether any narrowing exists.

\subsubsection{Transverse momentum dependence}
\label{sec:pPb_pT}

\begin{figure}
  \begin{center}
    \includegraphics[width=0.4\textwidth]{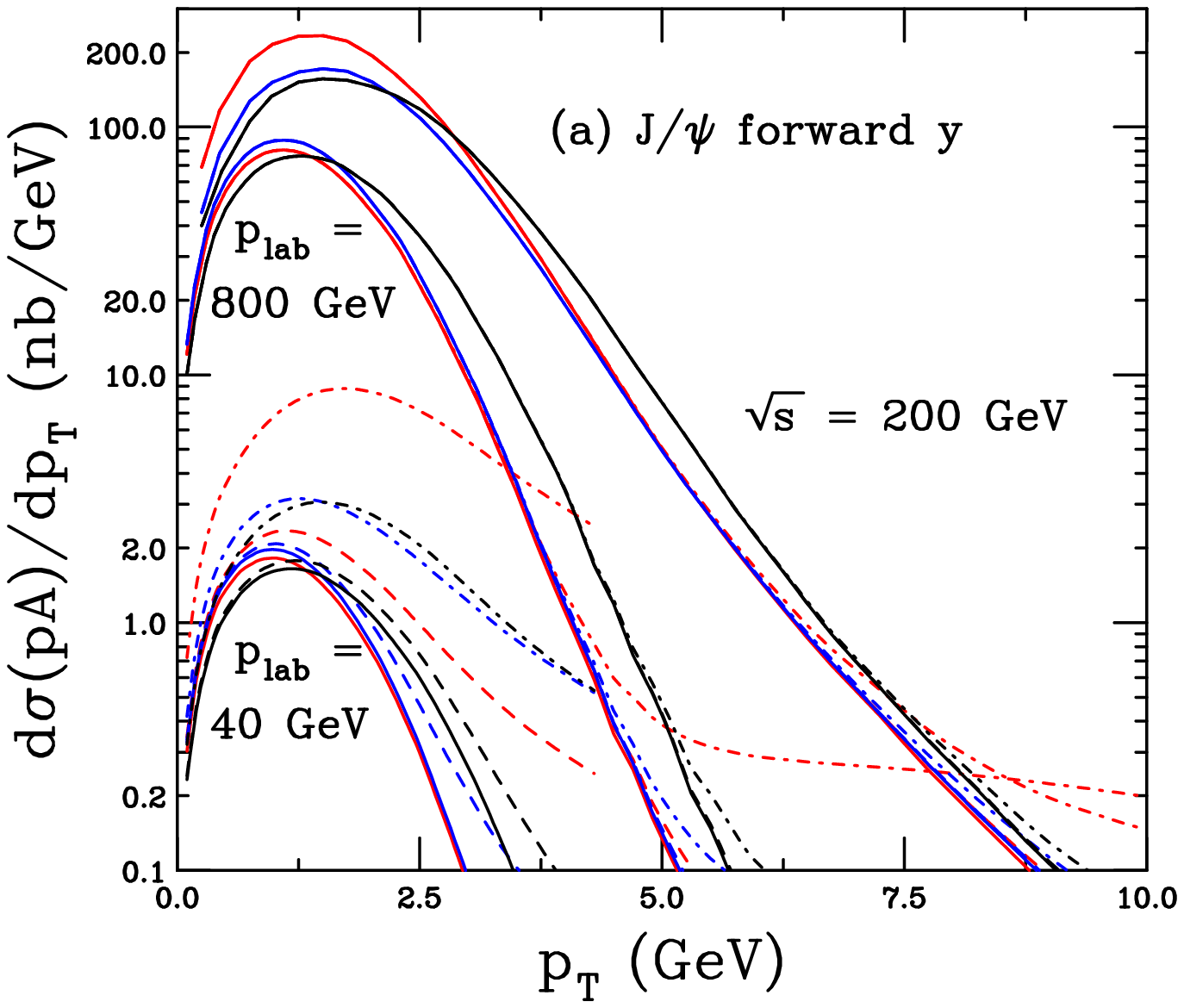} 
    \includegraphics[width=0.4\textwidth]{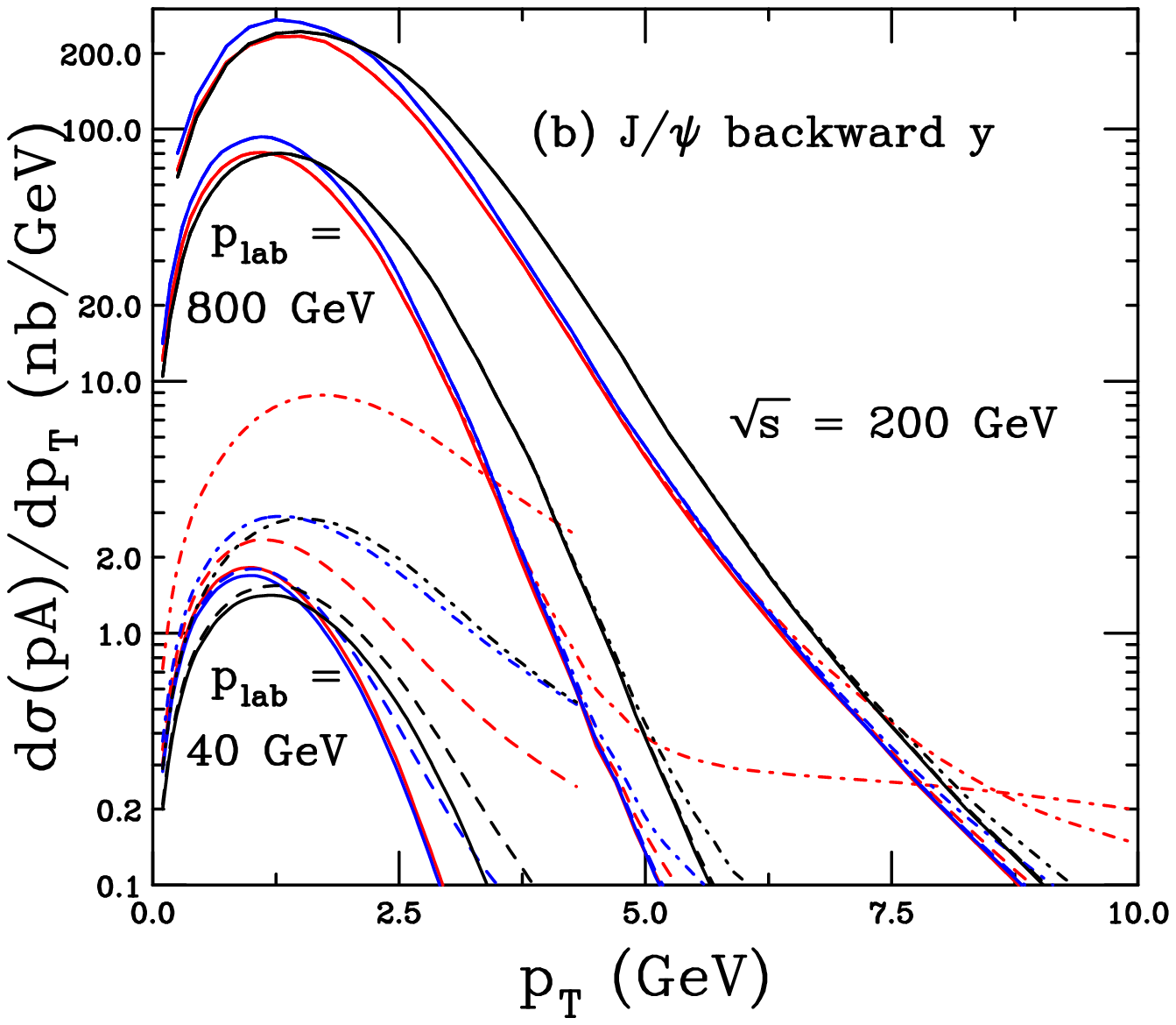}
  \end{center}
  \caption[]{ The $p+p$ and $p+{\rm Pb}$ (per nucleon)
    $J/\psi$ distributions at
    $p_{\rm lab} = 40$ and 800~GeV and $\sqrt{s} = 200$~GeV as a function of
    $p_T$ at forward (a) and backward (b) rapidity.
    The red curves show the results for $p+p$ collisions while the
    blue and black curves show the $p+{\rm Pb}$
    distributions without and with an enhanced intrinsic $k_T$ kick
    respectively.  Three curves are shown in each case: no intrinsic charm (pQCD
    only, solid); $P_{\rm ic \, 5}^0 = 0.1$\% (dashed);
    and $P_{\rm ic \, 5}^0 = 1$\% (dot-dashed).  No $J/\psi$ absorption by
    nucleons is considered in the $p+{\rm Pb}$ calculation.
    }
\label{pp_pPb_ptdists}
\end{figure}

The $J/\psi$ $p+p$ and $p+{\rm Pb}$ $p_T$ distributions at forward and backward
rapidity are shown in Fig.~\ref{pp_pPb_ptdists}.  At $p_{\rm lab} = 40$ and
800~GeV,
forward rapidity refers to $0 < y < 1$ whereas backward rapidity means
$-1 < y < 0$.  With $\sqrt{s_{NN}} = 200$~GeV, forward $y$ is $1.1 < y < 2.2$
while backward $y$ is $-2.2 < y < -1.1$.  Red curves are shown for the $p+p$
distribution, solid, without intrinsic charm and the dashed and dot-dashed
curves with $P_{\rm ic \, 5}^0 = 0.1$\% and 1\% respectively.  Now, however,
results for $p+{\rm Pb}$ are given both without (black) and with (blue)
enhanced $k_T$ broadening.
The resulting changes in the distributions due to the
different effects considered are discussed in turn.

At the energies shown, for low to moderate $p_T$, perturbative effects
dominate the shape of the $p_T$ distributions.  As shown in
Table~\ref{table:pT_averages}, the average $p_T$ from perturbative QCD increases
rather slowly with collision energy, growing 30\% between
$p_{\rm lab} = 40$~GeV and $\sqrt{s_{NN}} = 110.4$~GeV in the same rapidity
region and 44\% between $p_{\rm lab} = 40$~GeV and $\sqrt{s_{NN}} = 200$~GeV
even though the lower
energy calculation is made at central rapidity and the RHIC calculation
is for $1.1 < y < 2.2$.
The primary change is that the distribution becomes harder with
increasing energy.
Before any enhanced $k_T$ kick in the nucleus is included, one can see
antishadowing at low $p_T$ for $p_{\rm lab} = 40$ and 800~GeV
and shadowing at low
$p_T$ for $\sqrt{s_{NN}} = 200$~GeV at forward rapidity.
At backward rapidity,
on the other hand, the shape of the low $p_T$ distribution does not reveal a
significant modification due to the parton densities at the fixed-target
energies while a small enhancement due to antishadowing is seen for
$\sqrt{s_{NN}} = 200$~GeV.  At higher $p_T$, the effect due to employing the
EPPS16 set decreases due to the $Q^2$ evolution of the nuclear modification,
reducing the importance of the effect.  Indeed, at higher $p_T$ at both
energies, at backward rapidity the solid red and blue curves lie on top of each
other.  The modifications of the parton densities in the nucleus is the only
cold nuclear matter effect producing any differences in
the $p+ {\rm Pb}$ calculations (solid blue curves)
at forward and backward rapidity, see Fig.~\ref{pPb_yrats}.

Including the enhanced $k_T$ broadening, shown in the black curves of
Fig.~\ref{pp_pPb_ptdists}, has the effect of reducing the peak of the $p_T$
distribution at low $p_T$ and hardening the distribution at higher $p_T$.
The hardening is more significant at the lower energies where the average $p_T$
in perturbative QCD
increases by $\sim 15-20$\% while, at $\sqrt{s_{NN}} = 200$~GeV, the average
increase in $p_T$ is $\sim 9$\%.  Thus the ratio of $p+{\rm Pb}$ to $p+p$
with enhanced broadening
may be expected to be less than or close to unity at low $p_T$, depending on
whether forward or backward rapidity is considered, and grow significantly above
unity for $p_T > 1.25$~GeV at $p_{\rm lab} = 40$~GeV.  A similar but smaller
overall effect may be expected for the RHIC energy of $\sqrt{s_{NN}} = 200$~GeV.
This difference is because the enhanced $k_T$ kick due to the presence of the
nucleus is assumed to be effectively independent of incident energy with
$\delta k_T^2 \approx 0.45$~GeV$^2$.  This value is large relative to
$\langle k_T^2 \rangle_p^{1/2} < 1$~GeV$^2$ at $p_{\rm lab} = 40$~GeV where the
average $p_T$ in perturbative QCD is 1.23~GeV.  The higher overall average $p_T$
at $\sqrt{s_{NN}} = 200$~GeV reduces the effect of the enhanced broadening but
does not eliminate it: $\langle k_T^2 \rangle_p^{1/2} = 1.1$ at
$\sqrt{s_{NN}} = 200$~GeV while $\langle p_T \rangle = 1.77$~GeV for
$1.1 < y < 2.2$.  The additional increase in $\langle k_T^2 \rangle_A$ over
$\langle k_T^2 \rangle_p$ makes the enhanced broadening in lead almost equivalent to or greater than the average $p_T$ in $p+p$ collisions, even at collider
energies.  This rather large effect is particular to charm quarks relative
the heavier bottom quarks which show weaker modifications due to enhanced
broadening due to their larger mass and harder overall $p_T$ distributions
\cite{RV_azi1,RV_azi2}.

When intrinsic charm is also included in the calculation, the effect on the
$p_T$ distribution at $p_{\rm lab} = 40$~GeV is rather dramatic for the $p+p$
distribution, as already shown in Fig.~\ref{pp_pTdists_en}.
a long, much harder
tail is seen in the distribution, particularly at the lower energy.  As
previously discussed, the hardening of the distribution
above a certain $p_T$ is due to the
restriction of phase space at high $p_T$: only a fraction of the rapidity
integral will result in $x_F < 1$.  The end point of the calculation in $p_T$
is the value at which the criteria $x_F < 1$ can no longer be satisfied for
any part of the rapidity range.  The strong nuclear target suppression of
intrinsic charm relative to perturbative QCD is also seen here.  As was the
case for the rapidity distributions, the relative $A$ dependence will provide
a natural minimum of $R_{p{\rm Pb}}$ at high $p_T$ with intrinsic charm.

\begin{figure}
  \begin{center}
    \includegraphics[width=0.4\textwidth]{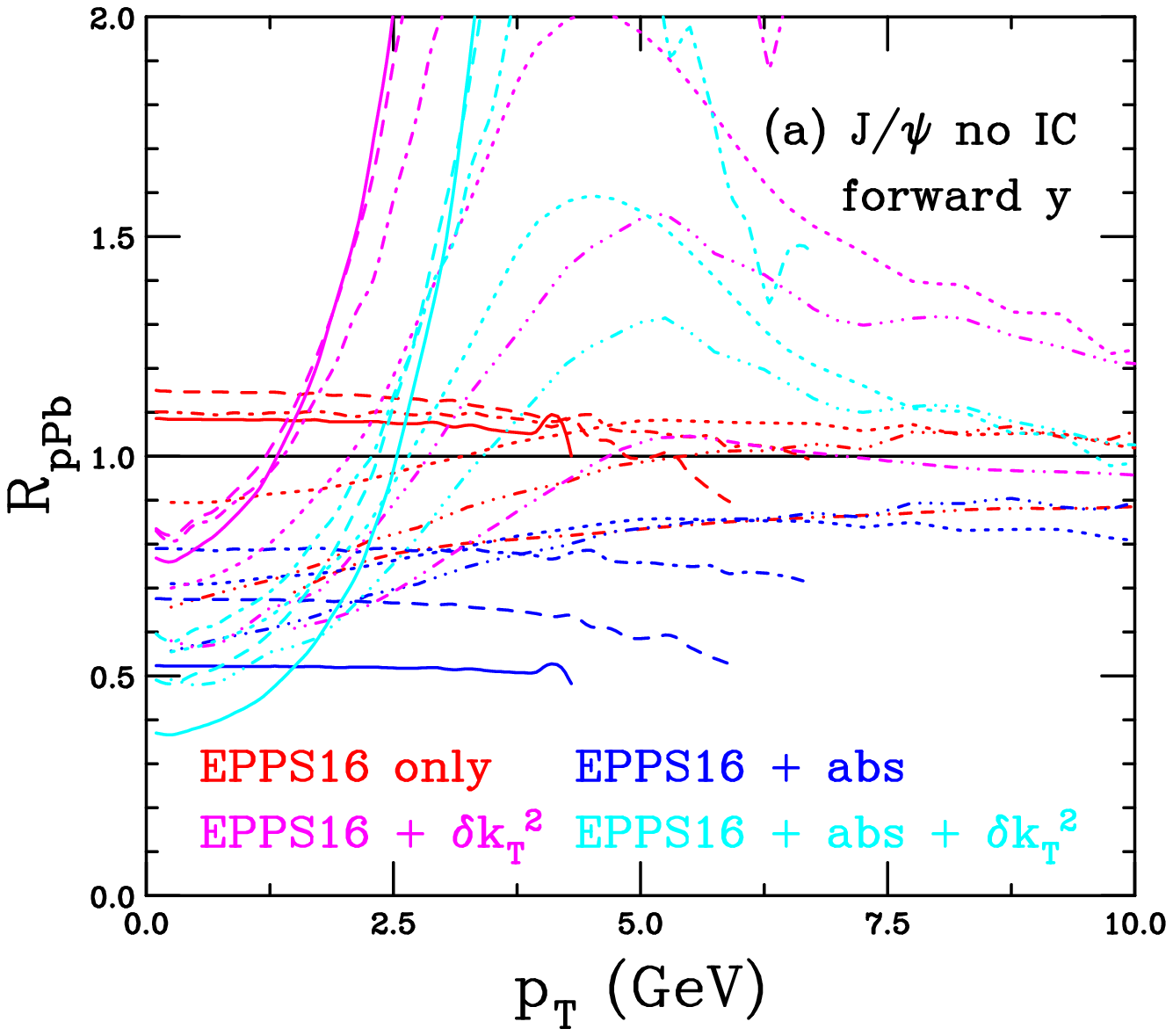}
    \includegraphics[width=0.4\textwidth]{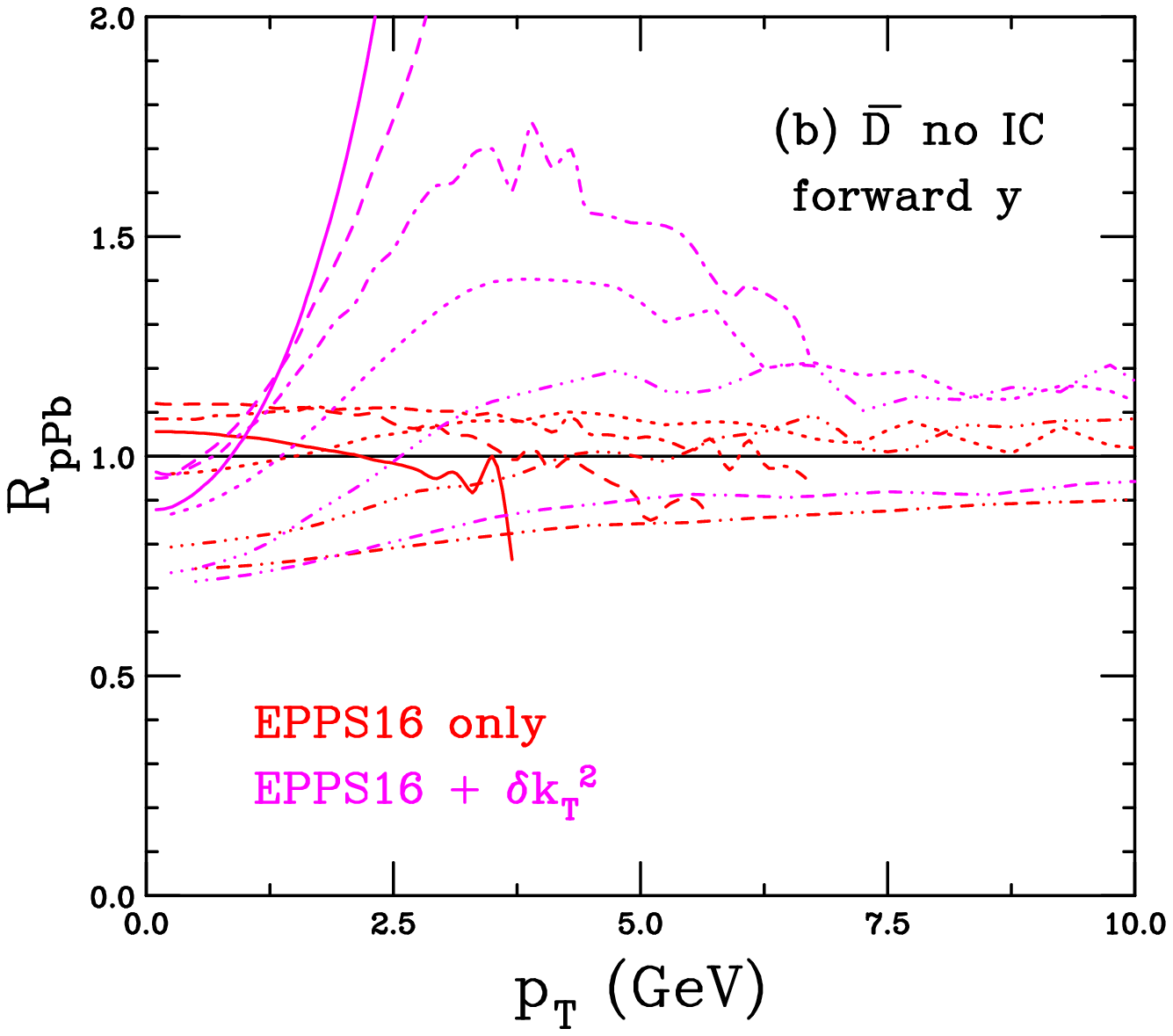} \\
    \includegraphics[width=0.4\textwidth]{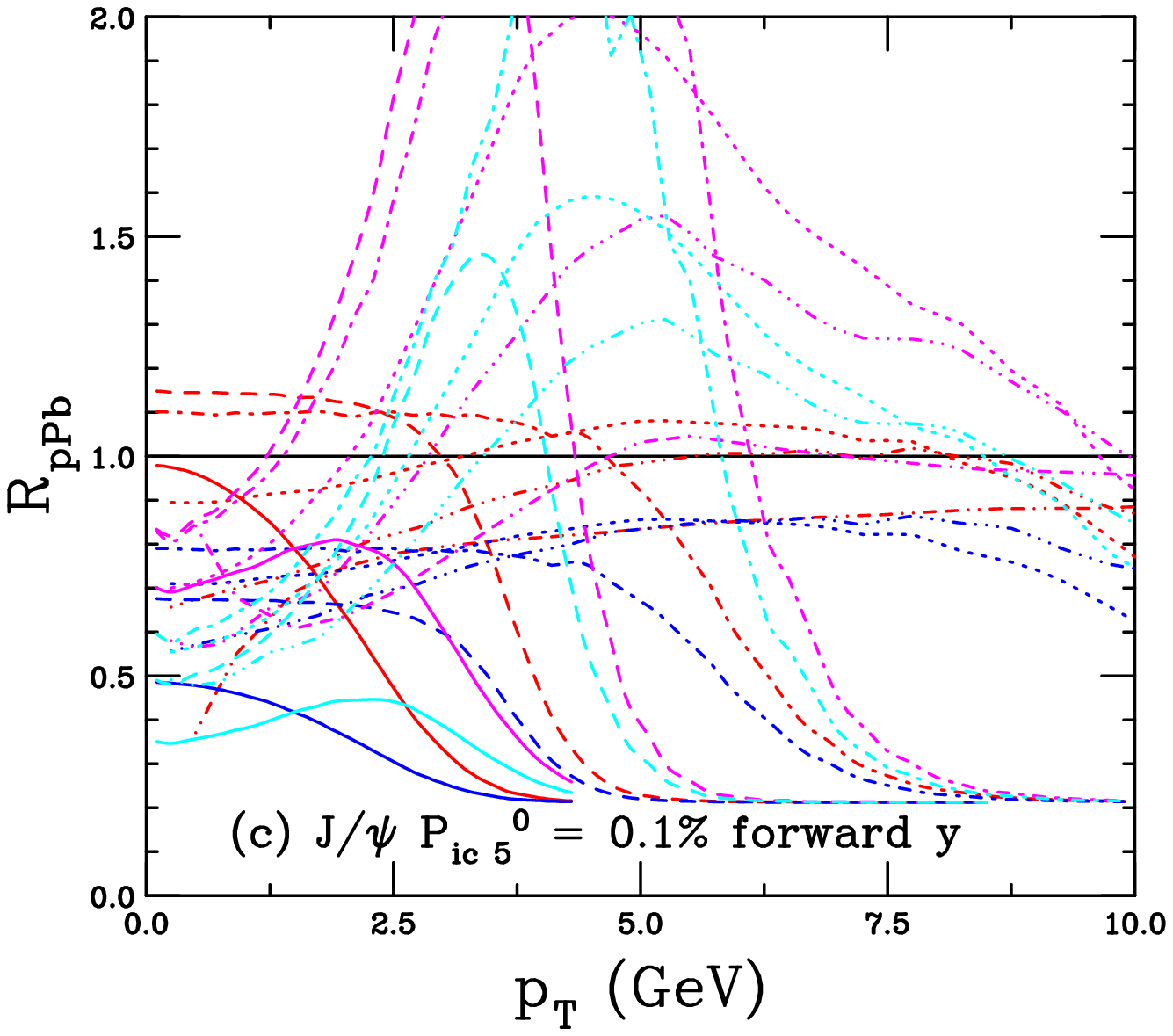}
    \includegraphics[width=0.4\textwidth]{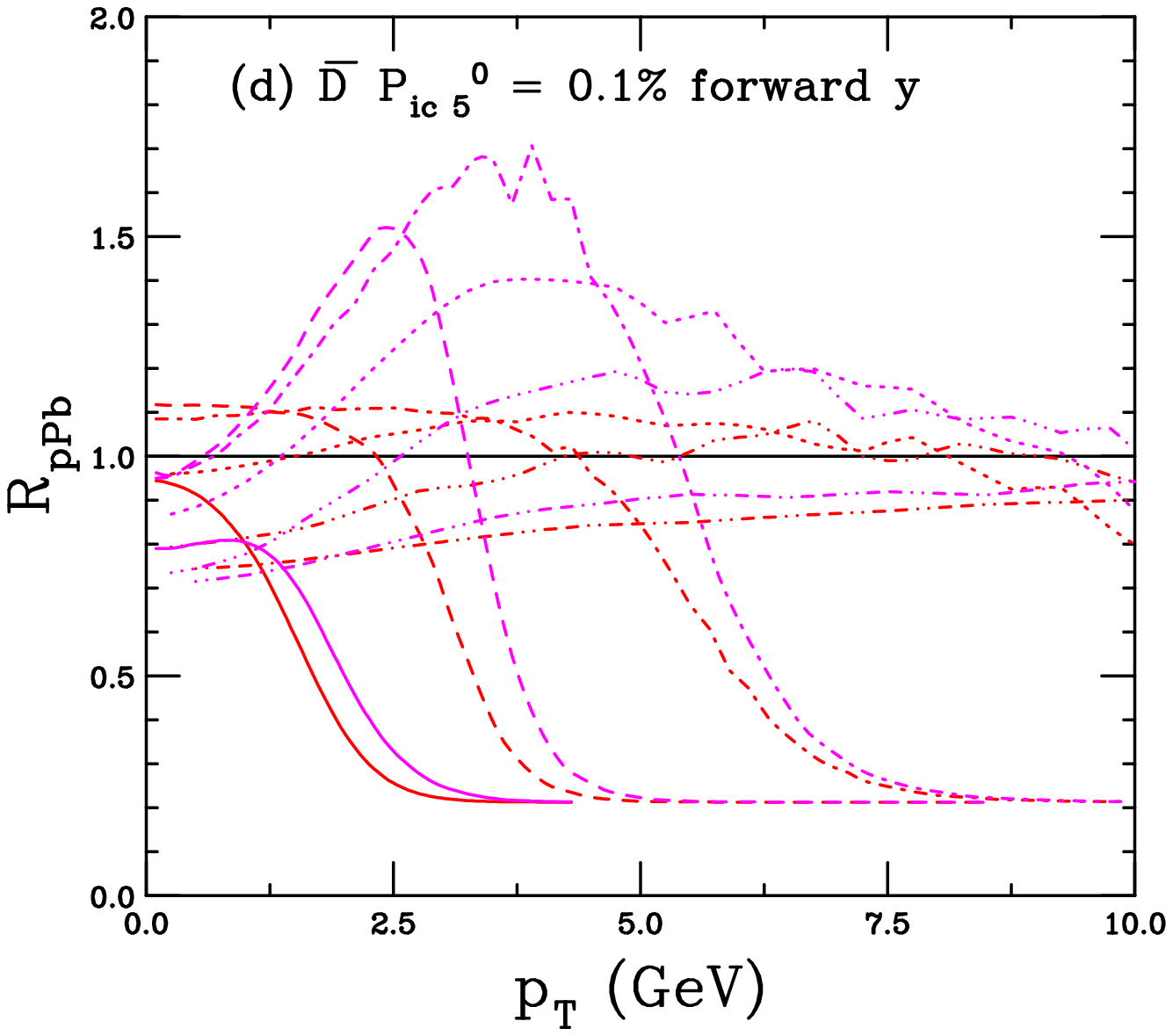} \\
    \includegraphics[width=0.4\textwidth]{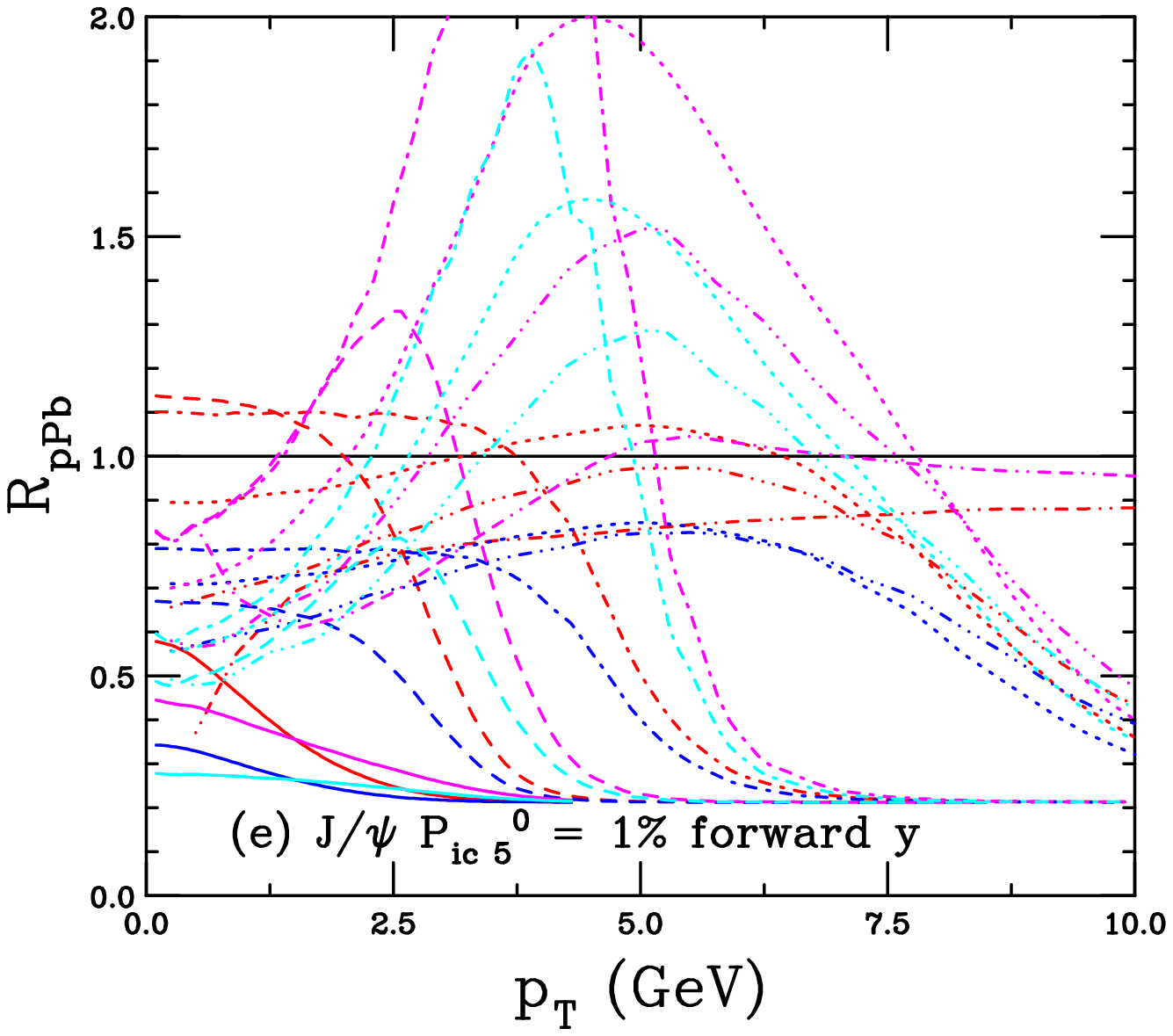}
    \includegraphics[width=0.4\textwidth]{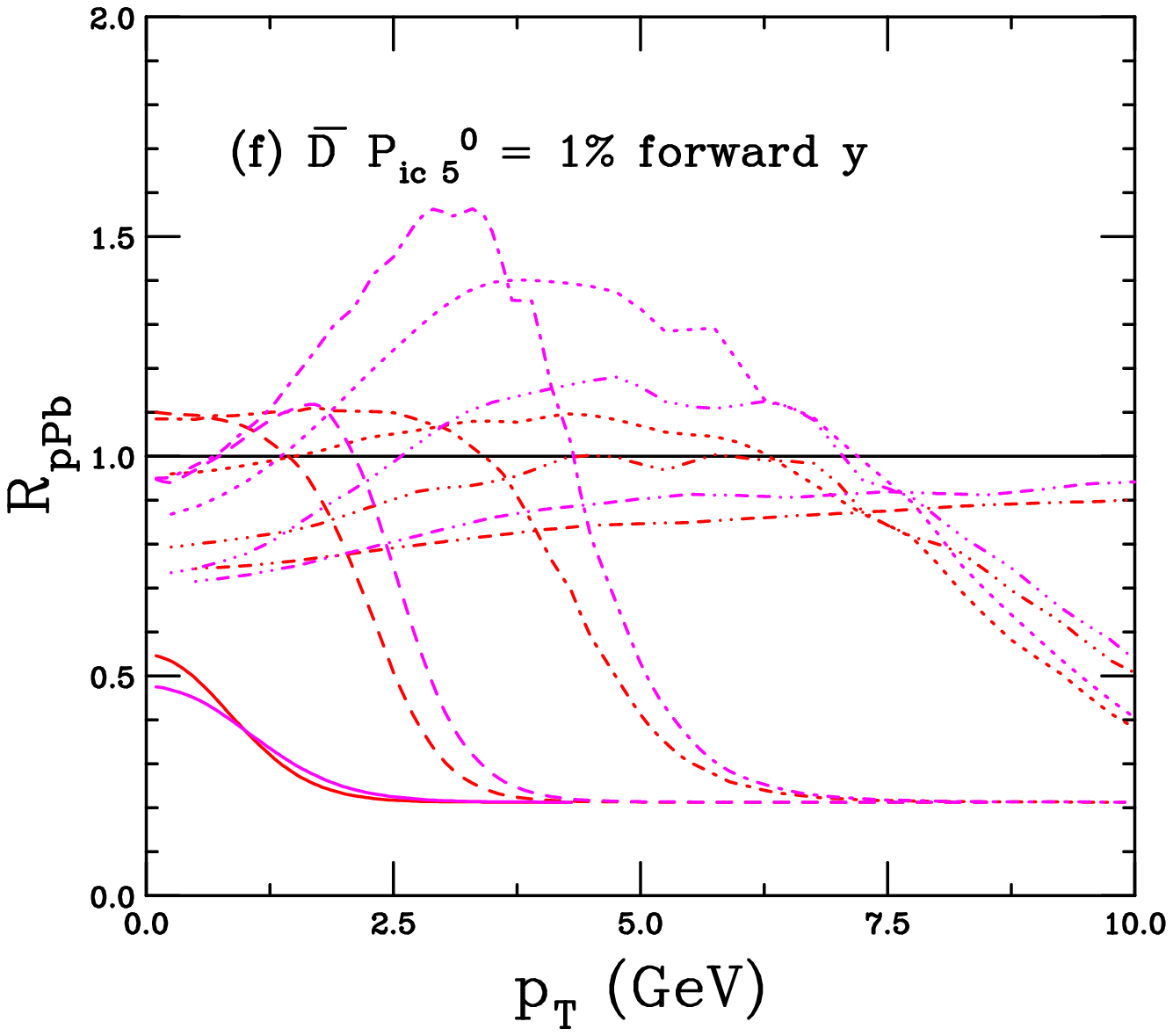}
  \end{center}
  \caption[]{ The nuclear suppression factor $R_{p {\rm Pb}}$ as a
    function of transverse momentum at forward rapidity for 
    $J/\psi$ (a), (c), (e) and $\overline D$ (b), (d), (f) mesons, including
    both the typical perturbative QCD contribution and intrinsic charm from a
    five-particle proton Fock state.  The intrinsic charm contribution is
    varied in the panels from no intrinsic charm (pQCD
    only) (a) and (b); $P_{\rm ic \, 5}^0 = 0.1$\% (c) and (d); and
    $P_{\rm ic \, 5}^0 = 1$\% (e) and (f).  The red curves include the EPPS16
    modifications of the parton densities only while the blue curves also
    include nuclear absorption of the $J/\psi$.  The magenta curves include
    the EPPS16 modifications as
    well as $k_T$ broadening while the cyan curves include EPPS16, nuclear
    absorption for the $J/\psi$, and $k_T$ broadening.
    (There is no absorption of the
    $\overline D$ mesons in cold nuclear matter.)  The line types denote
    different energies: $p_{\rm lab} = 40$~GeV
    (solid), 158~GeV (dashes), 800~GeV (dot-dashed), 
    $\sqrt{s_{NN}} = 87.7$~GeV (dotted), 200~GeV (dot-dot-dot-dashed) and 5~TeV
    (dot-dot-dash-dashed).  Note that the rapidity range is $0 < y < 1$ for
    all energies except the two highest where the rapidity range is
    $1.1 < y < 2.2$ for 200~GeV and $2.5 < y < 5$ for 5~TeV.
    }
\label{pPb_pTrats_fory}
\end{figure}

\begin{figure}
  \begin{center}
    \includegraphics[width=0.4\textwidth]{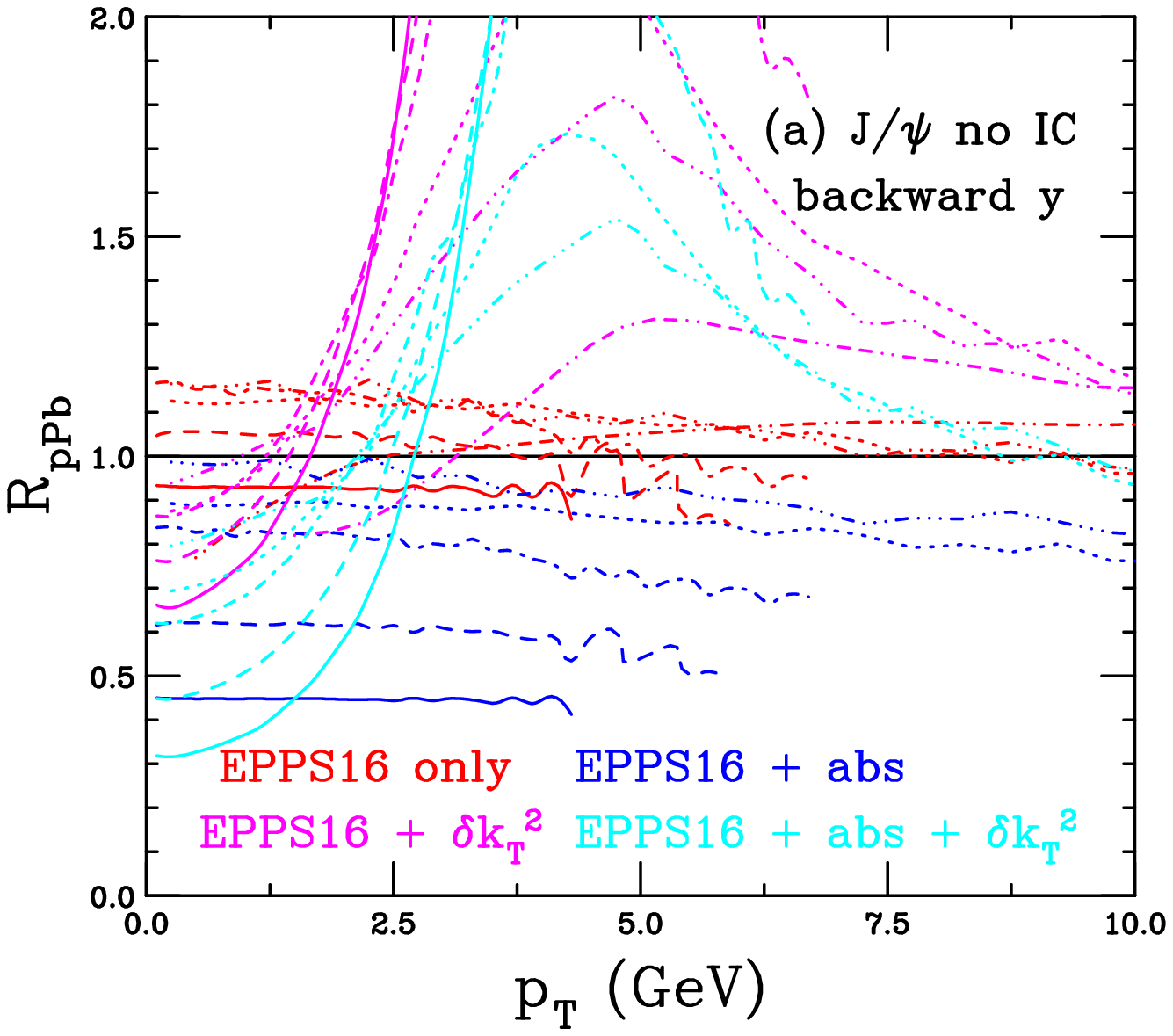}
    \includegraphics[width=0.4\textwidth]{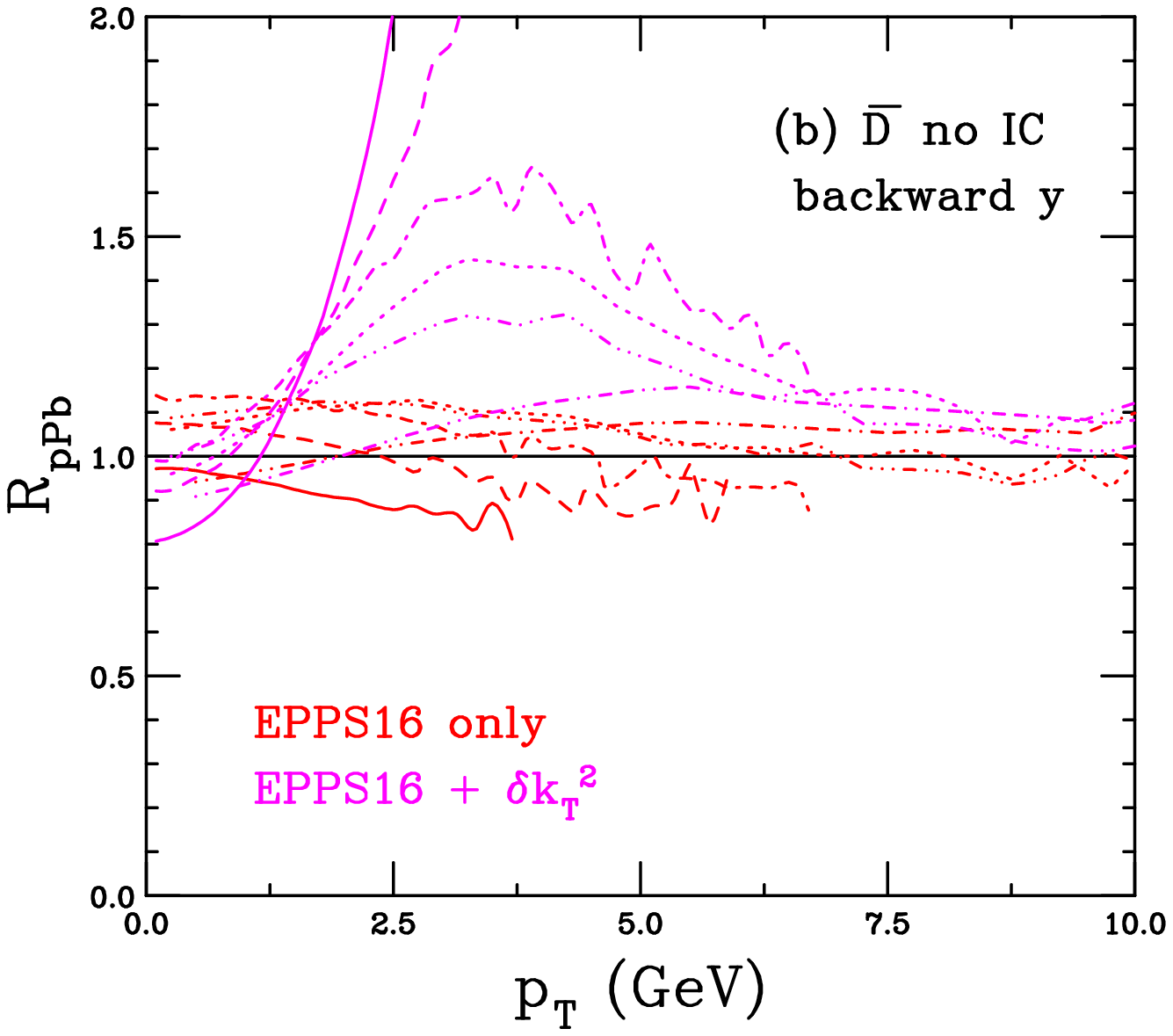} \\
    \includegraphics[width=0.4\textwidth]{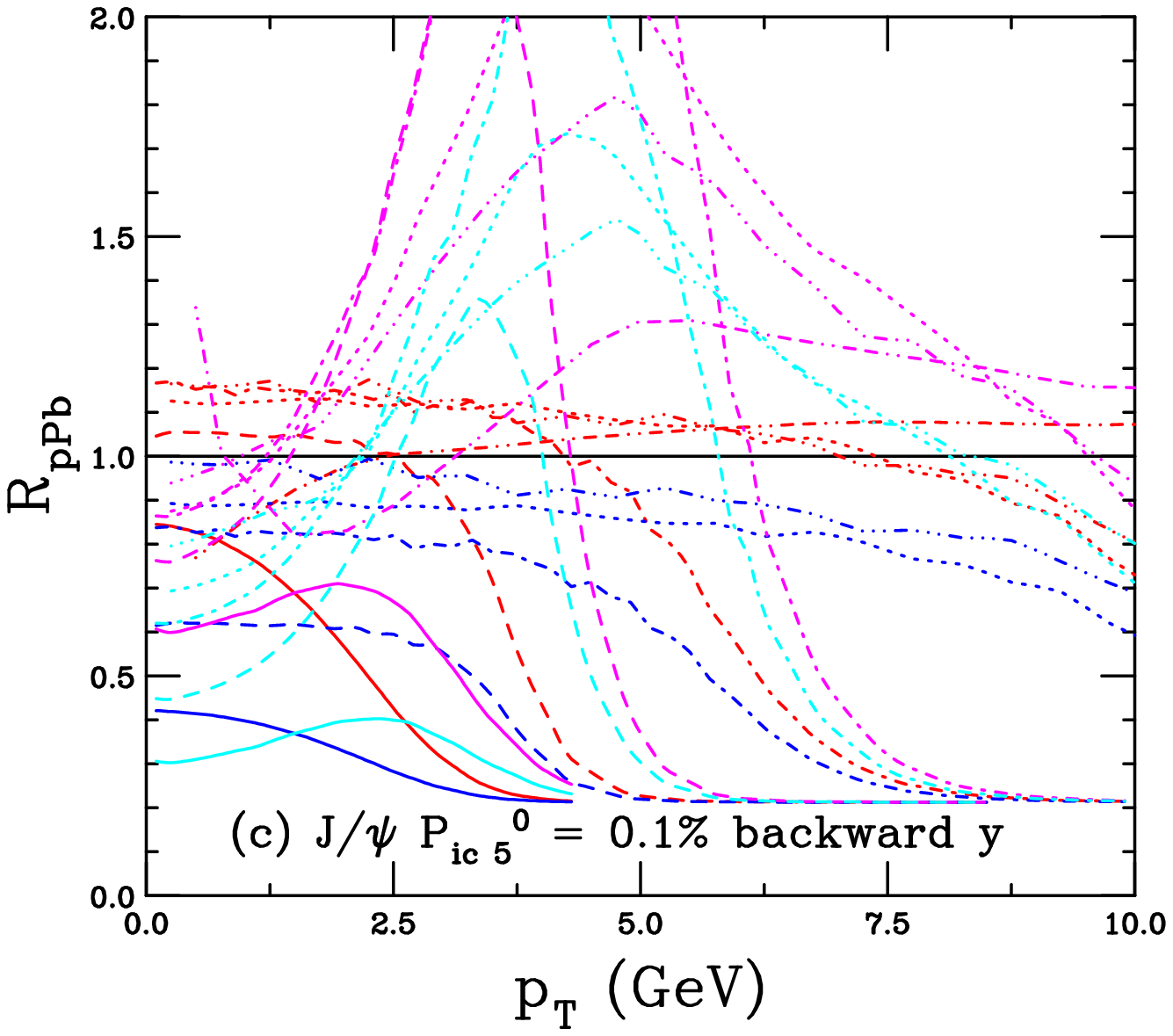}
    \includegraphics[width=0.4\textwidth]{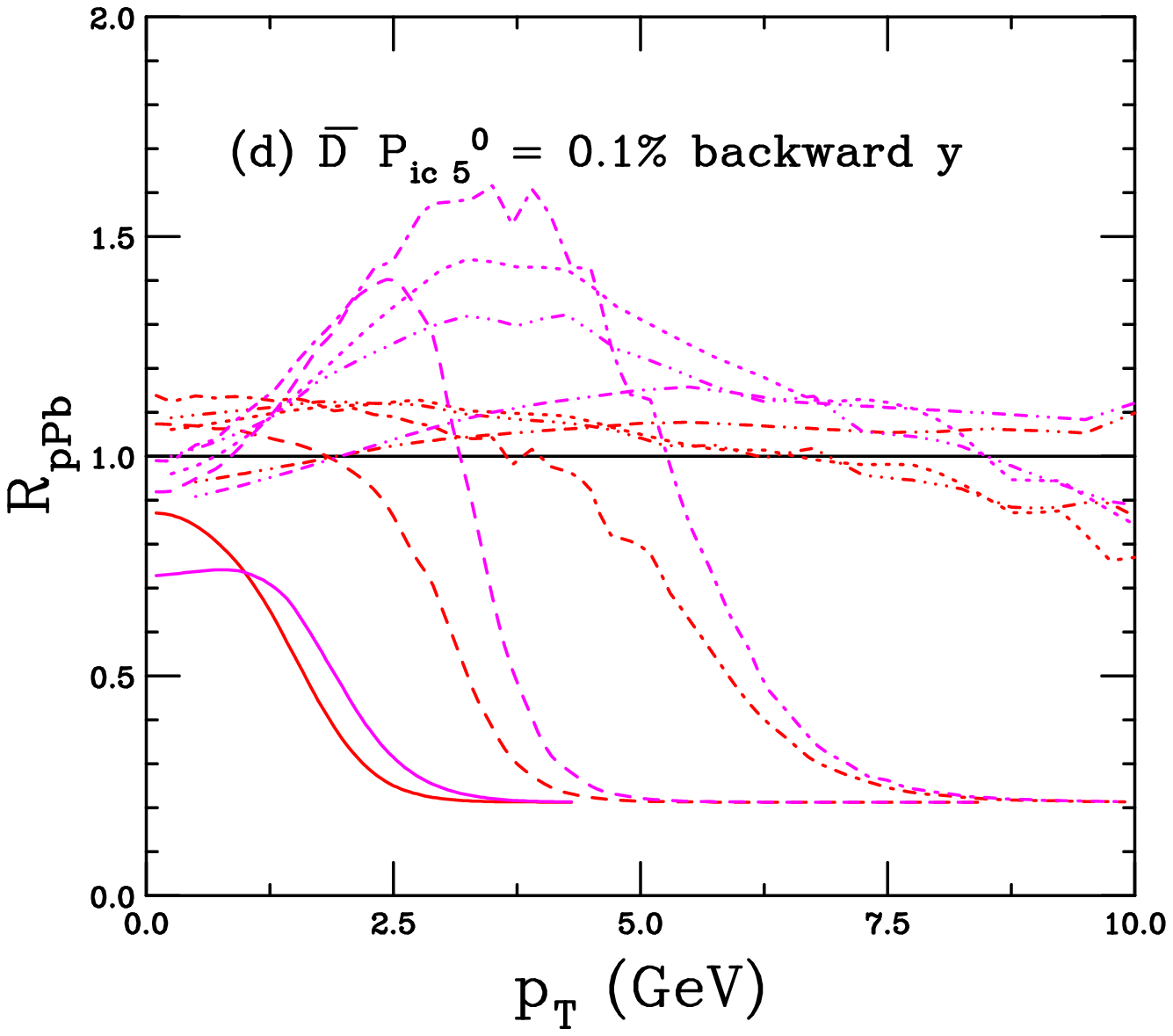} \\
    \includegraphics[width=0.4\textwidth]{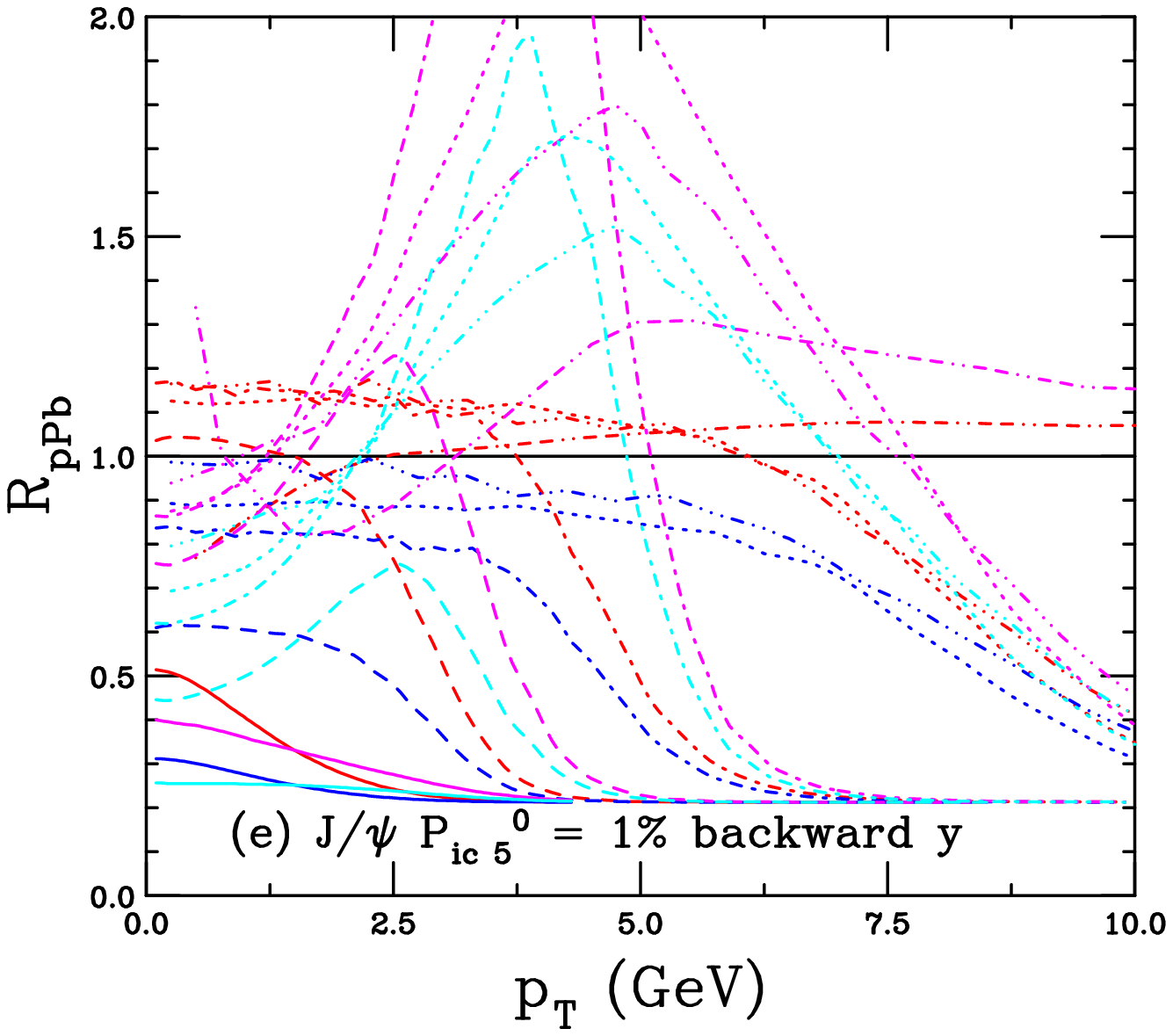}
    \includegraphics[width=0.4\textwidth]{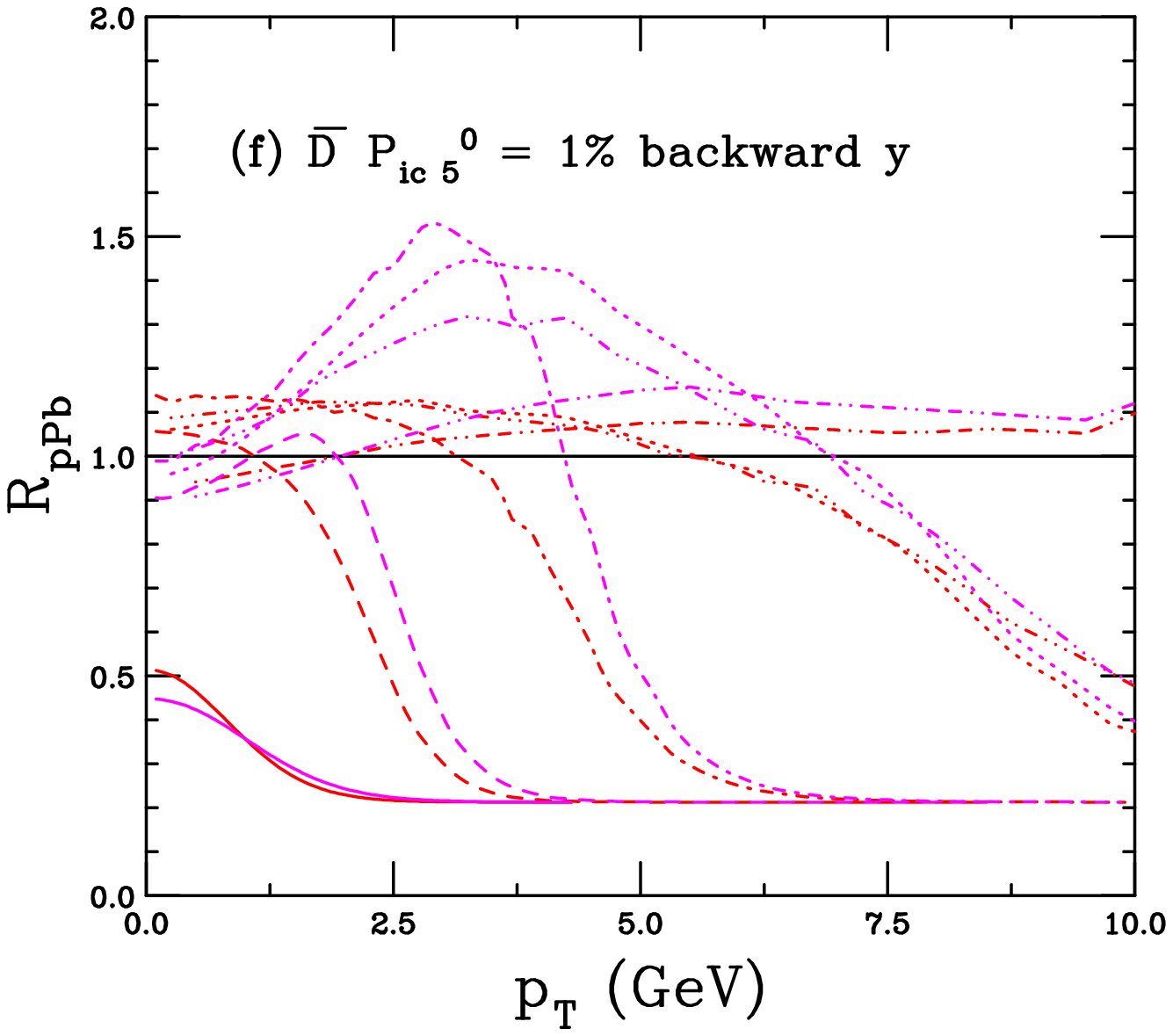}
  \end{center}
  \caption[]{ The nuclear suppression factor $R_{p {\rm Pb}}$ as a
    function of transverse momentum at backward rapidity for 
    $J/\psi$ (a), (c), (e) and $\overline D$ (b), (d), (f) mesons, including
    both the typical perturbative QCD contribution and intrinsic charm from a
    five-particle proton Fock state.  The intrinsic charm contribution is
    varied in the panels from no intrinsic charm (pQCD
    only) (a) and (b); $P_{\rm ic \, 5}^0 = 0.1$\% (c) and (d); and
    $P_{\rm ic \, 5}^0 = 1$\% (e) and (f).  The red curves include the EPPS16
    modifications of the parton densities only while the blue curves include
    nuclear absorption of the $J/\psi$.  The magenta curves include the
    EPPS16 modification as
    well as $k_T$ broadening while the cyan curves include EPPS16, nuclear
    absorption for the $J/\psi$, and $k_T$ broadening.
    (There is no absorption of the
    $\overline D$ mesons in cold nuclear matter.)  The line types denote
    different energies: $p_{\rm lab} = 40$~GeV
    (solid), 158~GeV (dashes), 800~GeV (dot-dashed), 
    $\sqrt{s_{NN}} = 87.7$~GeV (dotted), 200~GeV (dot-dot-dot-dashed) and 5~TeV
    (dot-dot-dash-dashed).  Note that the rapidity range is $-1 < y < 0$ for
    all energies except the two highest where the rapidity range is
    $-2.2 < y < -1.1$ for 200~GeV and $-5 < y < -2,5$ for 5~TeV.
    }
\label{pPb_pTrats_bcky}
\end{figure}

Now that the shape of the $p_T$ distributions due to the differences between
$p+p$ and $p+{\rm Pb}$ systems has been clarified in Fig.~\ref{pp_pPb_ptdists},
the behavior of
$R_{p{\rm Pb}}(p_T)$ can be better understood.  These ratios are shown in
Fig.~\ref{pPb_pTrats_fory} at forward rapidity and Fig.~\ref{pPb_pTrats_bcky}
at backward rapidity.  In addition to ``EPPS16 only'' and ``EPPS16 + abs'', as
shown in Fig.~\ref{pPb_yrats}, the ratios are now also shown for enhanced
$k_T$ broadening by the curves for
``EPPS16 + $\delta k_T^2$'' and ``EPPS16 + abs + $\delta k_T^2$''
because of the evident effects
of increased $k_T$ broadening shown in Fig.~\ref{pp_pPb_ptdists}.  The $J/\psi$
calculations are shown on the left-hand sides while the $\overline D$ results,
without absorption, are presented on the right-hand sides.  The calculations for
$p_{\rm lab} = 40$, 158, and 800~GeV as well as for $\sqrt{s_{NN}} = 87.7$~GeV
are all made for $0 < |y| < 1$ while the calculations at $\sqrt{s_{NN}} = 200$
and 5~TeV are made for $1.1 < |y| < 2.2$ and $2.5 < |y| < 5$ respectively.

The results without intrinsic charm are shown in the top two plots of both
figures.  Before absorption is included, the results without broadening are
above or close to unity for $\sqrt{s_{NN}} \leq 87.7$~GeV.
At collider energies and forward rapidity, a reduction due to shadowing is
seen for $\sqrt{s_{NN}} = 200$~GeV and 5~TeV for both $J/\psi$ and
$\overline D$.  At backward rapidity, antishadowing is seen instead at these
energies when no absorption is included.  Adding absorption makes $R_{p{\rm Pb}}$
less than unity at all energies.

The effect of enhanced $k_T$ broadening can perhaps most easily be seen for
$\overline D$ mesons because there are no absorption effects.
There is a very strong effect predicted for the lowest
energies in particular where the ratio goes from being nearly independent of
$p_T$ to a strong increase with $p_T$.  This is because,
as already mentioned, the
increase $\delta k_T^2$ in the lead nucleus is large compared to the overall
average $p_T$, about half the average $p_T$, and a total
$\langle k_T^2 \rangle_A$ similar to the charm quark mass itself.  Only at
higher energies does the ratio turn over and begin to approach unity from
above, as seen in typical results for the Cronin effect with lighter particles
\cite{Cronin}.  

Adding intrinsic charm reverses the trend and typically lowers the maximum of
peak due to enhanced $k_T$ broadening.  As was the case for the rapidity
distributions in
Fig.~\ref{pPb_yrats}, when intrinsic charm comes to dominate the
distribution, $R_{p{\rm Pb}}(p_T)$ reaches a minimum of $A^{\beta-1} = 0.213$ for
the ratio $p+{\rm Pb}$ to $p+p$.
That minimum is reached at increasingly higher $p_T$ for
higher energy reactions due to the larger perturbative QCD contribution to the
cross sections.  In the relatively low $p_T$ range of
Figs.~\ref{pPb_pTrats_fory}
and \ref{pPb_pTrats_bcky} compared to collider energies,
this minimum is reached only for $p_{\rm lab} = 40$,
158, and 800~GeV.  When enhanced $k_T$ broadening is included,
the perturbative QCD calculations extend to higher $p_T$, see
Fig.~\ref{pp_pPb_ptdists}, so that $R_{p{\rm Pb}}$ reaches its
minimum at increasingly higher $p_T$ as $\sqrt{s_{NN}}$ increases.
The value of $p_T$ where the
$R_{p{\rm Pb}}$ is minimized for $J/\psi$ does not depend on the absorption
cross section because this does not change the shape of the $p_T$ distribution.

Increasing the probability of intrinsic charm in the proton results in the
minimum of $R_{p {\rm Pb}}$ being achieved at lower $p_T$, compare {\it e.g.}
the results for $J/\psi$ in Fig.~\ref{pPb_pTrats_fory}(c) and (e).  Similar
results are found for $\overline D$ at forward rapidity in (b) and (f).
Because the nuclear dependence of the intrinsic charm contribution is assumed
to be the same at forward and backward rapidity, the behavior at large $p_T$
including intrinsic charm is the same for forward and backward rapidity, as
can be seen by comparing Fig.~\ref{pPb_pTrats_bcky}(c)-(f) with the same
calculations in Fig.~\ref{pPb_pTrats_fory}(c)-(f).

The calculations here as a function of $p_T$, as well as those as a function of
rapidity, are rather idealized because the only data that exist for
$R_{p{\rm Pb}}$ are for $\sqrt{s_{NN}} = 5$~TeV where the contribution from
intrinsic charm is negligible.  Calculations like these, without intrinsic
charm, have been compared to data from the LHC in
Refs.~\cite{RV_azi1,LHC_comp,LHC8TeV_pred} and have been shown to be in
relatively good agreement as a function of $y$ and $p_T$ although the
suppression in $R_{p{\rm Pb}}$ is underestimated at forward rapidity.

Using the same target here makes trends with energy more visible.
However, data exist for other nuclear targets at lower energies.  Calculations
for $p+ {\rm Au}$ collisions at RHIC agree well with the data \cite{Darren}
and calculations were recently compared favorably
to the E866 data \cite{e866} as a function of
$x_F$ and $p_T$ in different $x_F$ regions in Ref.~\cite{RV_SeaQuest}.
Collecting high statistics data at low center of mass energies to as high
$p_T$ as possible and binned into as many bins as feasible would test
these results.  The E866 data
were only available for $p_T$ up to $\sim 3$~GeV and, while the calculations
agreed well with the trend of the data in Ref.~\cite{RV_SeaQuest}, both as a
function of $x_F$ and $p_T$, more data at different energies, particularly as
a function of $p_T$ would be a stronger test of the calculations shown here.

\section{Conclusions}
\label{conclusions}

This work combines cold nuclear matter calculations in perturbative QCD with
intrinsic charm in the nucleon.  The rapidity and $p_T$ dependence of intrinsic
charm is explored in detail as a function of collision energy.  The $p_T$
dependence of intrinsic charm is shown to be very sensitive to the finite
rapidity acceptance of a real detector because the intrinsic charm rapidity
distribution is boosted along the beam direction.
The $p+p$ and $p+{\rm Pb}$ rapidity and $p_T$
distributions for the combined model are investigated over a wide range of
energies and trends are shown.

Comparison to previous fixed-target data as a function of $x_F$ shows some
preference for a contribution from intrinsic charm of up to 1\%, compatible
with recent forward results from LHCb \cite{LHCb_intc} on $Z + c$-jets relative
to $Z+{\rm jets}$ and the recent NNPDF global analysis \cite{NNPDF4}.
These earlier fixed-target data were, however,
all taken at higher energies than the proposed NA60+ energies of
$p_{\rm lab} = 40$, 80, and 120~GeV.
The boosted intrinsic charm rapidity distributions for $J/\psi$ and
$\overline D$ production suggest that lower energy, fixed-target experiments
such as NA60+ would provide the best laboratory for determining the presence of
intrinsic charm near midrapidity.  High statistics data as a function of
rapidity and $p_T$ at these energies
could be a significant test of the importance of intrinsic charm.

{\bf Acknowledgments}
I would like to thank R. Arnaldi, V. Cheung and E. Scomparin
for helpful discussions. This work was supported by the Office of Nuclear
Physics in the US Department of Energy under Contract DE-AC52-07NA27344 and
the LLNL-LDRD Program under Project No. 21-LW-034.

\end{document}